\documentclass[longauth]{aa}

\usepackage{graphicx}
\usepackage{natbib}
\usepackage{scalerel}

\usepackage[table]{xcolor}

\usepackage{txfonts}
\usepackage[pdfencoding=auto,psdextra]{hyperref}
\hypersetup{
    colorlinks=true,
    linkcolor=blue,
    filecolor=magenta,      
    urlcolor=blue,
    citecolor=blue
}
\makeatletter
\renewcommand*\aa@pageof{, page \thepage{} of \pageref*{LastPage}}
\makeatother

\usepackage[utf8]{inputenc}

\usepackage[switch, modulo]{lineno}

\usepackage{orcidlink} 
\newcommand{\orcid}[1]{\orcidlink{#1}}

\usepackage{lastpage}

\usepackage{euclid}

\begin{document}
%
%

\title{\Euclid: Early Release Observations -- Globular clusters in the Fornax galaxy cluster, from dwarf galaxies to the intracluster field\thanks{This paper is published on
       behalf of the Euclid Consortium}}

\author{T.~Saifollahi\orcid{0000-0002-9554-7660}\thanks{\email{Teymoor.saifollahi@astro.unistra.fr}}\inst{\ref{aff1},\ref{aff2}}
\and K.~Voggel\orcid{0000-0001-6215-0950}\inst{\ref{aff3}}
\and A.~Lan\c{c}on\orcid{0000-0002-7214-8296}\inst{\ref{aff1}}
\and Michele~Cantiello\orcid{0000-0003-2072-384X}\inst{\ref{aff4}}
\and M.~A.~Raj\orcid{0000-0002-8374-0340}\inst{\ref{aff2}}
\and J.-C.~Cuillandre\orcid{0000-0002-3263-8645}\inst{\ref{aff5}}
\and S.~S.~Larsen\orcid{0000-0003-0069-1203}\inst{\ref{aff6}}
\and F.~R.~Marleau\orcid{0000-0002-1442-2947}\inst{\ref{aff7}}
\and A.~Venhola\orcid{0000-0001-6071-4564}\inst{\ref{aff8}}
\and M.~Schirmer\orcid{0000-0003-2568-9994}\inst{\ref{aff9}}
\and D.~Carollo\orcid{0000-0002-0005-5787}\inst{\ref{aff10}}
\and P.-A.~Duc\orcid{0000-0003-3343-6284}\inst{\ref{aff3}}
\and A.~M.~N.~Ferguson\inst{\ref{aff11}}
\and L.~K.~Hunt\orcid{0000-0001-9162-2371}\inst{\ref{aff12}}
\and M.~K\"ummel\orcid{0000-0003-2791-2117}\inst{\ref{aff13}}
\and R.~Laureijs\inst{\ref{aff14}}
\and O.~Marchal\inst{\ref{aff1}}
\and A.~A.~Nucita\inst{\ref{aff15},\ref{aff16},\ref{aff17}}
\and R.~F.~Peletier\orcid{0000-0001-7621-947X}\inst{\ref{aff2}}
\and M.~Poulain\orcid{0000-0002-7664-4510}\inst{\ref{aff8}}
\and M.~Rejkuba\orcid{0000-0002-6577-2787}\inst{\ref{aff18}}
\and R.~S\'anchez-Janssen\orcid{0000-0003-4945-0056}\inst{\ref{aff19}}
\and M.~Urbano\orcid{0000-0001-5640-0650}\inst{\ref{aff1}}
\and Abdurro'uf\orcid{0000-0002-5258-8761}\inst{\ref{aff20}}
\and B.~Altieri\orcid{0000-0003-3936-0284}\inst{\ref{aff21}}
\and M.~Baes\orcid{0000-0002-3930-2757}\inst{\ref{aff22}}
\and M.~Bolzonella\orcid{0000-0003-3278-4607}\inst{\ref{aff23}}
\and C.~J.~Conselice\orcid{0000-0003-1949-7638}\inst{\ref{aff24}}
\and P.~Cote\orcid{0000-0003-1184-8114}\inst{\ref{aff25}}
\and P.~Dimauro\orcid{0000-0001-7399-2854}\inst{\ref{aff26},\ref{aff27}}
\and A.~H.~Gonzalez\orcid{0000-0002-0933-8601}\inst{\ref{aff28}}
\and R.~Habas\orcid{0000-0002-4033-3841}\inst{\ref{aff4}}
\and P.~Hudelot\inst{\ref{aff29}}
\and M.~Kluge\orcid{0000-0002-9618-2552}\inst{\ref{aff30}}
\and P.~Lonare\orcid{0009-0000-0028-0493}\inst{\ref{aff31},\ref{aff4}}
\and D.~Massari\orcid{0000-0001-8892-4301}\inst{\ref{aff23}}
\and E.~Romelli\orcid{0000-0003-3069-9222}\inst{\ref{aff10}}
\and R.~Scaramella\orcid{0000-0003-2229-193X}\inst{\ref{aff26},\ref{aff32}}
\and E.~Sola\orcid{0000-0002-2814-3578}\inst{\ref{aff33}}
\and C.~Stone\orcid{0000-0002-9086-6398}\inst{\ref{aff34}}
\and C.~Tortora\orcid{0000-0001-7958-6531}\inst{\ref{aff35}}
\and S.~E.~van~Mierlo\orcid{0000-0001-8289-2863}\inst{\ref{aff2}}
\and J.~H.~Knapen\orcid{0000-0003-1643-0024}\inst{\ref{aff36},\ref{aff37}}
\and J.~Mart\'{i}n-Fleitas\orcid{0000-0002-8594-569X}\inst{\ref{aff38}}
\and A.~Mora\orcid{0000-0002-1922-8529}\inst{\ref{aff38}}
\and J.~Rom\'an\orcid{0000-0002-3849-3467}\inst{\ref{aff37},\ref{aff36}}
\and N.~Aghanim\inst{\ref{aff39}}
\and A.~Amara\inst{\ref{aff40}}
\and S.~Andreon\orcid{0000-0002-2041-8784}\inst{\ref{aff41}}
\and N.~Auricchio\orcid{0000-0003-4444-8651}\inst{\ref{aff23}}
\and M.~Baldi\orcid{0000-0003-4145-1943}\inst{\ref{aff42},\ref{aff23},\ref{aff43}}
\and A.~Balestra\orcid{0000-0002-6967-261X}\inst{\ref{aff44}}
\and S.~Bardelli\orcid{0000-0002-8900-0298}\inst{\ref{aff23}}
\and A.~Basset\inst{\ref{aff45}}
\and R.~Bender\orcid{0000-0001-7179-0626}\inst{\ref{aff30},\ref{aff13}}
\and D.~Bonino\orcid{0000-0002-3336-9977}\inst{\ref{aff46}}
\and E.~Branchini\orcid{0000-0002-0808-6908}\inst{\ref{aff47},\ref{aff48},\ref{aff41}}
\and M.~Brescia\orcid{0000-0001-9506-5680}\inst{\ref{aff49},\ref{aff35},\ref{aff50}}
\and J.~Brinchmann\orcid{0000-0003-4359-8797}\inst{\ref{aff51}}
\and S.~Camera\orcid{0000-0003-3399-3574}\inst{\ref{aff52},\ref{aff53},\ref{aff46}}
\and V.~Capobianco\orcid{0000-0002-3309-7692}\inst{\ref{aff46}}
\and C.~Carbone\orcid{0000-0003-0125-3563}\inst{\ref{aff54}}
\and J.~Carretero\orcid{0000-0002-3130-0204}\inst{\ref{aff55},\ref{aff56}}
\and S.~Casas\orcid{0000-0002-4751-5138}\inst{\ref{aff57}}
\and M.~Castellano\orcid{0000-0001-9875-8263}\inst{\ref{aff26}}
\and S.~Cavuoti\orcid{0000-0002-3787-4196}\inst{\ref{aff35},\ref{aff50}}
\and A.~Cimatti\inst{\ref{aff58}}
\and G.~Congedo\orcid{0000-0003-2508-0046}\inst{\ref{aff11}}
\and L.~Conversi\orcid{0000-0002-6710-8476}\inst{\ref{aff59},\ref{aff21}}
\and Y.~Copin\orcid{0000-0002-5317-7518}\inst{\ref{aff60}}
\and F.~Courbin\orcid{0000-0003-0758-6510}\inst{\ref{aff61}}
\and H.~M.~Courtois\orcid{0000-0003-0509-1776}\inst{\ref{aff62}}
\and M.~Cropper\orcid{0000-0003-4571-9468}\inst{\ref{aff63}}
\and A.~Da~Silva\orcid{0000-0002-6385-1609}\inst{\ref{aff64},\ref{aff65}}
\and H.~Degaudenzi\orcid{0000-0002-5887-6799}\inst{\ref{aff66}}
\and A.~M.~Di~Giorgio\orcid{0000-0002-4767-2360}\inst{\ref{aff67}}
\and J.~Dinis\orcid{0000-0001-5075-1601}\inst{\ref{aff64},\ref{aff65}}
\and F.~Dubath\orcid{0000-0002-6533-2810}\inst{\ref{aff66}}
\and X.~Dupac\inst{\ref{aff21}}
\and S.~Dusini\orcid{0000-0002-1128-0664}\inst{\ref{aff68}}
\and M.~Fabricius\orcid{0000-0002-7025-6058}\inst{\ref{aff30},\ref{aff13}}
\and M.~Farina\orcid{0000-0002-3089-7846}\inst{\ref{aff67}}
\and S.~Farrens\orcid{0000-0002-9594-9387}\inst{\ref{aff5}}
\and S.~Ferriol\inst{\ref{aff60}}
\and P.~Fosalba\orcid{0000-0002-1510-5214}\inst{\ref{aff69},\ref{aff70}}
\and M.~Frailis\orcid{0000-0002-7400-2135}\inst{\ref{aff10}}
\and E.~Franceschi\orcid{0000-0002-0585-6591}\inst{\ref{aff23}}
\and M.~Fumana\orcid{0000-0001-6787-5950}\inst{\ref{aff54}}
\and S.~Galeotta\orcid{0000-0002-3748-5115}\inst{\ref{aff10}}
\and B.~Garilli\orcid{0000-0001-7455-8750}\inst{\ref{aff54}}
\and W.~Gillard\orcid{0000-0003-4744-9748}\inst{\ref{aff71}}
\and B.~Gillis\orcid{0000-0002-4478-1270}\inst{\ref{aff11}}
\and C.~Giocoli\orcid{0000-0002-9590-7961}\inst{\ref{aff23},\ref{aff72}}
\and P.~G\'omez-Alvarez\orcid{0000-0002-8594-5358}\inst{\ref{aff73},\ref{aff21}}
\and B.~R.~Granett\orcid{0000-0003-2694-9284}\inst{\ref{aff41}}
\and A.~Grazian\orcid{0000-0002-5688-0663}\inst{\ref{aff44}}
\and F.~Grupp\inst{\ref{aff30},\ref{aff13}}
\and L.~Guzzo\orcid{0000-0001-8264-5192}\inst{\ref{aff74},\ref{aff41}}
\and S.~V.~H.~Haugan\orcid{0000-0001-9648-7260}\inst{\ref{aff75}}
\and J.~Hoar\inst{\ref{aff21}}
\and H.~Hoekstra\orcid{0000-0002-0641-3231}\inst{\ref{aff76}}
\and W.~Holmes\inst{\ref{aff77}}
\and I.~Hook\orcid{0000-0002-2960-978X}\inst{\ref{aff78}}
\and F.~Hormuth\inst{\ref{aff79}}
\and A.~Hornstrup\orcid{0000-0002-3363-0936}\inst{\ref{aff80},\ref{aff81}}
\and K.~Jahnke\orcid{0000-0003-3804-2137}\inst{\ref{aff9}}
\and M.~Jhabvala\inst{\ref{aff82}}
\and E.~Keih\"anen\orcid{0000-0003-1804-7715}\inst{\ref{aff83}}
\and S.~Kermiche\orcid{0000-0002-0302-5735}\inst{\ref{aff71}}
\and A.~Kiessling\orcid{0000-0002-2590-1273}\inst{\ref{aff77}}
\and T.~Kitching\orcid{0000-0002-4061-4598}\inst{\ref{aff63}}
\and R.~Kohley\inst{\ref{aff21}}
\and B.~Kubik\orcid{0009-0006-5823-4880}\inst{\ref{aff60}}
\and K.~Kuijken\orcid{0000-0002-3827-0175}\inst{\ref{aff76}}
\and M.~Kunz\orcid{0000-0002-3052-7394}\inst{\ref{aff84}}
\and H.~Kurki-Suonio\orcid{0000-0002-4618-3063}\inst{\ref{aff85},\ref{aff86}}
\and O.~Lahav\orcid{0000-0002-1134-9035}\inst{\ref{aff87}}
\and D.~Le~Mignant\orcid{0000-0002-5339-5515}\inst{\ref{aff88}}
\and S.~Ligori\orcid{0000-0003-4172-4606}\inst{\ref{aff46}}
\and P.~B.~Lilje\orcid{0000-0003-4324-7794}\inst{\ref{aff75}}
\and V.~Lindholm\orcid{0000-0003-2317-5471}\inst{\ref{aff85},\ref{aff86}}
\and I.~Lloro\inst{\ref{aff89}}
\and D.~Maino\inst{\ref{aff74},\ref{aff54},\ref{aff90}}
\and E.~Maiorano\orcid{0000-0003-2593-4355}\inst{\ref{aff23}}
\and O.~Mansutti\orcid{0000-0001-5758-4658}\inst{\ref{aff10}}
\and O.~Marggraf\orcid{0000-0001-7242-3852}\inst{\ref{aff91}}
\and K.~Markovic\orcid{0000-0001-6764-073X}\inst{\ref{aff77}}
\and N.~Martinet\orcid{0000-0003-2786-7790}\inst{\ref{aff88}}
\and F.~Marulli\orcid{0000-0002-8850-0303}\inst{\ref{aff92},\ref{aff23},\ref{aff43}}
\and R.~Massey\orcid{0000-0002-6085-3780}\inst{\ref{aff93}}
\and S.~Maurogordato\inst{\ref{aff94}}
\and H.~J.~McCracken\orcid{0000-0002-9489-7765}\inst{\ref{aff29}}
\and E.~Medinaceli\orcid{0000-0002-4040-7783}\inst{\ref{aff23}}
\and S.~Mei\orcid{0000-0002-2849-559X}\inst{\ref{aff95}}
\and M.~Melchior\inst{\ref{aff96}}
\and Y.~Mellier\inst{\ref{aff97},\ref{aff29}}
\and M.~Meneghetti\orcid{0000-0003-1225-7084}\inst{\ref{aff23},\ref{aff43}}
\and G.~Meylan\inst{\ref{aff61}}
\and M.~Moresco\orcid{0000-0002-7616-7136}\inst{\ref{aff92},\ref{aff23}}
\and L.~Moscardini\orcid{0000-0002-3473-6716}\inst{\ref{aff92},\ref{aff23},\ref{aff43}}
\and E.~Munari\orcid{0000-0002-1751-5946}\inst{\ref{aff10},\ref{aff98}}
\and R.~Nakajima\inst{\ref{aff91}}
\and R.~C.~Nichol\inst{\ref{aff40}}
\and S.-M.~Niemi\inst{\ref{aff14}}
\and C.~Padilla\orcid{0000-0001-7951-0166}\inst{\ref{aff99}}
\and S.~Paltani\orcid{0000-0002-8108-9179}\inst{\ref{aff66}}
\and F.~Pasian\orcid{0000-0002-4869-3227}\inst{\ref{aff10}}
\and K.~Pedersen\inst{\ref{aff100}}
\and W.~J.~Percival\orcid{0000-0002-0644-5727}\inst{\ref{aff101},\ref{aff102},\ref{aff103}}
\and V.~Pettorino\inst{\ref{aff14}}
\and S.~Pires\orcid{0000-0002-0249-2104}\inst{\ref{aff5}}
\and G.~Polenta\orcid{0000-0003-4067-9196}\inst{\ref{aff104}}
\and M.~Poncet\inst{\ref{aff45}}
\and L.~A.~Popa\inst{\ref{aff105}}
\and L.~Pozzetti\orcid{0000-0001-7085-0412}\inst{\ref{aff23}}
\and G.~D.~Racca\inst{\ref{aff14}}
\and F.~Raison\orcid{0000-0002-7819-6918}\inst{\ref{aff30}}
\and R.~Rebolo\inst{\ref{aff36},\ref{aff37}}
\and A.~Refregier\inst{\ref{aff106}}
\and A.~Renzi\orcid{0000-0001-9856-1970}\inst{\ref{aff107},\ref{aff68}}
\and J.~Rhodes\orcid{0000-0002-4485-8549}\inst{\ref{aff77}}
\and G.~Riccio\inst{\ref{aff35}}
\and M.~Roncarelli\orcid{0000-0001-9587-7822}\inst{\ref{aff23}}
\and E.~Rossetti\orcid{0000-0003-0238-4047}\inst{\ref{aff42}}
\and R.~Saglia\orcid{0000-0003-0378-7032}\inst{\ref{aff13},\ref{aff30}}
\and D.~Sapone\orcid{0000-0001-7089-4503}\inst{\ref{aff108}}
\and B.~Sartoris\orcid{0000-0003-1337-5269}\inst{\ref{aff13},\ref{aff10}}
\and P.~Schneider\orcid{0000-0001-8561-2679}\inst{\ref{aff91}}
\and T.~Schrabback\orcid{0000-0002-6987-7834}\inst{\ref{aff7}}
\and A.~Secroun\orcid{0000-0003-0505-3710}\inst{\ref{aff71}}
\and G.~Seidel\orcid{0000-0003-2907-353X}\inst{\ref{aff9}}
\and S.~Serrano\orcid{0000-0002-0211-2861}\inst{\ref{aff69},\ref{aff109},\ref{aff110}}
\and C.~Sirignano\orcid{0000-0002-0995-7146}\inst{\ref{aff107},\ref{aff68}}
\and G.~Sirri\orcid{0000-0003-2626-2853}\inst{\ref{aff43}}
\and L.~Stanco\orcid{0000-0002-9706-5104}\inst{\ref{aff68}}
\and P.~Tallada-Cresp\'{i}\orcid{0000-0002-1336-8328}\inst{\ref{aff55},\ref{aff56}}
\and A.~N.~Taylor\inst{\ref{aff11}}
\and H.~I.~Teplitz\orcid{0000-0002-7064-5424}\inst{\ref{aff111}}
\and I.~Tereno\inst{\ref{aff64},\ref{aff112}}
\and R.~Toledo-Moreo\orcid{0000-0002-2997-4859}\inst{\ref{aff113}}
\and F.~Torradeflot\orcid{0000-0003-1160-1517}\inst{\ref{aff56},\ref{aff55}}
\and A.~Tsyganov\inst{\ref{aff114}}
\and I.~Tutusaus\orcid{0000-0002-3199-0399}\inst{\ref{aff115}}
\and E.~A.~Valentijn\inst{\ref{aff2}}
\and L.~Valenziano\orcid{0000-0002-1170-0104}\inst{\ref{aff23},\ref{aff116}}
\and T.~Vassallo\orcid{0000-0001-6512-6358}\inst{\ref{aff13},\ref{aff10}}
\and G.~Verdoes~Kleijn\orcid{0000-0001-5803-2580}\inst{\ref{aff2}}
\and A.~Veropalumbo\orcid{0000-0003-2387-1194}\inst{\ref{aff41},\ref{aff48},\ref{aff117}}
\and Y.~Wang\orcid{0000-0002-4749-2984}\inst{\ref{aff111}}
\and J.~Weller\orcid{0000-0002-8282-2010}\inst{\ref{aff13},\ref{aff30}}
\and O.~R.~Williams\orcid{0000-0003-0274-1526}\inst{\ref{aff114}}
\and G.~Zamorani\orcid{0000-0002-2318-301X}\inst{\ref{aff23}}
\and E.~Zucca\orcid{0000-0002-5845-8132}\inst{\ref{aff23}}
\and A.~Biviano\orcid{0000-0002-0857-0732}\inst{\ref{aff10},\ref{aff98}}
\and C.~Burigana\orcid{0000-0002-3005-5796}\inst{\ref{aff118},\ref{aff116}}
\and V.~Scottez\inst{\ref{aff97},\ref{aff119}}
\and P.~Simon\inst{\ref{aff91}}
\and M.~Balogh\orcid{0000-0003-4849-9536}\inst{\ref{aff102},\ref{aff101}}
\and D.~Scott\orcid{0000-0002-6878-9840}\inst{\ref{aff120}}}
										   
\institute{Observatoire Astronomique de Strasbourg (ObAS), Universit\'e de Strasbourg - CNRS, UMR 7550, Strasbourg, France\label{aff1}
\and
Kapteyn Astronomical Institute, University of Groningen, PO Box 800, 9700 AV Groningen, The Netherlands\label{aff2}
\and
Universit\'e de Strasbourg, CNRS, Observatoire astronomique de Strasbourg, UMR 7550, 67000 Strasbourg, France\label{aff3}
\and
INAF - Osservatorio Astronomico d'Abruzzo, Via Maggini, 64100, Teramo, Italy\label{aff4}
\and
Universit\'e Paris-Saclay, Universit\'e Paris Cit\'e, CEA, CNRS, AIM, 91191, Gif-sur-Yvette, France\label{aff5}
\and
Department of Astrophysics/IMAPP, Radboud University, PO Box 9010, 6500 GL Nijmegen, The Netherlands\label{aff6}
\and
Universit\"at Innsbruck, Institut f\"ur Astro- und Teilchenphysik, Technikerstr. 25/8, 6020 Innsbruck, Austria\label{aff7}
\and
Space physics and astronomy research unit, University of Oulu, Pentti Kaiteran katu 1, FI-90014 Oulu, Finland\label{aff8}
\and
Max-Planck-Institut f\"ur Astronomie, K\"onigstuhl 17, 69117 Heidelberg, Germany\label{aff9}
\and
INAF-Osservatorio Astronomico di Trieste, Via G. B. Tiepolo 11, 34143 Trieste, Italy\label{aff10}
\and
Institute for Astronomy, University of Edinburgh, Royal Observatory, Blackford Hill, Edinburgh EH9 3HJ, UK\label{aff11}
\and
INAF-Osservatorio Astrofisico di Arcetri, Largo E. Fermi 5, 50125, Firenze, Italy\label{aff12}
\and
Universit\"ats-Sternwarte M\"unchen, Fakult\"at f\"ur Physik, Ludwig-Maximilians-Universit\"at M\"unchen, Scheinerstrasse 1, 81679 M\"unchen, Germany\label{aff13}
\and
European Space Agency/ESTEC, Keplerlaan 1, 2201 AZ Noordwijk, The Netherlands\label{aff14}
\and
Department of Mathematics and Physics E. De Giorgi, University of Salento, Via per Arnesano, CP-I93, 73100, Lecce, Italy\label{aff15}
\and
INAF-Sezione di Lecce, c/o Dipartimento Matematica e Fisica, Via per Arnesano, 73100, Lecce, Italy\label{aff16}
\and
INFN, Sezione di Lecce, Via per Arnesano, CP-193, 73100, Lecce, Italy\label{aff17}
\and
European Southern Observatory, Karl-Schwarzschild Str. 2, 85748 Garching, Germany\label{aff18}
\and
UK Astronomy Technology Centre, Royal Observatory, Blackford Hill, Edinburgh EH9 3HJ, UK\label{aff19}
\and
Johns Hopkins University 3400 North Charles Street Baltimore, MD 21218, USA\label{aff20}
\and
ESAC/ESA, Camino Bajo del Castillo, s/n., Urb. Villafranca del Castillo, 28692 Villanueva de la Ca\~nada, Madrid, Spain\label{aff21}
\and
Sterrenkundig Observatorium, Universiteit Gent, Krijgslaan 281 S9, 9000 Gent, Belgium\label{aff22}
\and
INAF-Osservatorio di Astrofisica e Scienza dello Spazio di Bologna, Via Piero Gobetti 93/3, 40129 Bologna, Italy\label{aff23}
\and
Jodrell Bank Centre for Astrophysics, Department of Physics and Astronomy, University of Manchester, Oxford Road, Manchester M13 9PL, UK\label{aff24}
\and
NRC Herzberg, 5071 West Saanich Rd, Victoria, BC V9E 2E7, Canada\label{aff25}
\and
INAF-Osservatorio Astronomico di Roma, Via Frascati 33, 00078 Monteporzio Catone, Italy\label{aff26}
\and
Observatorio Nacional, Rua General Jose Cristino, 77-Bairro Imperial de Sao Cristovao, Rio de Janeiro, 20921-400, Brazil\label{aff27}
\and
Department of Astronomy, University of Florida, Bryant Space Science Center, Gainesville, FL 32611, USA\label{aff28}
\and
Institut d'Astrophysique de Paris, UMR 7095, CNRS, and Sorbonne Universit\'e, 98 bis boulevard Arago, 75014 Paris, France\label{aff29}
\and
Max Planck Institute for Extraterrestrial Physics, Giessenbachstr. 1, 85748 Garching, Germany\label{aff30}
\and
Dipartimento di Fisica, Universit\`a di Roma Tor Vergata, Via della Ricerca Scientifica 1, Roma, Italy\label{aff31}
\and
INFN-Sezione di Roma, Piazzale Aldo Moro, 2 - c/o Dipartimento di Fisica, Edificio G. Marconi, 00185 Roma, Italy\label{aff32}
\and
Institute of Astronomy, University of Cambridge, Madingley Road, Cambridge CB3 0HA, UK\label{aff33}
\and
Department of Physics, Universit\'{e} de Montr\'{e}al, 2900 Edouard Montpetit Blvd, Montr\'{e}al, Qu\'{e}bec H3T 1J4, Canada\label{aff34}
\and
INAF-Osservatorio Astronomico di Capodimonte, Via Moiariello 16, 80131 Napoli, Italy\label{aff35}
\and
Instituto de Astrof\'isica de Canarias, Calle V\'ia L\'actea s/n, 38204, San Crist\'obal de La Laguna, Tenerife, Spain\label{aff36}
\and
Departamento de Astrof\'isica, Universidad de La Laguna, 38206, La Laguna, Tenerife, Spain\label{aff37}
\and
Aurora Technology for European Space Agency (ESA), Camino bajo del Castillo, s/n, Urbanizacion Villafranca del Castillo, Villanueva de la Ca\~nada, 28692 Madrid, Spain\label{aff38}
\and
Universit\'e Paris-Saclay, CNRS, Institut d'astrophysique spatiale, 91405, Orsay, France\label{aff39}
\and
School of Mathematics and Physics, University of Surrey, Guildford, Surrey, GU2 7XH, UK\label{aff40}
\and
INAF-Osservatorio Astronomico di Brera, Via Brera 28, 20122 Milano, Italy\label{aff41}
\and
Dipartimento di Fisica e Astronomia, Universit\`a di Bologna, Via Gobetti 93/2, 40129 Bologna, Italy\label{aff42}
\and
INFN-Sezione di Bologna, Viale Berti Pichat 6/2, 40127 Bologna, Italy\label{aff43}
\and
INAF-Osservatorio Astronomico di Padova, Via dell'Osservatorio 5, 35122 Padova, Italy\label{aff44}
\and
Centre National d'Etudes Spatiales -- Centre spatial de Toulouse, 18 avenue Edouard Belin, 31401 Toulouse Cedex 9, France\label{aff45}
\and
INAF-Osservatorio Astrofisico di Torino, Via Osservatorio 20, 10025 Pino Torinese (TO), Italy\label{aff46}
\and
Dipartimento di Fisica, Universit\`a di Genova, Via Dodecaneso 33, 16146, Genova, Italy\label{aff47}
\and
INFN-Sezione di Genova, Via Dodecaneso 33, 16146, Genova, Italy\label{aff48}
\and
Department of Physics "E. Pancini", University Federico II, Via Cinthia 6, 80126, Napoli, Italy\label{aff49}
\and
INFN section of Naples, Via Cinthia 6, 80126, Napoli, Italy\label{aff50}
\and
Instituto de Astrof\'isica e Ci\^encias do Espa\c{c}o, Universidade do Porto, CAUP, Rua das Estrelas, PT4150-762 Porto, Portugal\label{aff51}
\and
Dipartimento di Fisica, Universit\`a degli Studi di Torino, Via P. Giuria 1, 10125 Torino, Italy\label{aff52}
\and
INFN-Sezione di Torino, Via P. Giuria 1, 10125 Torino, Italy\label{aff53}
\and
INAF-IASF Milano, Via Alfonso Corti 12, 20133 Milano, Italy\label{aff54}
\and
Centro de Investigaciones Energ\'eticas, Medioambientales y Tecnol\'ogicas (CIEMAT), Avenida Complutense 40, 28040 Madrid, Spain\label{aff55}
\and
Port d'Informaci\'{o} Cient\'{i}fica, Campus UAB, C. Albareda s/n, 08193 Bellaterra (Barcelona), Spain\label{aff56}
\and
Institute for Theoretical Particle Physics and Cosmology (TTK), RWTH Aachen University, 52056 Aachen, Germany\label{aff57}
\and
Dipartimento di Fisica e Astronomia "Augusto Righi" - Alma Mater Studiorum Universit\`a di Bologna, Viale Berti Pichat 6/2, 40127 Bologna, Italy\label{aff58}
\and
European Space Agency/ESRIN, Largo Galileo Galilei 1, 00044 Frascati, Roma, Italy\label{aff59}
\and
Universit\'e Claude Bernard Lyon 1, CNRS/IN2P3, IP2I Lyon, UMR 5822, Villeurbanne, F-69100, France\label{aff60}
\and
Institute of Physics, Laboratory of Astrophysics, Ecole Polytechnique F\'ed\'erale de Lausanne (EPFL), Observatoire de Sauverny, 1290 Versoix, Switzerland\label{aff61}
\and
UCB Lyon 1, CNRS/IN2P3, IUF, IP2I Lyon, 4 rue Enrico Fermi, 69622 Villeurbanne, France\label{aff62}
\and
Mullard Space Science Laboratory, University College London, Holmbury St Mary, Dorking, Surrey RH5 6NT, UK\label{aff63}
\and
Departamento de F\'isica, Faculdade de Ci\^encias, Universidade de Lisboa, Edif\'icio C8, Campo Grande, PT1749-016 Lisboa, Portugal\label{aff64}
\and
Instituto de Astrof\'isica e Ci\^encias do Espa\c{c}o, Faculdade de Ci\^encias, Universidade de Lisboa, Campo Grande, 1749-016 Lisboa, Portugal\label{aff65}
\and
Department of Astronomy, University of Geneva, ch. d'Ecogia 16, 1290 Versoix, Switzerland\label{aff66}
\and
INAF-Istituto di Astrofisica e Planetologia Spaziali, via del Fosso del Cavaliere, 100, 00100 Roma, Italy\label{aff67}
\and
INFN-Padova, Via Marzolo 8, 35131 Padova, Italy\label{aff68}
\and
Institut d'Estudis Espacials de Catalunya (IEEC),  Edifici RDIT, Campus UPC, 08860 Castelldefels, Barcelona, Spain\label{aff69}
\and
Institut de Ciencies de l'Espai (IEEC-CSIC), Campus UAB, Carrer de Can Magrans, s/n Cerdanyola del Vall\'es, 08193 Barcelona, Spain\label{aff70}
\and
Aix-Marseille Universit\'e, CNRS/IN2P3, CPPM, Marseille, France\label{aff71}
\and
Istituto Nazionale di Fisica Nucleare, Sezione di Bologna, Via Irnerio 46, 40126 Bologna, Italy\label{aff72}
\and
FRACTAL S.L.N.E., calle Tulip\'an 2, Portal 13 1A, 28231, Las Rozas de Madrid, Spain\label{aff73}
\and
Dipartimento di Fisica "Aldo Pontremoli", Universit\`a degli Studi di Milano, Via Celoria 16, 20133 Milano, Italy\label{aff74}
\and
Institute of Theoretical Astrophysics, University of Oslo, P.O. Box 1029 Blindern, 0315 Oslo, Norway\label{aff75}
\and
Leiden Observatory, Leiden University, Einsteinweg 55, 2333 CC Leiden, The Netherlands\label{aff76}
\and
Jet Propulsion Laboratory, California Institute of Technology, 4800 Oak Grove Drive, Pasadena, CA, 91109, USA\label{aff77}
\and
Department of Physics, Lancaster University, Lancaster, LA1 4YB, UK\label{aff78}
\and
Felix Hormuth Engineering, Goethestr. 17, 69181 Leimen, Germany\label{aff79}
\and
Technical University of Denmark, Elektrovej 327, 2800 Kgs. Lyngby, Denmark\label{aff80}
\and
Cosmic Dawn Center (DAWN), Denmark\label{aff81}
\and
NASA Goddard Space Flight Center, Greenbelt, MD 20771, USA\label{aff82}
\and
Department of Physics and Helsinki Institute of Physics, Gustaf H\"allstr\"omin katu 2, 00014 University of Helsinki, Finland\label{aff83}
\and
Universit\'e de Gen\`eve, D\'epartement de Physique Th\'eorique and Centre for Astroparticle Physics, 24 quai Ernest-Ansermet, CH-1211 Gen\`eve 4, Switzerland\label{aff84}
\and
Department of Physics, P.O. Box 64, 00014 University of Helsinki, Finland\label{aff85}
\and
Helsinki Institute of Physics, Gustaf H{\"a}llstr{\"o}min katu 2, University of Helsinki, Helsinki, Finland\label{aff86}
\and
Department of Physics and Astronomy, University College London, Gower Street, London WC1E 6BT, UK\label{aff87}
\and
Aix-Marseille Universit\'e, CNRS, CNES, LAM, Marseille, France\label{aff88}
\and
NOVA optical infrared instrumentation group at ASTRON, Oude Hoogeveensedijk 4, 7991PD, Dwingeloo, The Netherlands\label{aff89}
\and
INFN-Sezione di Milano, Via Celoria 16, 20133 Milano, Italy\label{aff90}
\and
Universit\"at Bonn, Argelander-Institut f\"ur Astronomie, Auf dem H\"ugel 71, 53121 Bonn, Germany\label{aff91}
\and
Dipartimento di Fisica e Astronomia "Augusto Righi" - Alma Mater Studiorum Universit\`a di Bologna, via Piero Gobetti 93/2, 40129 Bologna, Italy\label{aff92}
\and
Department of Physics, Institute for Computational Cosmology, Durham University, South Road, DH1 3LE, UK\label{aff93}
\and
Universit\'e C\^{o}te d'Azur, Observatoire de la C\^{o}te d'Azur, CNRS, Laboratoire Lagrange, Bd de l'Observatoire, CS 34229, 06304 Nice cedex 4, France\label{aff94}
\and
Universit\'e Paris Cit\'e, CNRS, Astroparticule et Cosmologie, 75013 Paris, France\label{aff95}
\and
University of Applied Sciences and Arts of Northwestern Switzerland, School of Engineering, 5210 Windisch, Switzerland\label{aff96}
\and
Institut d'Astrophysique de Paris, 98bis Boulevard Arago, 75014, Paris, France\label{aff97}
\and
IFPU, Institute for Fundamental Physics of the Universe, via Beirut 2, 34151 Trieste, Italy\label{aff98}
\and
Institut de F\'{i}sica d'Altes Energies (IFAE), The Barcelona Institute of Science and Technology, Campus UAB, 08193 Bellaterra (Barcelona), Spain\label{aff99}
\and
Department of Physics and Astronomy, University of Aarhus, Ny Munkegade 120, DK-8000 Aarhus C, Denmark\label{aff100}
\and
Waterloo Centre for Astrophysics, University of Waterloo, Waterloo, Ontario N2L 3G1, Canada\label{aff101}
\and
Department of Physics and Astronomy, University of Waterloo, Waterloo, Ontario N2L 3G1, Canada\label{aff102}
\and
Perimeter Institute for Theoretical Physics, Waterloo, Ontario N2L 2Y5, Canada\label{aff103}
\and
Space Science Data Center, Italian Space Agency, via del Politecnico snc, 00133 Roma, Italy\label{aff104}
\and
Institute of Space Science, Str. Atomistilor, nr. 409 M\u{a}gurele, Ilfov, 077125, Romania\label{aff105}
\and
Institute for Particle Physics and Astrophysics, Dept. of Physics, ETH Zurich, Wolfgang-Pauli-Strasse 27, 8093 Zurich, Switzerland\label{aff106}
\and
Dipartimento di Fisica e Astronomia "G. Galilei", Universit\`a di Padova, Via Marzolo 8, 35131 Padova, Italy\label{aff107}
\and
Departamento de F\'isica, FCFM, Universidad de Chile, Blanco Encalada 2008, Santiago, Chile\label{aff108}
\and
Satlantis, University Science Park, Sede Bld 48940, Leioa-Bilbao, Spain\label{aff109}
\and
Institute of Space Sciences (ICE, CSIC), Campus UAB, Carrer de Can Magrans, s/n, 08193 Barcelona, Spain\label{aff110}
\and
Infrared Processing and Analysis Center, California Institute of Technology, Pasadena, CA 91125, USA\label{aff111}
\and
Instituto de Astrof\'isica e Ci\^encias do Espa\c{c}o, Faculdade de Ci\^encias, Universidade de Lisboa, Tapada da Ajuda, 1349-018 Lisboa, Portugal\label{aff112}
\and
Universidad Polit\'ecnica de Cartagena, Departamento de Electr\'onica y Tecnolog\'ia de Computadoras,  Plaza del Hospital 1, 30202 Cartagena, Spain\label{aff113}
\and
Centre for Information Technology, University of Groningen, P.O. Box 11044, 9700 CA Groningen, The Netherlands\label{aff114}
\and
Institut de Recherche en Astrophysique et Plan\'etologie (IRAP), Universit\'e de Toulouse, CNRS, UPS, CNES, 14 Av. Edouard Belin, 31400 Toulouse, France\label{aff115}
\and
INFN-Bologna, Via Irnerio 46, 40126 Bologna, Italy\label{aff116}
\and
Dipartimento di Fisica, Universit\`a degli studi di Genova, and INFN-Sezione di Genova, via Dodecaneso 33, 16146, Genova, Italy\label{aff117}
\and
INAF, Istituto di Radioastronomia, Via Piero Gobetti 101, 40129 Bologna, Italy\label{aff118}
\and
Junia, EPA department, 41 Bd Vauban, 59800 Lille, France\label{aff119}
\and
Department of Physics and Astronomy, University of British Columbia, Vancouver, BC V6T 1Z1, Canada\label{aff120}}    
										   
\institute{Observatoire Astronomique de Strasbourg (ObAS), Universit\'e de Strasbourg - CNRS, UMR 7550, Strasbourg, France\label{aff1}
\and
Kapteyn Astronomical Institute, University of Groningen, PO Box 800, 9700 AV Groningen, The Netherlands\label{aff2}
\and
Universit\'e de Strasbourg, CNRS, Observatoire astronomique de Strasbourg, UMR 7550, 67000 Strasbourg, France\label{aff3}
\and
INAF - Osservatorio Astronomico d'Abruzzo, Via Maggini, 64100, Teramo, Italy\label{aff4}
\and
Universit\'e Paris-Saclay, Universit\'e Paris Cit\'e, CEA, CNRS, AIM, 91191, Gif-sur-Yvette, France\label{aff5}
\and
Department of Astrophysics/IMAPP, Radboud University, PO Box 9010, 6500 GL Nijmegen, The Netherlands\label{aff6}
\and
Universit\"at Innsbruck, Institut f\"ur Astro- und Teilchenphysik, Technikerstr. 25/8, 6020 Innsbruck, Austria\label{aff7}
\and
Space physics and astronomy research unit, University of Oulu, Pentti Kaiteran katu 1, FI-90014 Oulu, Finland\label{aff8}
\and
Max-Planck-Institut f\"ur Astronomie, K\"onigstuhl 17, 69117 Heidelberg, Germany\label{aff9}
\and
INAF-Osservatorio Astronomico di Trieste, Via G. B. Tiepolo 11, 34143 Trieste, Italy\label{aff10}
\and
Institute for Astronomy, University of Edinburgh, Royal Observatory, Blackford Hill, Edinburgh EH9 3HJ, UK\label{aff11}
\and
INAF-Osservatorio Astrofisico di Arcetri, Largo E. Fermi 5, 50125, Firenze, Italy\label{aff12}
\and
Universit\"ats-Sternwarte M\"unchen, Fakult\"at f\"ur Physik, Ludwig-Maximilians-Universit\"at M\"unchen, Scheinerstrasse 1, 81679 M\"unchen, Germany\label{aff13}
\and
European Space Agency/ESTEC, Keplerlaan 1, 2201 AZ Noordwijk, The Netherlands\label{aff14}
\and
Department of Mathematics and Physics E. De Giorgi, University of Salento, Via per Arnesano, CP-I93, 73100, Lecce, Italy\label{aff15}
\and
INAF-Sezione di Lecce, c/o Dipartimento Matematica e Fisica, Via per Arnesano, 73100, Lecce, Italy\label{aff16}
\and
INFN, Sezione di Lecce, Via per Arnesano, CP-193, 73100, Lecce, Italy\label{aff17}
\and
European Southern Observatory, Karl-Schwarzschild Str. 2, 85748 Garching, Germany\label{aff18}
\and
UK Astronomy Technology Centre, Royal Observatory, Blackford Hill, Edinburgh EH9 3HJ, UK\label{aff19}
\and
Johns Hopkins University 3400 North Charles Street Baltimore, MD 21218, USA\label{aff20}
\and
ESAC/ESA, Camino Bajo del Castillo, s/n., Urb. Villafranca del Castillo, 28692 Villanueva de la Ca\~nada, Madrid, Spain\label{aff21}
\and
Sterrenkundig Observatorium, Universiteit Gent, Krijgslaan 281 S9, 9000 Gent, Belgium\label{aff22}
\and
INAF-Osservatorio di Astrofisica e Scienza dello Spazio di Bologna, Via Piero Gobetti 93/3, 40129 Bologna, Italy\label{aff23}
\and
Jodrell Bank Centre for Astrophysics, Department of Physics and Astronomy, University of Manchester, Oxford Road, Manchester M13 9PL, UK\label{aff24}
\and
NRC Herzberg, 5071 West Saanich Rd, Victoria, BC V9E 2E7, Canada\label{aff25}
\and
INAF-Osservatorio Astronomico di Roma, Via Frascati 33, 00078 Monteporzio Catone, Italy\label{aff26}
\and
Observatorio Nacional, Rua General Jose Cristino, 77-Bairro Imperial de Sao Cristovao, Rio de Janeiro, 20921-400, Brazil\label{aff27}
\and
Department of Astronomy, University of Florida, Bryant Space Science Center, Gainesville, FL 32611, USA\label{aff28}
\and
Institut d'Astrophysique de Paris, UMR 7095, CNRS, and Sorbonne Universit\'e, 98 bis boulevard Arago, 75014 Paris, France\label{aff29}
\and
Max Planck Institute for Extraterrestrial Physics, Giessenbachstr. 1, 85748 Garching, Germany\label{aff30}
\and
Dipartimento di Fisica, Universit\`a di Roma Tor Vergata, Via della Ricerca Scientifica 1, Roma, Italy\label{aff31}
\and
INFN-Sezione di Roma, Piazzale Aldo Moro, 2 - c/o Dipartimento di Fisica, Edificio G. Marconi, 00185 Roma, Italy\label{aff32}
\and
Institute of Astronomy, University of Cambridge, Madingley Road, Cambridge CB3 0HA, UK\label{aff33}
\and
Department of Physics, Universit\'{e} de Montr\'{e}al, 2900 Edouard Montpetit Blvd, Montr\'{e}al, Qu\'{e}bec H3T 1J4, Canada\label{aff34}
\and
INAF-Osservatorio Astronomico di Capodimonte, Via Moiariello 16, 80131 Napoli, Italy\label{aff35}
\and
Instituto de Astrof\'isica de Canarias, Calle V\'ia L\'actea s/n, 38204, San Crist\'obal de La Laguna, Tenerife, Spain\label{aff36}
\and
Departamento de Astrof\'isica, Universidad de La Laguna, 38206, La Laguna, Tenerife, Spain\label{aff37}
\and
Aurora Technology for European Space Agency (ESA), Camino bajo del Castillo, s/n, Urbanizacion Villafranca del Castillo, Villanueva de la Ca\~nada, 28692 Madrid, Spain\label{aff38}
\and
Universit\'e Paris-Saclay, CNRS, Institut d'astrophysique spatiale, 91405, Orsay, France\label{aff39}
\and
School of Mathematics and Physics, University of Surrey, Guildford, Surrey, GU2 7XH, UK\label{aff40}
\and
INAF-Osservatorio Astronomico di Brera, Via Brera 28, 20122 Milano, Italy\label{aff41}
\and
Dipartimento di Fisica e Astronomia, Universit\`a di Bologna, Via Gobetti 93/2, 40129 Bologna, Italy\label{aff42}
\and
INFN-Sezione di Bologna, Viale Berti Pichat 6/2, 40127 Bologna, Italy\label{aff43}
\and
INAF-Osservatorio Astronomico di Padova, Via dell'Osservatorio 5, 35122 Padova, Italy\label{aff44}
\and
Centre National d'Etudes Spatiales -- Centre spatial de Toulouse, 18 avenue Edouard Belin, 31401 Toulouse Cedex 9, France\label{aff45}
\and
INAF-Osservatorio Astrofisico di Torino, Via Osservatorio 20, 10025 Pino Torinese (TO), Italy\label{aff46}
\and
Dipartimento di Fisica, Universit\`a di Genova, Via Dodecaneso 33, 16146, Genova, Italy\label{aff47}
\and
INFN-Sezione di Genova, Via Dodecaneso 33, 16146, Genova, Italy\label{aff48}
\and
Department of Physics "E. Pancini", University Federico II, Via Cinthia 6, 80126, Napoli, Italy\label{aff49}
\and
INFN section of Naples, Via Cinthia 6, 80126, Napoli, Italy\label{aff50}
\and
Instituto de Astrof\'isica e Ci\^encias do Espa\c{c}o, Universidade do Porto, CAUP, Rua das Estrelas, PT4150-762 Porto, Portugal\label{aff51}
\and
Dipartimento di Fisica, Universit\`a degli Studi di Torino, Via P. Giuria 1, 10125 Torino, Italy\label{aff52}
\and
INFN-Sezione di Torino, Via P. Giuria 1, 10125 Torino, Italy\label{aff53}
\and
INAF-IASF Milano, Via Alfonso Corti 12, 20133 Milano, Italy\label{aff54}
\and
Centro de Investigaciones Energ\'eticas, Medioambientales y Tecnol\'ogicas (CIEMAT), Avenida Complutense 40, 28040 Madrid, Spain\label{aff55}
\and
Port d'Informaci\'{o} Cient\'{i}fica, Campus UAB, C. Albareda s/n, 08193 Bellaterra (Barcelona), Spain\label{aff56}
\and
Institute for Theoretical Particle Physics and Cosmology (TTK), RWTH Aachen University, 52056 Aachen, Germany\label{aff57}
\and
Dipartimento di Fisica e Astronomia "Augusto Righi" - Alma Mater Studiorum Universit\`a di Bologna, Viale Berti Pichat 6/2, 40127 Bologna, Italy\label{aff58}
\and
European Space Agency/ESRIN, Largo Galileo Galilei 1, 00044 Frascati, Roma, Italy\label{aff59}
\and
Universit\'e Claude Bernard Lyon 1, CNRS/IN2P3, IP2I Lyon, UMR 5822, Villeurbanne, F-69100, France\label{aff60}
\and
Institute of Physics, Laboratory of Astrophysics, Ecole Polytechnique F\'ed\'erale de Lausanne (EPFL), Observatoire de Sauverny, 1290 Versoix, Switzerland\label{aff61}
\and
UCB Lyon 1, CNRS/IN2P3, IUF, IP2I Lyon, 4 rue Enrico Fermi, 69622 Villeurbanne, France\label{aff62}
\and
Mullard Space Science Laboratory, University College London, Holmbury St Mary, Dorking, Surrey RH5 6NT, UK\label{aff63}
\and
Departamento de F\'isica, Faculdade de Ci\^encias, Universidade de Lisboa, Edif\'icio C8, Campo Grande, PT1749-016 Lisboa, Portugal\label{aff64}
\and
Instituto de Astrof\'isica e Ci\^encias do Espa\c{c}o, Faculdade de Ci\^encias, Universidade de Lisboa, Campo Grande, 1749-016 Lisboa, Portugal\label{aff65}
\and
Department of Astronomy, University of Geneva, ch. d'Ecogia 16, 1290 Versoix, Switzerland\label{aff66}
\and
INAF-Istituto di Astrofisica e Planetologia Spaziali, via del Fosso del Cavaliere, 100, 00100 Roma, Italy\label{aff67}
\and
INFN-Padova, Via Marzolo 8, 35131 Padova, Italy\label{aff68}
\and
Institut d'Estudis Espacials de Catalunya (IEEC),  Edifici RDIT, Campus UPC, 08860 Castelldefels, Barcelona, Spain\label{aff69}
\and
Institut de Ciencies de l'Espai (IEEC-CSIC), Campus UAB, Carrer de Can Magrans, s/n Cerdanyola del Vall\'es, 08193 Barcelona, Spain\label{aff70}
\and
Aix-Marseille Universit\'e, CNRS/IN2P3, CPPM, Marseille, France\label{aff71}
\and
Istituto Nazionale di Fisica Nucleare, Sezione di Bologna, Via Irnerio 46, 40126 Bologna, Italy\label{aff72}
\and
FRACTAL S.L.N.E., calle Tulip\'an 2, Portal 13 1A, 28231, Las Rozas de Madrid, Spain\label{aff73}
\and
Dipartimento di Fisica "Aldo Pontremoli", Universit\`a degli Studi di Milano, Via Celoria 16, 20133 Milano, Italy\label{aff74}
\and
Institute of Theoretical Astrophysics, University of Oslo, P.O. Box 1029 Blindern, 0315 Oslo, Norway\label{aff75}
\and
Leiden Observatory, Leiden University, Einsteinweg 55, 2333 CC Leiden, The Netherlands\label{aff76}
\and
Jet Propulsion Laboratory, California Institute of Technology, 4800 Oak Grove Drive, Pasadena, CA, 91109, USA\label{aff77}
\and
Department of Physics, Lancaster University, Lancaster, LA1 4YB, UK\label{aff78}
\and
Felix Hormuth Engineering, Goethestr. 17, 69181 Leimen, Germany\label{aff79}
\and
Technical University of Denmark, Elektrovej 327, 2800 Kgs. Lyngby, Denmark\label{aff80}
\and
Cosmic Dawn Center (DAWN), Denmark\label{aff81}
\and
NASA Goddard Space Flight Center, Greenbelt, MD 20771, USA\label{aff82}
\and
Department of Physics and Helsinki Institute of Physics, Gustaf H\"allstr\"omin katu 2, 00014 University of Helsinki, Finland\label{aff83}
\and
Universit\'e de Gen\`eve, D\'epartement de Physique Th\'eorique and Centre for Astroparticle Physics, 24 quai Ernest-Ansermet, CH-1211 Gen\`eve 4, Switzerland\label{aff84}
\and
Department of Physics, P.O. Box 64, 00014 University of Helsinki, Finland\label{aff85}
\and
Helsinki Institute of Physics, Gustaf H{\"a}llstr{\"o}min katu 2, University of Helsinki, Helsinki, Finland\label{aff86}
\and
Department of Physics and Astronomy, University College London, Gower Street, London WC1E 6BT, UK\label{aff87}
\and
Aix-Marseille Universit\'e, CNRS, CNES, LAM, Marseille, France\label{aff88}
\and
NOVA optical infrared instrumentation group at ASTRON, Oude Hoogeveensedijk 4, 7991PD, Dwingeloo, The Netherlands\label{aff89}
\and
INFN-Sezione di Milano, Via Celoria 16, 20133 Milano, Italy\label{aff90}
\and
Universit\"at Bonn, Argelander-Institut f\"ur Astronomie, Auf dem H\"ugel 71, 53121 Bonn, Germany\label{aff91}
\and
Dipartimento di Fisica e Astronomia "Augusto Righi" - Alma Mater Studiorum Universit\`a di Bologna, via Piero Gobetti 93/2, 40129 Bologna, Italy\label{aff92}
\and
Department of Physics, Institute for Computational Cosmology, Durham University, South Road, DH1 3LE, UK\label{aff93}
\and
Universit\'e C\^{o}te d'Azur, Observatoire de la C\^{o}te d'Azur, CNRS, Laboratoire Lagrange, Bd de l'Observatoire, CS 34229, 06304 Nice cedex 4, France\label{aff94}
\and
Universit\'e Paris Cit\'e, CNRS, Astroparticule et Cosmologie, 75013 Paris, France\label{aff95}
\and
University of Applied Sciences and Arts of Northwestern Switzerland, School of Engineering, 5210 Windisch, Switzerland\label{aff96}
\and
Institut d'Astrophysique de Paris, 98bis Boulevard Arago, 75014, Paris, France\label{aff97}
\and
IFPU, Institute for Fundamental Physics of the Universe, via Beirut 2, 34151 Trieste, Italy\label{aff98}
\and
Institut de F\'{i}sica d'Altes Energies (IFAE), The Barcelona Institute of Science and Technology, Campus UAB, 08193 Bellaterra (Barcelona), Spain\label{aff99}
\and
Department of Physics and Astronomy, University of Aarhus, Ny Munkegade 120, DK-8000 Aarhus C, Denmark\label{aff100}
\and
Waterloo Centre for Astrophysics, University of Waterloo, Waterloo, Ontario N2L 3G1, Canada\label{aff101}
\and
Department of Physics and Astronomy, University of Waterloo, Waterloo, Ontario N2L 3G1, Canada\label{aff102}
\and
Perimeter Institute for Theoretical Physics, Waterloo, Ontario N2L 2Y5, Canada\label{aff103}
\and
Space Science Data Center, Italian Space Agency, via del Politecnico snc, 00133 Roma, Italy\label{aff104}
\and
Institute of Space Science, Str. Atomistilor, nr. 409 M\u{a}gurele, Ilfov, 077125, Romania\label{aff105}
\and
Institute for Particle Physics and Astrophysics, Dept. of Physics, ETH Zurich, Wolfgang-Pauli-Strasse 27, 8093 Zurich, Switzerland\label{aff106}
\and
Dipartimento di Fisica e Astronomia "G. Galilei", Universit\`a di Padova, Via Marzolo 8, 35131 Padova, Italy\label{aff107}
\and
Departamento de F\'isica, FCFM, Universidad de Chile, Blanco Encalada 2008, Santiago, Chile\label{aff108}
\and
Satlantis, University Science Park, Sede Bld 48940, Leioa-Bilbao, Spain\label{aff109}
\and
Institute of Space Sciences (ICE, CSIC), Campus UAB, Carrer de Can Magrans, s/n, 08193 Barcelona, Spain\label{aff110}
\and
Infrared Processing and Analysis Center, California Institute of Technology, Pasadena, CA 91125, USA\label{aff111}
\and
Instituto de Astrof\'isica e Ci\^encias do Espa\c{c}o, Faculdade de Ci\^encias, Universidade de Lisboa, Tapada da Ajuda, 1349-018 Lisboa, Portugal\label{aff112}
\and
Universidad Polit\'ecnica de Cartagena, Departamento de Electr\'onica y Tecnolog\'ia de Computadoras,  Plaza del Hospital 1, 30202 Cartagena, Spain\label{aff113}
\and
Centre for Information Technology, University of Groningen, P.O. Box 11044, 9700 CA Groningen, The Netherlands\label{aff114}
\and
Institut de Recherche en Astrophysique et Plan\'etologie (IRAP), Universit\'e de Toulouse, CNRS, UPS, CNES, 14 Av. Edouard Belin, 31400 Toulouse, France\label{aff115}
\and
INFN-Bologna, Via Irnerio 46, 40126 Bologna, Italy\label{aff116}
\and
Dipartimento di Fisica, Universit\`a degli studi di Genova, and INFN-Sezione di Genova, via Dodecaneso 33, 16146, Genova, Italy\label{aff117}
\and
INAF, Istituto di Radioastronomia, Via Piero Gobetti 101, 40129 Bologna, Italy\label{aff118}
\and
Junia, EPA department, 41 Bd Vauban, 59800 Lille, France\label{aff119}
\and
Department of Physics and Astronomy, University of British Columbia, Vancouver, BC V6T 1Z1, Canada\label{aff120}}

\abstract{We present an analysis of \Euclid observations of a 0.5\,deg$^2$ field in the central region of the Fornax galaxy cluster that were acquired during the performance verification phase. With these data, we investigate the potential of \Euclid for identifying GCs at 20\,Mpc, and validate the search methods using artificial GCs and known GCs within the field from the literature. Our analysis of artificial GCs injected into the data shows that \Euclid's data in $\IE$ band is 80\% complete at about $I_\sfont{E} \sim 26.0$\,mag ($M_{V\rm } \sim -5.0$\,mag), and resolves GCs as small as $r_{\rm h} = 2.5$\,pc. In the $\IE$ band, we detect more than 95\% of the known GCs from previous spectroscopic surveys and GC candidates of the ACS Fornax Cluster Survey, of which more than 80\% are resolved. We identify more than 5000 new GC candidates within the field of view down to $I_\sfont{E} = 25.0$\,mag, about 1.5\,mag fainter than the typical GC luminosity function turn-over magnitude, and investigate their spatial distribution within the intracluster field. We then focus on the GC candidates around dwarf galaxies and investigate their numbers, stacked luminosity distribution and stacked radial distribution. While the overall GC properties are consistent with those in the literature, an interesting over-representation of relatively bright candidates is found within a small number of relatively GC-rich dwarf galaxies. Our work confirms the capabilities of \Euclid data in detecting GCs and separating them from foreground and background contaminants at a distance of 20\,Mpc, particularly for low-GC count systems such as dwarf galaxies.}
%
%
%
%
\titlerunning{\Euclid: ERO -- A view of GCs in the Fornax galaxy cluster}
\authorrunning{Saifollahi et al.}
   
\maketitle

\section{\label{sc:Intro}Introduction}

It is well-established that globular clusters (GCs) mostly formed during the earliest stages of star and galaxy formation, and co-evolved with their host galaxies during their lifetimes (\citealp{Pfeffer-2018,creasey,massari2019,horta}). This view of GC formation, while not completely understood, implies a connection between the properties of GCs and those of galaxies (\citealp{bastian}). There are several known scaling relations between the properties of GCs and their host galaxies (\citealp{Spitler2009,misgeld-2011,harris2013,forbes2018,hudson2018,kruijssen,burkert2020}), as well as the environment that those galaxies are inhabiting (\citealp{peng2011,harris2017}), which overall are the main evidence of a \textit{galaxy-GC-halo connection}. Currently, it has become possible to observe GCs at higher redshifts \citep{AlamoMartinez2013, Lee2022jwst, Faisst2022, Harris2023} and examine directly what the GC progenitors at redshifts $z > 5$ may be \citep{Vanzella2017a,ChenStark2023}. Combining the information from high-$z$ observations with a wide-field census of the GC populations in the local Universe, we will gather unprecedented empirical constraints on the models for processes that determine how these compact stellar clusters are formed and distributed in space over time.
 
Massive galaxies have collected their GCs over time during their mass assembly. Therefore, the overall GC properties of massive galaxies, such as GC number count, radial distribution, and average colour, are valuable tracers of their hosts' mass-assembly process (\citealp{BK17,beasley2018,elbadry,lucas}). However, mergers cannot have played a major role in the assembly of low-mass galaxies and the present-time GCs of dwarf galaxies are expected to have formed with the galaxy. Additionally, because of the lower mass of their host galaxy, the dynamical mass loss and evaporation (\citealp{marta2020,gieles2023}) had a minimal effect on their evolution. Therefore, the GCs of dwarf galaxies potentially reflect the conditions and the environment at the time they formed together with their host (\citealp{soren2022,aaron2022,anastasia}). However, the majority of dwarf galaxies host few or no GCs (\citealp{Georgiev,gc-lg5,prole2018,gc-lg6,francine,roman2021,lamarca2}), with some exceptions (\citealp{gc-lg2,gc-lg1,gc-lg3,gc-lg4,vd17,saifollahi2021a,muller21,matlas-danieli}). Obtaining a general census of GCs of dwarf galaxies requires studying large samples, over a wide range of galaxy masses and environments (\citealp{durrell1996,miller1998,lotz2004,basino,Gerogiev2}). 

Several deep ground-based surveys have collected data on extragalactic GCs within the local Universe (\citealp{kissler,durrell2014,Lim2020,Cantiello2020}). However, extragalactic GC detection with ground-based surveys is limited to the depth and the spatial resolution that can be achieved from ground. The deep optical ground-based surveys typically detect the brightest GCs, brighter than the turnover magnitude of the GC luminosity function (GCLF) at $M_{V}=-7.5$ (\citealp{rejkuba2012}) for galaxies at distances further than 10\,Mpc. This is the distance regime that two nearest galaxy clusters, the Virgo galaxy cluster and the Fornax galaxy clusters are located, at 16\,Mpc and 20\,Mpc respectively. While these surveys can detect bright GCs at such distances, separating GCs from non-GCs is very challenging. Given that the typical size of GCs is about 3\,pc (\citealp{king,masters2010,Baumgardt,hilker2020}), in the seeing-limited imaging data of ground-based telescopes GCs hosted by galaxies more distant than about $5$~Mpc appear as point sources. In the meantime, they share the same optical colours as the foreground stars and distant background galaxies which also appear as point-sources. Therefore, any attempt to photometrically identify GCs results in a sample highly contaminated by foregrounds stars and background galaxies. Having a low rate of contamination in GC samples is crucial for studying systems with low GC count such as dwarf galaxies. Separating GCs and non-GCs can be improved by combining deep optical and near-infrared imaging data (\citealp{Munoz-2014,Liu-2020,Saifollahi2021b}). However, deep near-infrared observations deep enough to reach the GCLF turn-over magnitude at a few tens of Mpc require long exposure times and have not been feasible on large spatial scales.

Because of the above-mentioned limitations of ground-based surveys, the \textit{Hubble} Space Telescope (HST) has been used frequently for GC studies (\citealp{miller,Georgiev,peng2011}). With the HST, it is possible to resolve GCs in the local Universe within a few tens of Mpc (\citealp{jordan2007,jordan2015}). Even for more distant objects up to 100 Mpc, where GCs are unresolved in HST images, HST data are valuable for GC identification: GCs appear as point sources while most of the background objects are resolved (\citealp{peng2011,harris2020}). However, HST observations are limited to a small field of view (FoV) of a few arcmin$^2$ for a limited number of targets selected for an observing programme. This issue will be tackled in the next years using the 6-year wide survey with the recently launched \Euclid mission.

The Euclid Wide Survey (EWS, \citealt{Scaramella-EP1}) will cover an area of $14\,000\, {\rm deg}^2$, and will enhance our understanding of GCs and their connection to their host galaxy and environment. It is in particular expected that \Euclid will observe a large number of dwarf galaxies and their GC systems (\citealp{EuclidSkyOverview}). \Euclid is equipped with two instruments: VIS for imaging at red optical wavelengths with a broad bandwidth (one filter,  $\IE$); and NISP for imaging in the near infrared (three filters,  $\YE$,  $\JE$, and $\HE$; \citealp{Schirmer-EP18}), as well as low-resolution near-infrared spectroscopy. The imaging data of the EWS will reach a depth of $I_\sfont{E}=26.2$, $Y_\sfont{E}=24.3$, $J_\sfont{E}=24.5$, and $H_\sfont{E}=24.4$ (5$\sigma$ detection for point-like sources, \citealp{Scaramella-EP1}). The spatial resolution of VIS images and the inferred \Euclid colours can distinguish GCs and non-GCs (foreground stars and background galaxies), which makes the EWS a unique survey for identifying GCs around galaxies in the local Universe (Euclid Collaboration: Voggel et al. in prep.)  and for further studies on their properties (\citealp{ERONearbyGals,EROPerseusDGs}). 

\begin{figure*}[htbp!]
  \begin{center}
    \includegraphics[trim={0 0 0 0},clip,width=1.0\linewidth]{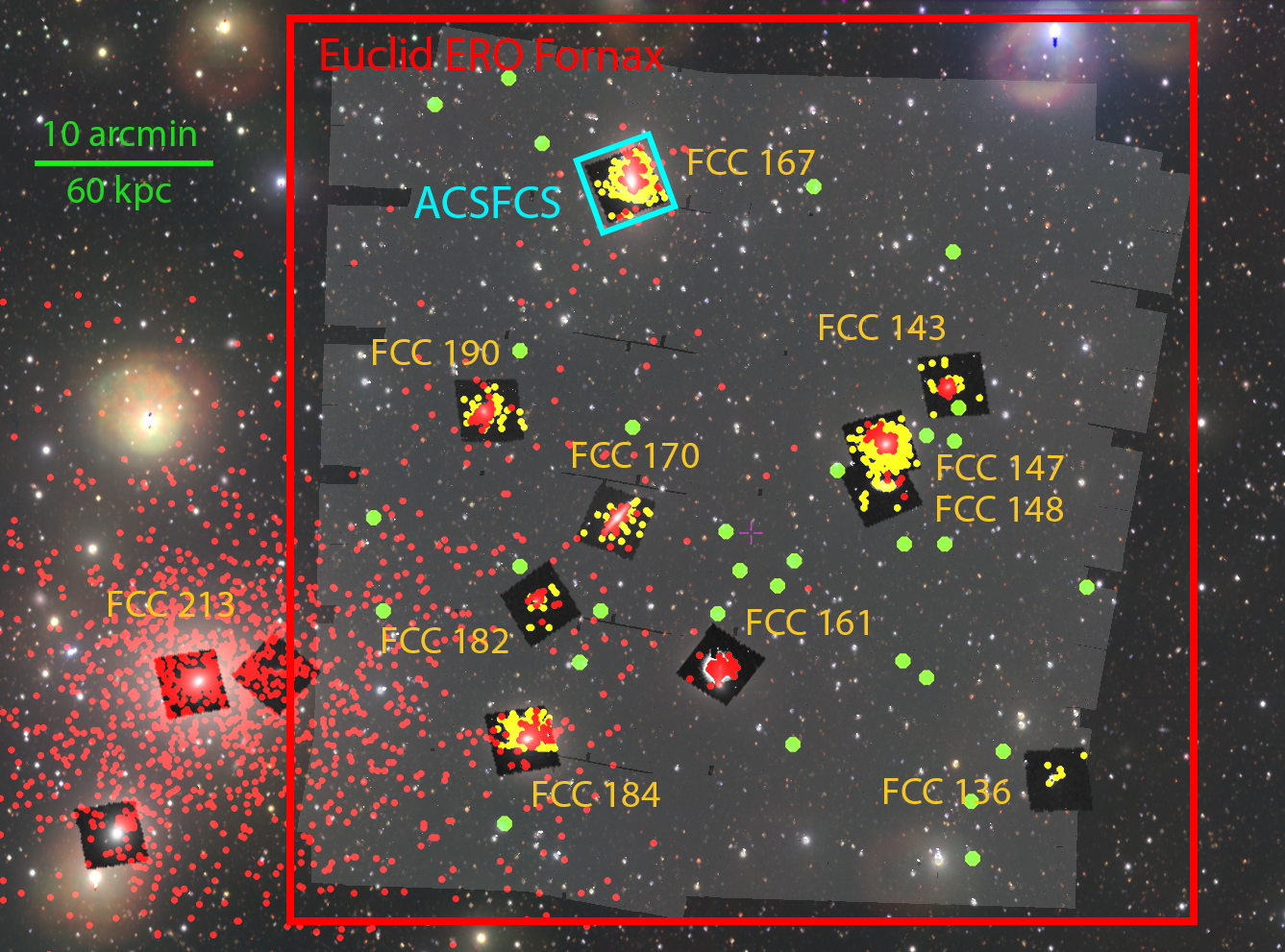}
   \end{center}
\caption{FoV of the \Euclid observations of the ERO Fornax cluster (ERO-F) presented in this paper. North is up and east to the left. The stack of four VIS images is shown in translucent grey above a colour image from the Fornax Deep Survey (FDS), which shows the brightest cluster galaxy NGC\,1399 just off the \Euclid\ pointing in the south-east. The ERO-F field includes ten massive galaxies previously targeted by the ACSFCS (small inset images) around which about 900 GC candidates were identified by that survey (yellow dots: GC probability $p_{\rm GC}>0.95$; \citealp{jordan2015}). The galaxy FCC161 (NGC1379) is among the galaxies in the ACSFCS survey; however, its final HST data are corrupted. Therefore, no ACSFCS GC catalogue for this galaxy has been published. Additionally, the \Euclid FoV overlaps with the position of about 30 dwarf galaxies (large green dots; \citealp{venhola2018,venhola2022}) and about 602 spectroscopically confirmed GCs (red dots; references in the text).}
\label{fornax-data}
\end{figure*}

Motivated by this, we selected a 0.5\,deg$^2$ field in the Fornax galaxy cluster as a target for \Euclid's Early Release Observations (ERO) programme (\citealp{EROcite}). We assess the capabilities of \Euclid in identifying and studying GCs around dwarf galaxies as well as the intracluster GCs. 
The Fornax galaxy cluster, located at 20\,Mpc (\citealp{Blakeslee2009}), is the second nearest massive galaxy cluster. It has an estimated virial mass of $M_{\rm virial}=7 \times 10^{13} M_{\rm \odot}$ and a virial radius of $R_{\rm virial} = 700$\,kpc (\citealp{drinkwater2001}). It has been the target of many past ground-based surveys, the most recent deep ones being the Fornax Deep Survey (FDS, \citealp{venhola2018}) and the Next Generation Fornax Survey (NGFS, \citealp{eigenthaler}), which led to several studies on compact sources and GCs (\citealp{prole2019,Cantiello2020,Saifollahi2021b}). Additionally, the ACS Fornax Cluster Survey (ACSFCS, \citealt{jordan2007}) has targeted 43 galaxies within the cluster with HST (galaxies with $M_B < -16.0$\footnote{Expressed in the Vega magnitude system here.}) and identified the GCs around these. Figure\,\ref{fornax-data} shows the FoV of the ERO Fornax data (referred to as ERO-F) in comparison to ACSFCS. This FoV was selected for the \Euclid ERO programme because of the availability of a number of massive galaxies and dwarf galaxies, as well as several hundreds of known GCs from the previous spectroscopic surveys and ACSFCS. 

The rich archival data within ERO-F allows us to make a detailed assessment of the power of the EWS for GC identification around galaxies at a distance of 20\,Mpc. This paper presents the results of our assessment. The structure of this paper is as follows. In Sect.\,\ref{sc:Data}, \Euclid data and the complementary archival data are described. Section \ref{sc:Methods} presents the methodology for data analysis and GCs identification. This includes several steps, such as point-spread function modelling, source detection, photometry, simulations of GCs, and GC selection. Using the output of this analysis, Sect.\,\ref{sc:Results} examines the performance of GC identification and studies the properties of the GCs within the cluster and around dwarf galaxies. Section\,\ref{sc:Summary} summarizes the findings. In this paper, we adopt a cosmology with $H_0 = 68\,{\rm km\,s^{-1}\,Mpc^{-1}}$, ${\rm {\Omega}_m = 0.3}$ and ${\rm {\Omega}_{\Lambda} = 0.7}$. All magnitudes are in the AB system, unless otherwise specified.

\section{\label{sc:Data}Data} 

This work uses the imaging data of the \Euclid VIS (\citealp{EuclidSkyVIS}) and NISP (\citealp{EuclidSkyNISP}) instruments, in filters $\IE$, $\YE$, $\JE$, and $\HE$, and the galaxy and GC catalogues available in the literature. We describe these data sets below.

\subsection{\label{sc:Data-vis-nisp}\Euclid VIS and NISP}

The \Euclid observations for ERO-F were carried out during the performance verification (PV) phase of the mission in August and September 2023. The observations are centred at RA~$ = \ra{03;36;08.759}$ and Dec~$ = \ang{-35;16;00.38}$ in the Fornax cluster and cover an area of 0.5\,deg$^2$. In total four 560\,s exposures in $\IE$, three 112\,s exposures in $\YE$, and four 112\,s in each of $\JE$ and $\HE$ were acquired that satisfied our requirements for scientific exploitation. The EWS will provide similar total exposure times with four exposures per filter. While the dither pattern of the EWS was designed to fill gaps between detectors, the ERO-F images were taken at two different dates in the PV sequence with two different orientations on the sky, which leaves small areas uncovered after stacking. The data reduction and stacking procedures are described in \citet{EROData}. The resulting stacked frames have the original pixel scales of $\ang{;;0.1}$ for $\IE$ and $\ang{;;0.3}$ for the NISP bands. The $\IE$ stacked frame, based on four exposures, is reserved for the examination of extended sources. For the study of GCs, we produce an additional stacked frame with only the two $\IE$ images with the best point-spread function (PSF), using the  \texttt{SWarp} code (\citealp{swarp}). The other two $\IE$ frames are excluded here because of PSF imperfections due to \Euclid's tracking issue in the first month of the PV phase. In the two-image stacked frame, about 65\% and 35\% of the area is covered, respectively, by two and one exposures. Such a distribution is clearly not ideal, and in particular, it makes it impossible to correct all the pixels affected by cosmic ray hits. This leads to increased errors in measurements compared to expectations in the standard EWS data. In this work, we analyse 100\% of the covered area.

\begin{figure*}[htbp!]
  \begin{center}
    \includegraphics[trim={0 0 0 0},clip,width=0.99\linewidth]{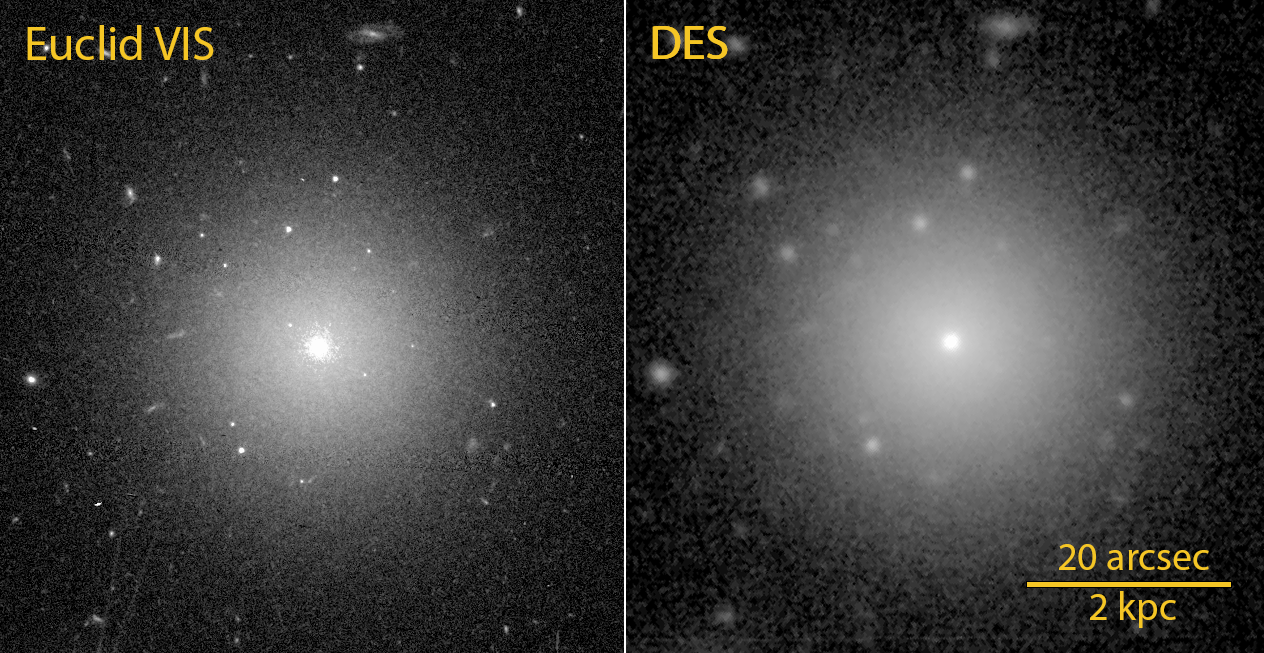}
   \end{center}
\caption{\Euclid $\IE$ data (two-image stacked frame) versus the Dark Energy Survey (DES, \citealp{des}) in the $r$ band for a cutout around dwarf galaxy FCC188 (FDS11$\_$DWARF155). The galaxy is surrounded by several small point-like sources, some of which are the GCs of the galaxy. Given the high-resolution images of \Euclid in $\IE$, we are able to resolve the majority of GCs around similar objects and distinguish them from foreground stars. These images have a total exposure time of 1120\,s and 720\,s for \Euclid VIS and DES, respectively.}
\label{fig:fornax-data+}
\end{figure*}

The full-width at half-maximum (FWHM) of the average PSF in the two-image $\IE$ stacked frame is $\ang{;;0.19}$  (1.9\,pixels). The PSF of that $\IE$ stack is not typical of the EWS. It is affected by the resampling of only two images that are initially under-sampled and have different orientations,  and it is 20\% larger than the native FWHM of the initial frames, which is $\ang{;;0.16}$ (1.6\,pixels). Nevertheless, as is seen in Fig.\,\ref{fig:fornax-data+}, this is a considerable improvement in spatial resolution with respect to the ground-based optical wide-field surveys. The FWHM of the PSF in the $\YE$,  $\JE$, and $\HE$ stacked frames are $\ang{;;0.50}$ (1.7\,pixels),  $\ang{;;0.54}$ (1.7\,pixels), and  $\ang{;;0.55}$  (1.8\,pixels), respectively. For point sources, the stacked frames reach a $10\sigma$ depth of $I_\sfont{E}=25.5$, $Y_{\rm E}=23.4$, $J_{\rm E}=23.6$, and $H_{\rm E}=23.5$.

At the distance of the Fornax cluster, 1\,pixel of the VIS instrument ($\ang{;;0.1}$) corresponds to a physical size of about 10\,pc. Based on the previous experience with HST (\citealp{jordan2015}), we expect to resolve bright (high signal-to-noise ratio) GCs of one-fifth of the pixel-scale. At 20\,Mpc, this corresponds to a FWHM of 2\,pc and (for typical GC light profiles) a half-light radius of about $r_{\rm h} =3$\,pc, which is the average size for Galactic GCs. 

\subsection{\label{sc:Data-galaxy}Galaxy catalogues}
The FoV of ERO-F was chosen to overlap with the location of 10 major galaxies (Fig.\,\ref{fornax-data}), of which nine have published ACSFCS data\footnote{Note that FCC161 (NGC1379) is among the galaxies in the ACSFCS survey; however, its HST data are corrupted and have not been used in ACSFCS publications.} (\citealp{jordan2007,jordan2015}). Additionally, the FoV contains 48 galaxies listed as dwarf or low-surface brightness (LSB) galaxies in the Fornax Deep Survey (FDS) dwarf catalogues (\citealp{venhola2018,venhola2022}). Here, we consider the 45 galaxies with $M_r < -17\,$ as dwarf galaxies. The remaining three galaxies in the FDS dwarf catalogue are relatively brighter than the rest of the sample and are among the 10 massive galaxies observed by ACSFCS. After visual inspection of the $\IE$ frame, we exclude from further analysis some objects from the FDS list: those that appear to be spiral galaxies; duplicates in the catalogues; and dwarf galaxies that are not fully covered by stacked frames in one of the filters. This selection leads to a sample of 30 dwarf galaxies for further analysis of their GCs. Tables~\ref{gal-catalogue-m} and \ref{gal-catalogue-d} present the galaxy samples used in this paper. In these tables, properties of galaxies are adopted from \citet{venhola2018,venhola2022} and \citet{spavone}. The stellar masses of the dwarf galaxies in Table\,\ref{gal-catalogue-d} are derived based on \citet{taylor} from the total $r$-band magnitudes and $g-i$ colours in \citet{venhola2018,venhola2022}.

\begin{table*}[ht]
\centering
\small
\caption{Properties of the 10 most massive galaxies located within the footprint of the  \Euclid ERO-F images. Columns 1 and 2 present the FCC and other names of the galaxies (NGC and/or FDS names from \citealp{venhola2018}). Columns 3, 4, and 5 present the total absolute magnitude in the $r$-band, effective radius, and stellar mass of galaxies (\citealp{spavone,venhola2018}). When available, the distances derived from the surface-brightness fluctuations studies of \citet{Blakeslee2009} are provided in Col.\,(6). Throughout this paper, we use the FCC names of massive galaxies to refer to them.}
\begin{tabular}{llcccc}
\hline\hline
\noalign{\vskip 2pt}
Galaxy & Alternative name(s) & $M_{r}$ [mag] & $R_{\rm e}$ [kpc] & $\logten(M_{*}/M_{\odot})$ & $D$ [Mpc]\\
(1) & (2) & (3) & (4) & (5) & (6) \\
\hline
\noalign{\vskip 2pt}
FCC167 & NGC\,1380 & $-$22.36 & 5.80 & 10.99 & $21.2 \pm 0.7$\\  
FCC184 & NGC\,1387 & $-$21.43 & 3.22 & 10.67 & $19.9 \pm 0.8$\\  
FCC161 & NGC\,1379 & $-$21.02 & 2.76 & 10.42 & -- \\  
FCC147  & NGC\,1374 & $-$20.96 & 2.36 & 10.38 & $19.6 \pm 0.6$\\
FCC170  & NGC\,1381 & $-$20.71 & 1.69 & 10.35 & $21.9 \pm 0.8$\\  
FCC148 & NGC\,1375 & $-$19.79 & 2.73 & 9.76 & $19.6 \pm 0.7$ \\
FCC190 & NGC\,1380B & $-$19.28 & 2.36 & 9.73 & $20.3 \pm 0.7$\\ 
FCC143  & NGC\,1373, FDS16$\_$DWARF002 & $-$18.77 &  1.03 & 9.45 & $19.3 \pm 0.8$\\
FCC182 & FDS11$\_$DWARF279 & $-$17.88 & 0.97 & 9.10 & $19.6 \pm 0.8$\\ 
FCC136 & FDS16$\_$DWARF159 & $-$17.76 & 1.75 & 9.01 & $18.8 \pm 0.7$\\ 
\hline
\end{tabular}
\label{gal-catalogue-m}
\end{table*}


\begin{table*}[ht]
\centering
\small
\caption{Similar to Table\,\ref{gal-catalogue-m} but for the 30 dwarf galaxies overlapped with the \Euclid ERO data of the Fornax cluster. Columns 1 and 2 present the FCC and other names of the galaxies (FDS names from \citealp{venhola2018} and/or NGFS names from \citealp{eigenthaler}). Columns 3, 4, and 5 present the total absolute magnitude in the $r$ band, effective radius, and stellar mass of galaxies (\citealp{venhola2018,venhola2022}). Throughout this paper, we use the FDS names of dwarf galaxies to refer to them. }
\begin{tabular}{llccc}
\hline\hline
\noalign{\vskip 2pt}
Galaxy FCC name & Alternative name(s) & $M_{r}$ [mag] & $R_{\rm e}$ [kpc] & $\logten(M_{*}/M_{\odot})$ \\
(1) & (2) & (3) & (4) & (5) \\
\hline
\noalign{\vskip 2pt}
FCC188 & FDS11$\_$DWARF155, NGFS\,J033705$-$353525 & $-16.25$ & 1.22 & 8.34 \\
FCC195 & FDS10$\_$DWARF014, NGFS\,J033723$-$345401, & $-$15.43 & 1.28 & 8.04 \\
FCC181 & FDS10$\_$DWARF003, NGFS\,J033653$-$345619 & $-$14.95 & 0.97 & 7.78 \\
FCC160 & FDS11$\_$DWARF289, NGFS\,J033604$-$352320 & $-$14.81 & 1.36& 7.72 \\
FCC133 & FDS16$\_$DWARF253, NGFS\,J033420$-$352145 & $-$14.61 & 1.13 & 7.70\\
FCC156 & FDS16$\_$DWARF257, NGFS\,J033543$-$352018 & $-$14.98 & 1.54 & 7.67 \\

FCC171 & FDS11$\_$DWARF294, NGFS\,J033637$-$352309 & $-$14.25 & 2.05 & 7.46 \\
FCC157 & FDS16$\_$DWARF185a, NGFS\,J033543$-$353051 & $-$13.75 & 1.19 & 7.33 \\
FCC175 & FDS11$\_$DWARF246, NGFS\,J033643$-$352609 & $-$13.80 & 1.61 & 7.19 \\
FCC140 & FDS16$\_$DWARF346, NGFS\,J033456$-$351127 & $-$13.39 & 0.95 & 7.16 \\
FCC142 & FDS16$\_$DWARF441, NGFS\,J033458$-$350235  & $-$13.38 & 0.85 & 7.09 \\
FCC145 &  FDS16$\_$DWARF330, NGFS\,J033505$-$351307 & $-$12.88 & 0.54 & 7.07 \\
FCC168 & FDS11$\_$DWARF365, NGFS\,J033628$-$351239 & $-$13.43 & 0.86 & 7.02 \\
-- &  FDS11$\_$DWARF306, NGFS\,J033700$-$352035 & $-$13.48  & 1.45  & 7.01 \\

FCC197 & FDS11$\_$DWARF330, NGFS\,J033741$-$351746 & $-$12.92 & 0.62 & 6.89 \\
FCC185 & FDS10$\_$DWARF034, NGFS\,J033703$-$345233 & $-$12.54 & 0.53 & 6.82 \\
FCC144 & FDS16$\_$DWARF280, NGFS\,J033500$-$351920 & $-$12.68 & 0.57 & 6.77 \\
-- & FDS11$\_$DWARF299, NGFS\,J033738$-$352308 & $-$11.62 & 0.60 & 6.69 \\
-- & FDSLSB56 & $-$12.70 & 0.93 & 6.68 \\
-- & FDS16$\_$DWARF172, NGFS\,J033453$-$353411 & $-$12.16 & 0.42 & 6.67 \\
FCC146 & FDS16$\_$DWARF281, NGFS\,J033512$-$351923 & $-$12.41 & 0.54 & 6.59 \\
-- & FDS16$\_$DWARF272, NGFS\,J033558$-$352053 & $-$12.05 & 0.40 & 6.49 \\
-- & FDS16$\_$DWARF329, NGFS\,J033458$-$351324 & $-$11.76 & 0.43 & 6.32 \\
-- & FDSLSB220 & $-$11.80 & 1.06 & 6.30\\

-- & FDS16$\_$DWARF141 & $-$11.29 & 0.29 & 6.03 \\
-- & FDSLSB45 & $-$10.60 & 0.30 & 5.89 \\
-- & FDSLSB43 & $-$10.60 & 0.39 & 5.83 \\
-- & FDS16$\_$DWARF230b, NGFS\,J033512$-$352603 & $-$10.67 & 0.59 & 5.78 \\
-- & FDS16$\_$DWARF227, NGFS\,J033505$-$352703 & $-$11.50 & 0.85 & 5.77 \\
-- & FDSLSB36 & $-$10.60 & 0.39 & 5.76 \\
\hline
\end{tabular}
\label{gal-catalogue-d}
\end{table*}

\subsection{\label{sc:Data-gc}GC catalogues}

Our searches for GCs in the Fornax cluster take advantage of previous surveys of GCs and compact sources in the cluster, conducted from the ground and from space. ACSFCS (\citealp{jordan2015}) used the HST Advanced Camera for Surveys (ACS) to acquire $202\arcsec \times 202\arcsec$ images centred on galaxies within the ERO FoV (Fig.\,\ref{fig:fornax-data+}). The images were taken in the F$475$W band with an exposure time of 760\,s, and in the F$850$LP band with a total exposure of 1220\,s ($2 \times 565\,\mathrm{s} + 90\,\mathrm{s}$), and have an initial pixel scale of $\ang{;;0.05}$. The ACSFCS team exploited a combination of colour, magnitude, compactness, and distance to the host-galaxy centre to identify GC candidates and compute the probability of the candidates being a GC \citep{jordan2015,Liu_etal19_ACSFCS}. This probability is larger than 95\% for 906 ACSFCS catalogue objects in our field. In the remainder of this paper, we always use this 95\% probability cut when exploiting the ACSFCS catalogue. Furthermore, the Fornax cluster has been the target of several spectroscopic surveys and more than 2\,800 GCs are spectroscopically confirmed (\citealp{Hilker-1999,drinkwater2000,Mieske2004,bergond2007,firth2007,firth2008,gregg2009,schuberth2010,Pota-2018,Fahrion_etal20_GCcat,Chaturvedi2022}).\footnote{Compilation of these catalogues are provided by \citet{wittmann-2016} and \citet{Maddox-2019}, with ground-based optical and near-infrared photometry in \citet{Saifollahi2021b}.} About 602 of these GCs overlap with the \Euclid observations, of which 225 GCs are in common with the ACSFCS GC catalogue. We use these samples of previously known GCs as a reference set to establish our methodology for GC identification in Sect.\,\ref{sc:GCselection}.

\section{\label{sc:Methods}Data processing}

Given the small angular sizes of GCs at the distance of the Fornax cluster, and the similarity of their photometric properties to some of the foreground stars and background galaxies, GC detection and identification requires careful analysis of the detected sources.\footnote{The source code for the analysis described in this paper is available at \url{https://github.com/teymursaif/GCEx}.} To do this, firstly, we model the PSF in all bands (Sect.\,\ref{sc:Methods-psf}). Then we perform source detection and photometry to construct a source catalogue (Sect.\,\ref{sc:Methods-photom}). The performance of source detection is then investigated using GC simulations (Sect.\,\ref{sc:Methods-gcsim}). Subsequently, we examine the photometric properties of the known GCs, namely ACSFCS GC candidates and spectroscopically confirmed GCs (Sect.\,\ref{sc:Methods-knowngc}). Using the observed properties of known GCs combined with the outcome of the GC simulations, we finally search for GC candidates within the data (Sect.\,\ref{sc:Methods-unknowngc}). GC selection is done using the compactness indices and colours of the sources. The resulting GC sample will be used in Sect.\,\ref{sc:Results}, where we examine the distribution of intracluster GCs (ICGCs) and GC properties of Fornax cluster dwarf galaxies.

\subsection{\label{sc:Methods-psf}Point-spread function (PSF)}

PSF models are constructed from the stacked frames and they are used to estimate aperture corrections (Sect.\,\ref{sc:Methods-photom}) as well as to simulate GC images (Sect.\,\ref{sc:Methods-gcsim}). One PSF model is produced per filter. For $\IE$, the PSF model is made from the two-image stacked frame. To produce these models for a given filter, an initial catalogue is produced with \texttt{SExtractor} (\citealp{sex}) in its default configuration. Subsequently, non-saturated bright points sources are selected based on \texttt{MAG$\_$AUTO}, \texttt{FWHM$\_$IMAGE}, and \texttt{ELLIPTICITY}, as well as the \texttt{FLAGS} parameters of \texttt{SExtractor}. We select objects with 19 $<$ \texttt{MAG$\_$AUTO} $<$ 21 (for the $\IE$ PSF model), 18 $<$ \texttt{MAG$\_$AUTO} $<$ 20 (for the $\YE$, $\JE$, and $\HE$ PSF models), $\mathtt{ELLIPTICITY}<0.1$, $\mathtt{FLAGS}<4$ and a range in \texttt{FWHM$\_$IMAGE} that is determined from the point-source sequence in the $\mathtt{FWHM\_IMAGE}-\mathtt{MAG\_AUTO}$ diagram of a given filter. Next, cutouts of 40\,pixels $\times$ 40\,pixels ($4\arcsec \times 4\arcsec$ in $\IE$\ and $12\arcsec \times 12\arcsec$ in $\YE$, $\JE$, and $\HE$) are made around the selected point sources (about 1000 sources in each filter). We update the centroid of sources by running \texttt{SExtractor} on the cutouts. Subsequently, taking into account the new centroids, all the cutouts are normalised and stacked using \texttt{SWarp} (\citealp{swarp}). The output of the stacking is the PSF model. The stacking is done with an over-sampling factor of 10 for each filter. 

\subsection{\label{sc:Methods-photom}Source detection and photometry}

We now produce a detection frame from the $\IE$ image and subsequently measure the photometry of compact sources. We make the detection frame by applying a ring filter with inner and outer radii of 4 and 8 pixels to the data in $\IE$. This procedure subtracts the light of galaxies, which improves the detection of GCs around galaxies, in particular for the most massive galaxies. This step is particularly important because the majority of the known GCs that we target to validate our methodologies are located in such massive galaxies.

For source detection, we run \textsc{SExtractor} on the detection frame. We modify some of the \texttt{SExtractor} parameters and use \texttt{tophat$\_$1.5$\_$3x3.conv} kernel for filtering the stacked frames to maximize point-source detection. In total 93\,995 and 207\,842 sources  brighter than $\IE = 25$ and $\IE = 26$, respectively, are present in the final source catalogue. Table\,\ref{sexparams} summarizes the \texttt{SExtractor} parameters that are used for source detection.

\begin{table}[htbp!]
\centering
\small
\caption{\texttt{SExtractor} parameters that were used for making detection frames and source extraction.}
\begin{tabular}{ l c c c }
\hline\hline
\noalign{\vskip 2pt}
\texttt{SExtractor} parameter& Value \\
\noalign{\vskip 1pt}
\hline
\noalign{\vskip 2pt}
\texttt{DETECT$\_$MINAREA} & 4\\
\texttt{DETECT$\_$MAXAREA} & 200\\
\texttt{DETECT$\_$THRESH} & 1.5  \\
\texttt{ANALYSIS$\_$THRESH} & 1.5  \\
\texttt{DEBLEND$\_$NTHRESH} &  32 \\
\texttt{DEBLEND$\_$MINCONT} &  0.0005 \\
\texttt{BACKPHOTO$\_$TYPE} &  GLOBAL \\
\texttt{BACK$\_$SIZE} &  32 \\
\texttt{BACK$\_$FILTERSIZE} & 1\\
\hline 
\end{tabular}
\label{sexparams}
\end{table}

Given the detections in $\IE$ from the previous step, we perform (forced) aperture photometry at these positions (now using the \texttt{aperture} task of \texttt{photutils} because this circumvents the need for identical pixel coordinates in all images). Fluxes are measured within an aperture radius of 1.5 times the FWHM of the PSF for each filter (2.8\,pixels in $\IE$, and between 2.4\,pixels and 2.8\, pixels in  $\YE$,  $\JE$, and \HE). Afterwards, for each source, the background is estimated within an annulus with an inner radius of 5 times the FWHM of the PSF and a thickness of 20\,pixels.

We correct the measured aperture magnitudes for the aperture size using the PSF models. Flux corrections are 8\% in $\IE$ and 5\% in  $\YE$,  $\JE$, and $\HE$ stacked frames, consistent with the measurements of \citet{EROGalGCs}. Considering that the majority of the GCs at the distance of the Fornax cluster, though compact, are not strictly speaking point sources in $\IE$, we expect that our aperture-corrected photometry misses a small fraction of the total flux of GCs in that band. We evaluate this fraction for $\IE$ (this effect is negligible for the near-infrared bands) by aperture photometry of ACSFCS GC candidates with a diameter of 1.5 times and 20 times the FWHM (2.8\,pixels and 40\,pixels) and find an average magnitude offset less than 5\%. The larger aperture diameter encloses more than 0.998 of the total flux of the PSF (\citealp{EROData}).  Additionally, photometry of the artificial GCs injected into the data (Sect.\,\ref{sc:Methods-gcsim}) shows that for a typical GC with $r_{\rm h} = 3$\,pc, we lose up to 5\% of the total flux. 

Additionally, we perform aperture photometry within 2, 4, and 8\,pixel aperture diameters. We use these aperture magnitudes to set up proxies for source compactness, which will allow us to exploit the spatial resolution of the VIS images when selecting GC candidates. Differences between two aperture magnitudes are a widely used and simple way of characterising the light profile of compact sources in this context \citep{peng2011,Powalka2016,lim2018,harris2020}. Here we define two compactness indices in \IE, namely $C_{\rm 2-4}$ (for aperture diameters of 2 and 4 pixels) and $C_{\rm 4-8}$ (4 and 8\,pixels).\footnote{With this definition, which agrees with common practice, large values of $C_{\rm 2-4}$ and $C_{\rm 4-8}$ imply less concentrated (more extended) profiles, which in the literature has led some authors to re-name the same quantities ``inverse concentration indices''} The first compactness index measures the concentration of the light in the inner parts of light profiles, while the second is sensitive to the outer parts. 

\begin{figure}[htbp!]
\begin{center}
\includegraphics[trim={0 0 0 0},angle=0,width=1\linewidth]{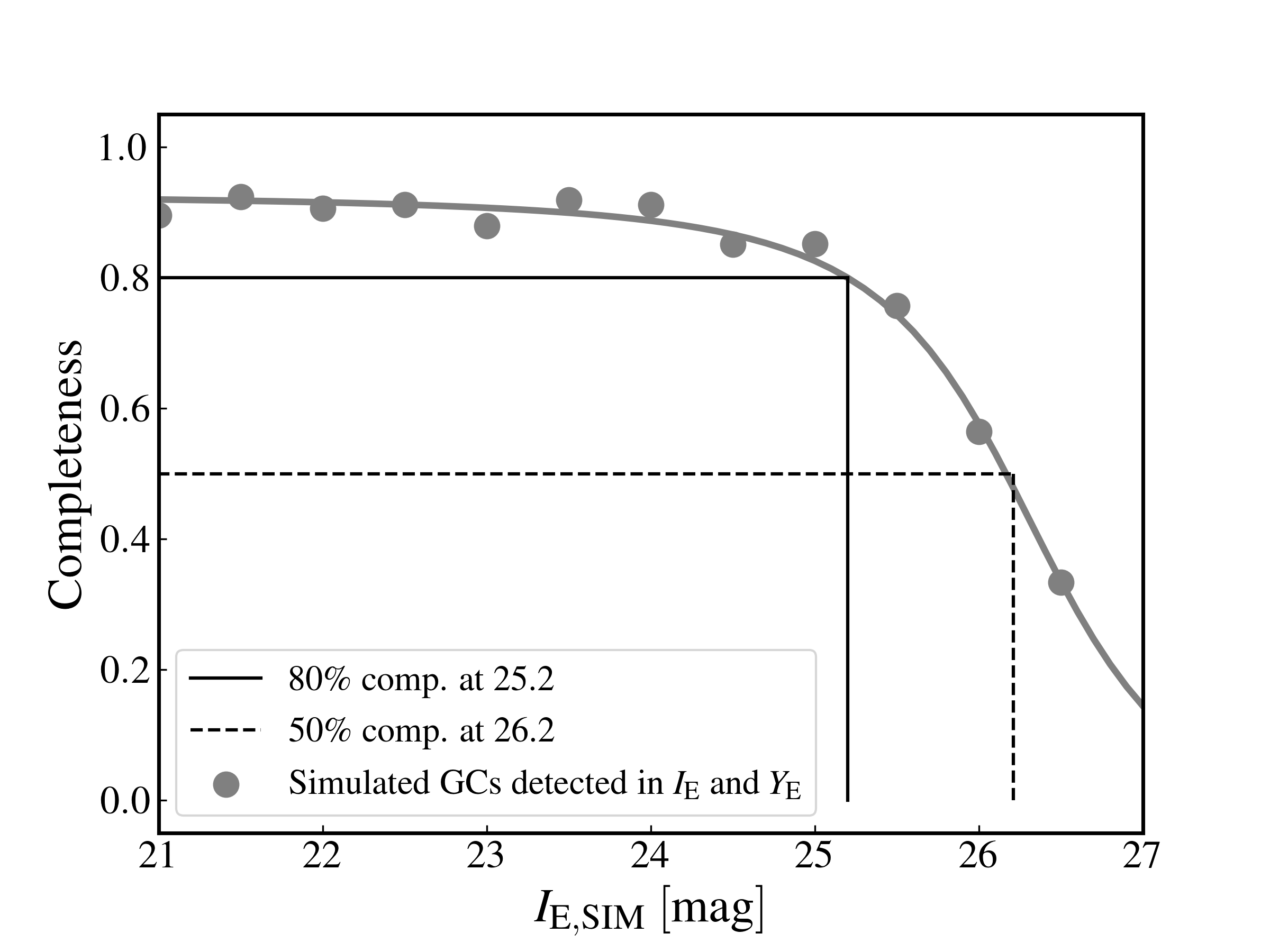}
\includegraphics[trim={0 0 0 0},angle=0,width=1\linewidth]{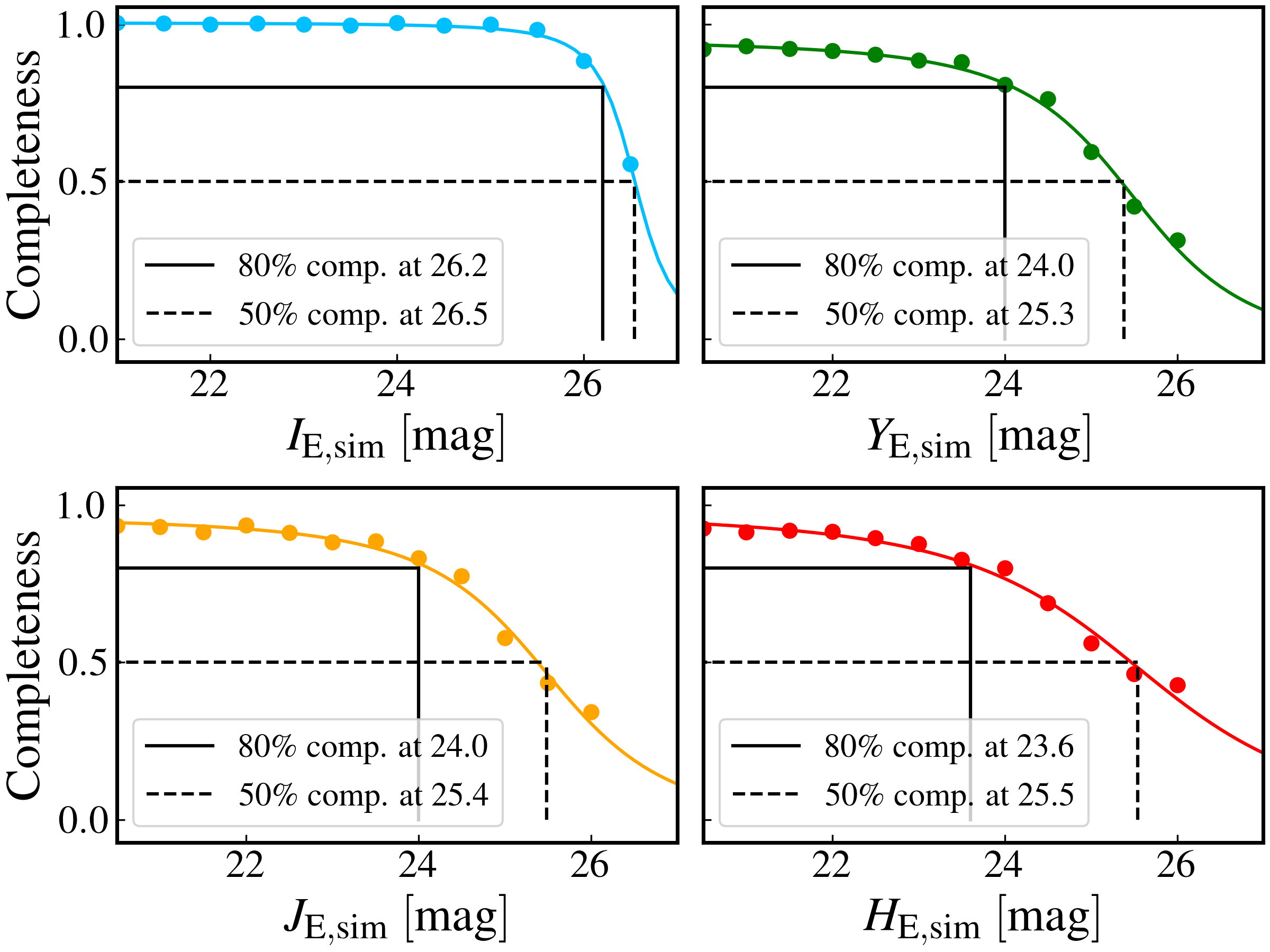}
\end{center}
\caption{Completeness expected in the ERO-Fornax catalogue data based on artificial GCs. \textit{Top panel}: The combined completeness, for sources detected in both $I_\sfont{E}$ and $Y_\sfont{E}$, as a function of simulated input $\IE$ magnitude. \textit{Bottom}: The detection completeness in the four individual filters as a function of the respective input magnitude. For old GCs with sub-Solar metallicities, the GCLF turn-over magnitude is expected to be at $\IE = 23.5$.}
\label{completeness-all}
\end{figure}

\begin{figure*}[htbp!]
\begin{center}
\includegraphics[trim={50 50 50 80},angle=0,width=\linewidth]{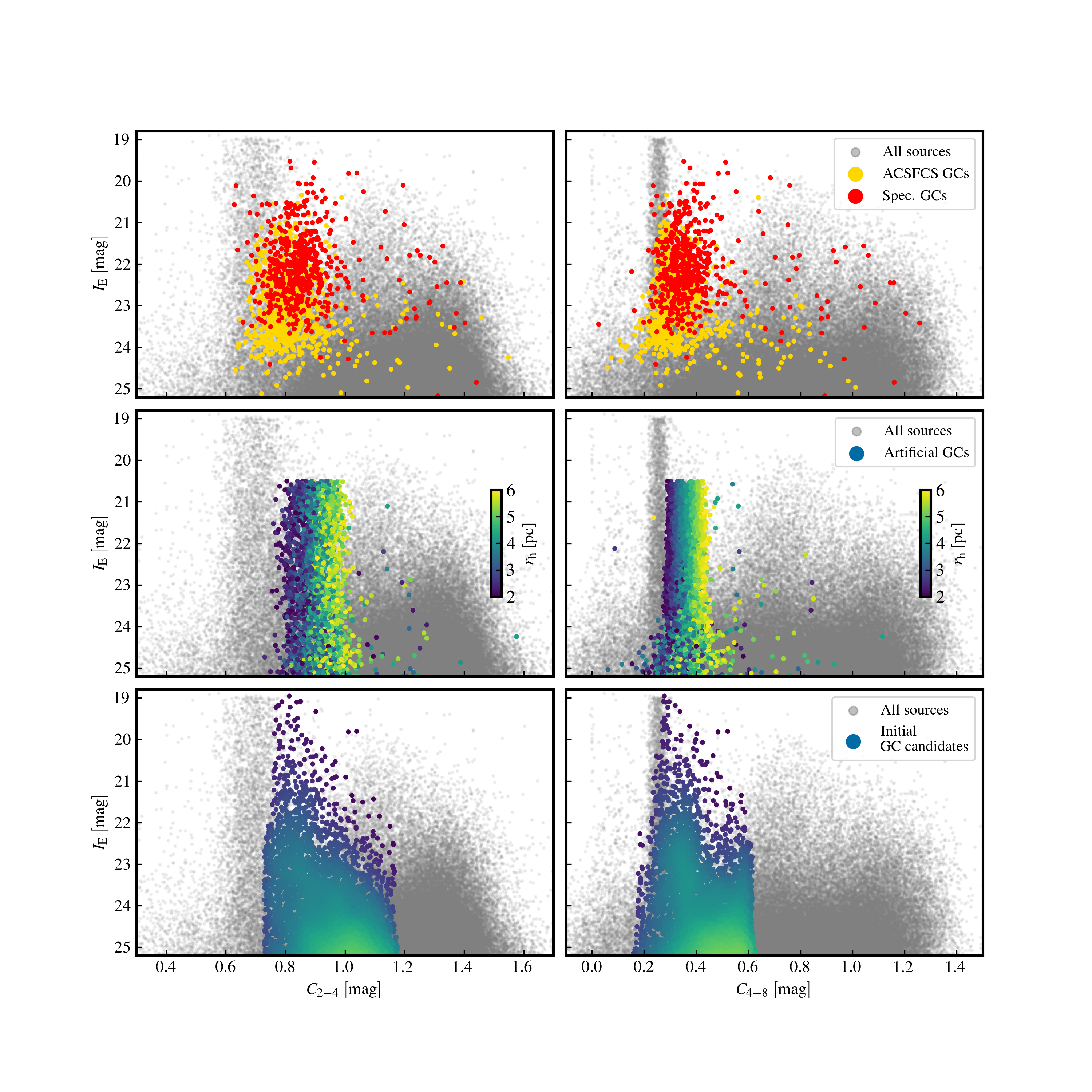}
\caption{Compactness indices measured in $\IE$, and first step of the GC-selection. The sources in the full catalogue are displayed in grey. \textit{Top panels}: Compactness indices of the spectroscopically confirmed GCs (red points), and of GCs in the ACSFCS catalogue (yellow points). These two samples serve as empirical references. The vertical sequences at $C_{2-4}=0.70$ and $C_{4-8}=0.26$ correspond to point sources. Objects on the right side of this sequence with a larger compactness index are extended sources. Objects with smaller compactness index are mostly artefacts (e.g., cosmic rays) in the data.
\textit{Middle panels}: Compactness indices of the 3000 artificial GCs with half-light radii between 2 and 6\,pc, which were injected into the images (black points).  \textit{Lower panels}: Compactness indices of the sources in the initial GC sample after selection based on their compactness indices, as described in Sect.\,\ref{sc:Methods-unknowngc} (green points). }
\label{compactness}
\end{center}
\end{figure*}

\subsection{\label{sc:Methods-gcsim}Artificial GCs and expected completenes}

We assess the performance of GC detection by injecting artificial GCs into the frames and applying the same source detection and photometry procedure as in Sect.\,\ref{sc:Methods-photom}. The light profiles of the artificial GC images follow King profiles (\citealp{King1962}), which are characterised by a core radius $r_{\rm c}$, a tidal radius $r_{\rm t}$ and a concentration index $\logten(r_{\rm t}/r_{\rm c})$. We set the latter to 1.4, a central value in the range 0.5--2.4 that is observed for Milky Way GCs \citep{Harris1996}. 

We vary the half-light radii $r_{\rm h}$ between 2\,pc and 6\,pc ($0.8 \leqslant r_{\rm c}/\mathrm{pc} \leqslant 2.4$), and the absolute $\IE$ magnitudes between $-11$ and $-5$, and we set the colours to $\IE-\YE=0.45$, $\IE-\JE=0.45$, and $\IE-\HE=0.45$ (see Appendix\,\ref{ap:colors} and Euclid Collaboration: Voggel et al. (in prep.) for typical spectral energy distributions of GCs over the spectral range relevant to \Euclid). Each King model is then convolved with the PSF. We produce 3000 artificial GCs in total, making sure to avoid placing them inside the image gaps. We then run the detection and measurement pipeline of Sect.\,\ref{sc:Methods-photom} on the new images. Since that carries out forced photometry in the NISP bands at the position of the $\IE$ detections, we declare a GC detected in one of the infrared bands if the difference between input and output magnitude is smaller than 1.0.

We show in Fig.\,\ref{completeness-all} that the $I_\sfont{E}$, $Y_\sfont{E}$, $J_\sfont{E}$, and $H_\sfont{E}$ detections are 80\% (50\%) complete down to magnitudes 26.2 (26.5), 24.0 (25.3), 24.0 (25.4), and 23.6 (25.5), respectively. The 50\,\% and 80\,\% detection-completeness magnitudes are estimated by fitting a modified version of the interpolation function of \citealp{fleming}):
\begin{equation}
    f (m) = \frac{1}{2} \left[ {c_0 - \frac{\alpha(m - m_0)}{\sqrt{1 + \alpha^2 (m - m_0)^2}}} \right],
    \label{eq:Fleming}
\end{equation}
where $c_0$ is an additional free parameter compared to the original equation that corresponds to the completeness at the brightest magnitude. The $I_\sfont{E}-Y_\sfont{E}$ colour will be used later for GC identification. When both $I_\sfont{E}$ and $Y_\sfont{E}$ detections are required, we expect a completeness of about 80\% (50\%) for objects brighter than $I_\sfont{E}=25.2$ (26.2). These magnitudes are about $1.7$ and $2.7$ fainter than the typical turn-over magnitude of the GCLF, located at $\IE = 23.5$ (i.e., $M_V = -7.5$ with $ V - \IE= 0.5$, as described in Sect.\,\ref{sc:Methods-unknowngc}.).

\section{\label{sc:GCselection} Selection of GC candidates}

Our selection of GCs among all the detected sources is based on the properties of previously known GCs (Sect.\,\ref{sc:Data-gc}), and on the characteristics of the simulated GCs (Sect.\,\ref{sc:Methods-gcsim}). Here we first examine the characteristics of the known GCs in the $\IE$ stacked frame before defining a selection procedure. 

\subsection{\label{sc:Methods-knowngc}Compactness of the known GCs and of simulated GCs}

The top panel of Fig.\,\ref{compactness} shows the compactness indices measured for the two reference sets of known GCs in the 2-image VIS stacked frame. We match catalogues using a cross-match radius of $\ang{;;0.5}$. In the ACSFCS catalogue, we consider only the objects with a GC probability ($P_{\rm GC}$) larger than 0.95. The scatter in compactness originates from the natural range of sizes of globular clusters, combined with noise that mainly affects the candidates located very close to their host galaxy (closer than $\ang{;;30}$), for which the photometry is affected by the light of their host galaxies.

The compactness indices of the simulated clusters are displayed in the middle panel of Fig.\,\ref{compactness}. For index $C_{2-4}$ (left panel), the dispersion in the empirical stellar sequence is larger than the dispersion among artificial clusters with small radii. The simulated GCs are injected in the stacked frame rather than into raw frames, with a single PSF model; hence the point-source scatter shows that much of the dispersion in the empirical $C_{2-4}$ is likely due to source-to-source variations of the core of the PSF in our 2-image stacks, themselves resulting from a combination of intrinsic effects (spatial and colour-dependent variations of the PSF in the raw frames and changes in the PSF between the two combined exposures), effects of interpolation during the astrometric transformation, and possible weighting effects while stacking. All these effects are not typical of the EWS, and developing more detailed specific software for one particular non-standard image set was deemed too costly. The middle row of Fig.\,\ref{compactness} nevertheless indicates the expected location of clusters and shows that our reference samples (top panels), both spectroscopically confirmed and from ACSFCS, mostly contain objects with half-light radii smaller than about 5\,pc, with a sparsely populated tail to larger values. 

From Fig.\,\ref{compactness} (upper panel) we see that the majority of the known GCs have a larger compactness index than the vertical sequence of point sources. This is clear in particular for $C_{\rm 2-4}$. Point sources sources have average $C_{\rm 2-4}$ and $C_{\rm 4-8}$ values of 0.70 and 0.26. The majority of known GCs have $C_{\rm 2-4}$ in the range of 0.7--1.0, and $C_{\rm 4-8}$ between 0.2 and 0.5 This range of compactness is also consistent with the outcome of the GC simulation, as is shown in the middle panels of Fig.\,\ref{compactness}. Based on this figure, objects with $C_{\rm 2-4} > 0.8$ are resolved. This compactness threshold corresponds to a GC half-light radius of 2.5\,pc measured by ACSFCS.

\begin{figure*}[htbp!]
\begin{center}
\includegraphics[trim={0 0 0 0},angle=0,width=\linewidth]{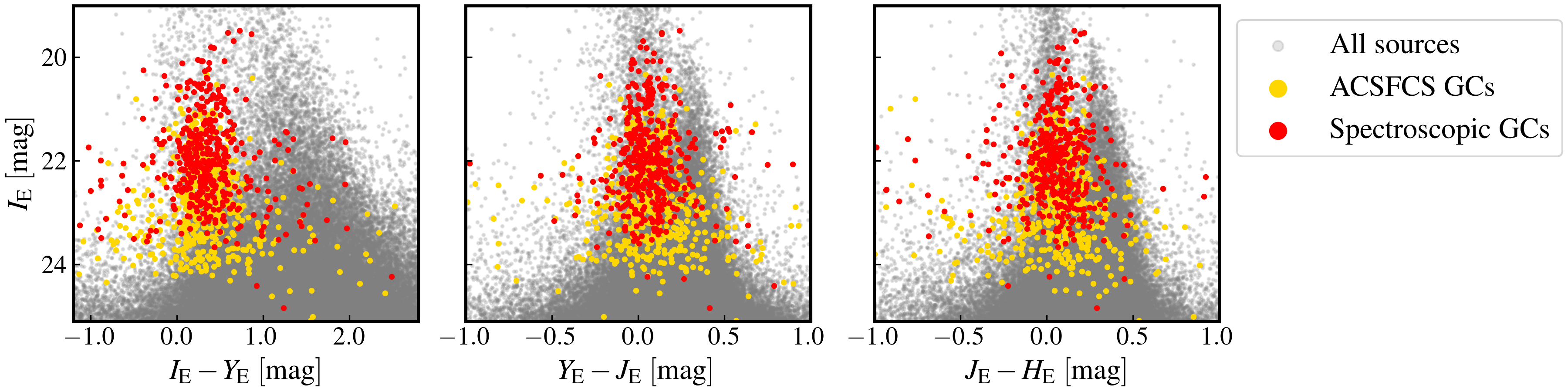}
\caption{Colour-magnitude diagram of the detected sources (grey points) versus the spectroscopically confirmed GCs (red points), and GCs in the ACSFCS catalogue (yellow points). Considering the photometric uncertainties of the fainter GCs (fainter than $\IE = 22$), the majority of the known GCs have $0.0<\IE-\YE<0.8$, $-0.3<\YE-\JE<0.5$, and $-0.3<\JE-\HE<0.5$. In this work, we use these colour ranges and apply a colour cut for GC selection.}
\label{cmd}
\end{center}
\end{figure*}

\begin{figure*}[htbp!]
\begin{center}
\includegraphics[trim={0 0 0 0},angle=0,width=\linewidth]{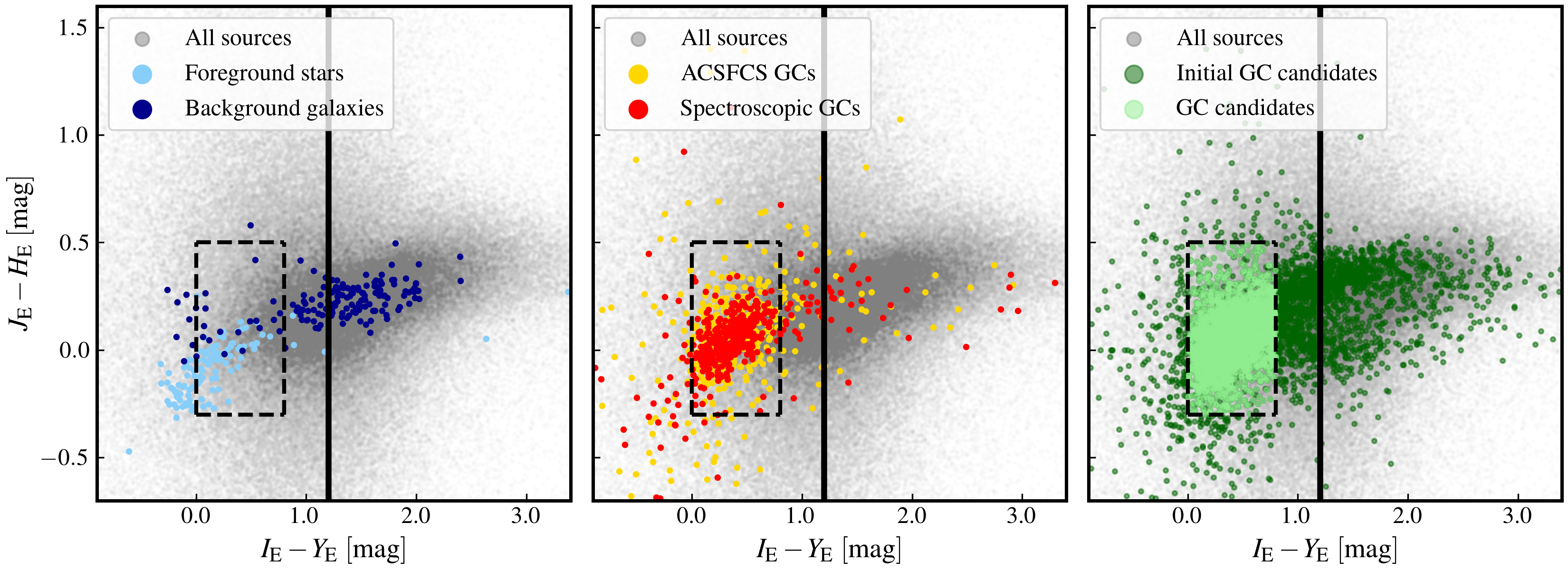}
\caption{The three panels show the \Euclid colour-colour diagram with ($\IE-\YE$) plotted against the ($\IE-\HE$) colour index. In each panel, the grey dots show the distribution of all photometric sources. 
In the left panel, we show the location of foreground stars and background galaxies in cyan and blue, respectively. The middle panels show spectroscopically confirmed GCs as red points, and GCs in the ACSFCS catalogue as yellow points. In the third panel, we show the initial GC candidates as dark green points and the retained GC candidates in light green colours. The applied colour cuts are shown as a dashed box. These colour cuts are used to identify the intracluster GC candidates in the ERO-F data. However, we use a more relaxed colour cut for identifying GC candidates around dwarf galaxies. For that, we apply an upper limit in \IE-\YE, shown by the vertical solid black line.}
\label{cc}
\end{center}
\end{figure*}

\subsection{\label{sc:Methods-unknowngc}GC identification}

We identify candidate GCs in the source catalogue in two steps, the first based on the compactness indices, and the second on colour. In the first step, we select marginally resolved sources by requesting compactness indices that are broadly within the range expected from known GCs and simulations. We select sources with compactness indices within 99\% quantile from the median value of $C_{\rm 2-4}$ and $C_{\rm 4-8}$ in a given magnitude range for the artificial GCs. Additionally, to take into account any other effects in GC compactness not included in the simulations (see Sect.\ref{sc:Methods-knowngc}), we extend the upper limits on compactness by 0.1\,mag. This value corresponds to half of the width of the stellar sequence in $C_{\rm 2-4}$. The resulting sample, after applying the compactness criteria, is shown in the lowest panels of Fig.\,\ref{compactness}. This selection picks up more than 80\% of the spectroscopic GCs, as well as the GC candidates in the ACSFCS catalogue. The majority of the remaining 20\% of GCs are objects that are in close vicinity to the bright galaxies and therefore their photometry is strongly affected by the galaxy; this effect is stronger for fainter GCs. Selection in this step includes 17\,596 sources brighter than $\IE = 25$ out of the initial 93\,995 sources (18.7\%).

Subsequently, in the second step of GC identification, we apply $\IE-\YE$ colour cuts to the initial GC sample. The range for these colour cuts is determined using the observed colours of known GCs in the ERO-F data, as shown in Fig.\,\ref{cmd}. With colour selection, we mainly aim to clean the initial GC sample from background galaxies and artefacts. Figure\,\ref{cc} shows the  $\IE-\YE$ colour of GCs, as well as foreground stars and background galaxies in the ERO-F. The stars and galaxies are selected based on their radial velocities (\citealp{Maddox-2019,Saifollahi2021b}). The background galaxies have $\IE-\YE>0.8$ while GCs have $\IE-\YE<0.8$. Therefore, we select objects with $\IE-\YE$ bluer than 0.8. We also apply a lower limit at $\IE-\YE=0$ given the colours of the known GCs in Fig.\,\ref{cc}. The colour-colour diagrams show that the known GCs and foreground stars have similar $\IE-\YE$ colour and therefore a careful GC selection based on the compactness of sources is essential (previous step). In addition to the $\IE-\YE$ colour cut, we apply $\YE-\JE$ and $\JE-\HE$ colour cuts. Here we aim to increase the purity of the sample while retaining as much completeness as possible. Therefore, we apply these colour cuts to objects with positive flux (from forced photometry at fixed positions) in $\JE$ and \HE. This means that we keep sources without detection in $\JE$ or \HE. The colour-ranges in $\YE-\JE$ and $\JE-\HE$ are between $-0.3$ and $0.5$ (for both colours). The GC colour range found here is consistent with the synthetic photometry of star cluster models (cf. Appendix \ref{ap:colors}). These models show that GCs with old stellar populations (older than 7\,Gyr) and sub-Solar metallicities ($Z<0.02$) have \Euclid colours $0.3<\IE-\YE<0.8$, $0.05<\YE-\JE<0.13$, and $-0.02<\JE-\HE<0.16$. In addition to the colour cuts, here we only include sources with \texttt{ELLIPTICITY} smaller than 0.5 (in \IE) as GC candidates. This is the observed upper limit for the known GCs as well as the artificial GCs. The procedure described above results in 5449 GC candidates as faint as $\IE = 25$, out of the 17\,241 sources in the initial GC sample (31.6\%).

When it comes to GC selection, there is always a trade-off between the completeness and purity (cleanness) of the final GC sample. The above-mentioned limits on colours are strict and are applied in order to have a clean GC catalogue. However, for identifying GCs around dwarfs, a more complete sample is desired. Therefore, for selecting GCs of dwarf galaxies we extend the colour range to take into account the scatter from photometric uncertainties, in particular for the fainter GCs. We select sources with $\IE-\YE < 1.2$ with no lower limit, and $\YE-\JE$ and $\YE-\JE$ between $-1.0$ and $1.0$. 

\section{\label{sc:Results}Results}

In the previous section, we performed source detection and photometry, produced multi-wavelength source catalogues, and selected GCs based on their compactness, colours, and ellipticity. Here, we assess the performance of our methodology and the completeness of the GC selection based on archival GC catalogues. Then, we investigate the properties of GCs within the Fornax cluster and around its dwarf galaxies.

\begin{figure*}[htbp!]
\begin{center}
\includegraphics[angle=0,width=\linewidth]{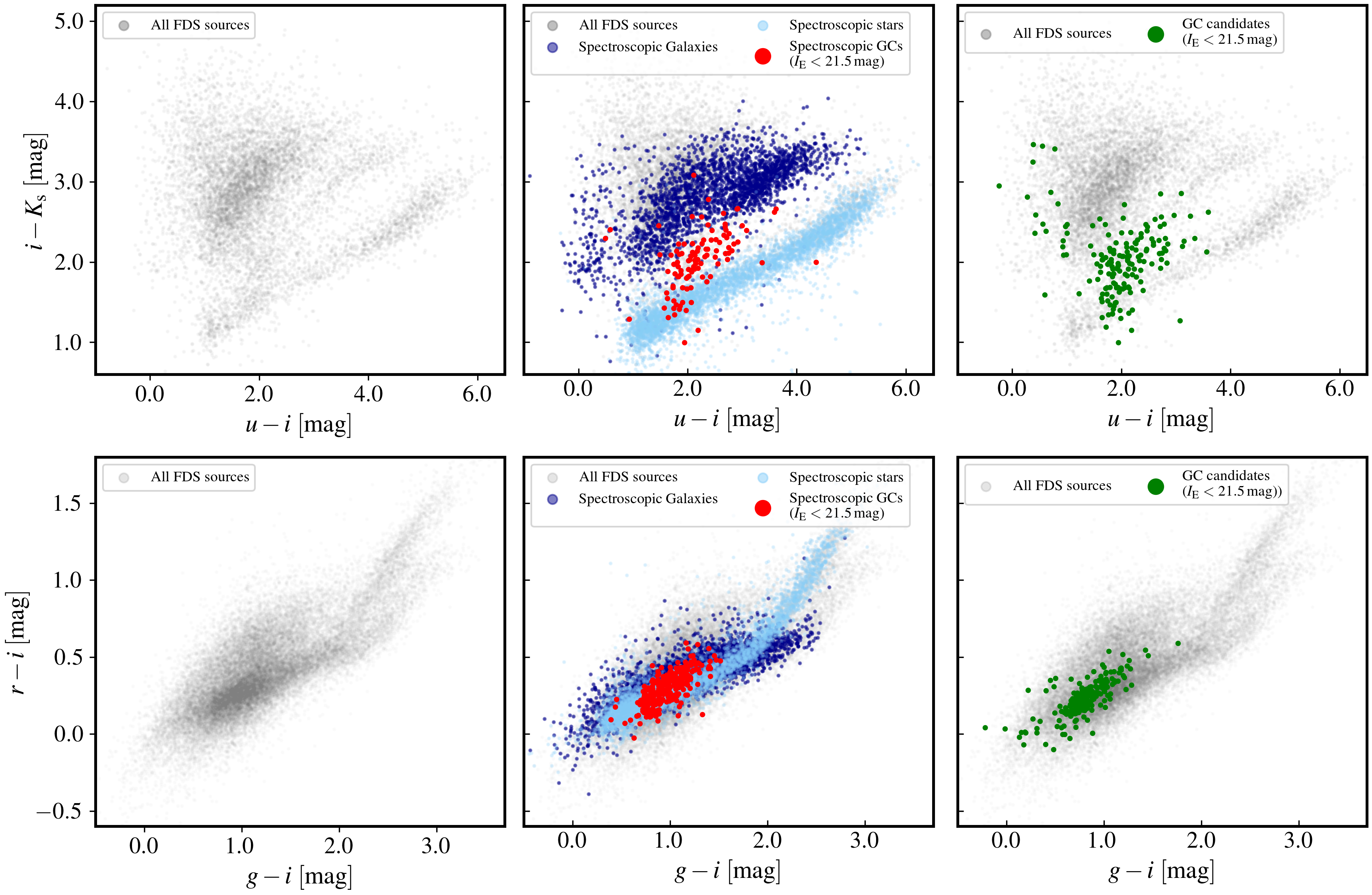}
\caption{Colour-colour diagrams showing $uiK_{\rm s}$  (upper panel) and $gri$ (lower panel) for the detected sources in ERO-F data (grey points), using the photometry of the Fornax Deep Survey (FDS, \citealp{fds}) provided in \citet{Saifollahi2021b}. It is known that GCs (red points) show a well-defined sequence in $uiK_{\rm s}$, separated from the stars except at the bluest end (light blue points) and from galaxies (dark blue points). More than 90\% of the bright GC candidates with $\IE < 21.5$ (green points) selected in this work are located on this diagram while less than 10\% are consistent with being a foreground star or background galaxy, respectively.}
\label{gc-ugrik}
\end{center}
\end{figure*}

\subsection{\label{sc:Results-vs-archive}The \Euclid GC sample compared to archive samples}

We compare our own source catalogues and GC (candidate) catalogues with the existing ACSFCS and spectroscopically confirmed GC catalogues (Sect.\,\ref{sc:Data-gc}).  For simplicity, we refer to the objects in these reference sets as known GCs. Our first aim is to estimate the completeness of our sample relative to the reference sets. There are 602 spectroscopically confirmed GCs in the FoV of our observations, out of which 600 (more than 99.5\%) are initially detected by \texttt{SExtractor} in \IE. In the case of the ACSFCS GC candidates, 906 candidates overlap with the data, of which 888 (98\%) are detected. This is consistent with the expectations from the completeness assessment based on artificial GCs. As seen in Fig.\,\ref{completeness-all}, GC detection in \IE\ is more than 95\% complete by magnitude $\IE=25.0$. GC selection based on compactness indices retains 80\% of the known GC sample. About half of the non-selected GCs are close to the centres of major galaxies, within about 30\,arcsec (2.9\,kpc), in regions where the light of the bright galaxies affects their photometric properties. 

When adding the cuts in colour space to our GC selection we find that our completeness decreases. Our colour selection includes only the main locus of the GC colours (see middle panel of Fig.\,\ref{cc}), and we have traded completeness for higher purity. Here we identify 70\% of the known GCs, including ACSFCS GC candidates and spectroscopically confirmed GCs, farther than 90\,arcsec from the massive galaxies. This rate is expected to be valid elsewhere in the data, within the cluster. This rate is about 80\% for dwarf galaxies considering the less strict colour cut applied for GC selection around them. This is consistent with the estimated completeness of combined $\IE$ and $\YE$ detections from GC simulations; that GC detection rate is 80\% down to $\IE = 25.2$, which is 1.7\,mag fainter than the typical turn-over magnitude of GCLF. After applying colour cuts, we identify 5631 and 1923 GC candidates in the Fornax cluster brighter than $\IE=25$ and $\IE=23.5$ (the typical GCLF turn-over magnitude), respectively, of which 4691 and 1556 are not in any of the previous spectroscopic catalogues and the ACSFCS candidates (with high GC probability).

The purity of the final GC sample is harder to evaluate. Normally, the purity of a photometric GC sample can be evaluated by comparing the sample with a complete sample of spectroscopically confirmed GCs and/or assessing the identified GC candidates in a field where no GCs are expected. Neither of these two approaches applies to the ERO-F given the incomplete spectroscopic GC samples and the high GC density environments in the central region of the Fornax cluster. Instead, we employ the $uiK_{\rm s}$ diagram 
($u-i$ versus $i-K_{\rm s}$) to make an assessment of the GC candidates, assuming that the true GCs lie on a specific GC sequence in this colour-colour space (\citealp{Munoz-2014}). For this purpose, we cross-match the GC candidates with the optical/near-infrared photometry of the Fornax cluster provided in \citet{Saifollahi2021b}. Given the limited depth of the ground-based data in $u$ and $K_{\rm s}$, we can only assess GC candidates brighter than $\IE = 21.5$ with this method. In this magnitude range, 90\% of the candidates have photometry in $u$ and $K_{\rm s}$. The upper panel of Fig.\,\ref{gc-ugrik} shows the $uiK_{\rm s}$ diagram for the detected sources in the data, including spectroscopic GCs, spectroscopic stars and galaxies, as well as the GC candidates in this work. By visually assessing this plot, more than 90\% of the GC candidates seem to be located on the GC sequence, while less than 10\% are consistent with being foreground stars and background galaxies, respectively. 

While the $uiK_{\rm s}$ diagram is powerful in distinguishing GCs and non-GCs, collecting sufficiently deep (ground-based) data in $u$ and $K_{\rm s}$ is very challenging and observationally expensive. In the meantime, the $gri$ colour-colour diagram ($g-r$ vs. $r-i$), even though it does not provide a clear separation between GCs and non-GCs (Fig.\,\ref{gc-ugrik} in the lower panel), can be assessed for GCs 2\,mag fainter than what is shown in the $uiK_{\rm s}$ diagram. Here we do not apply any colour cuts based on the ground-based optical/near-infrared photometry; however, these additional photometric data from deep ground-based surveys (e.g., FDS, DES, NGFS for the Fornax cluster) are complementary to \Euclid data and could be used in future for further cleaning the GC samples. For the EWS, the data from \Euclid ground segment will provide complementary deep optical imaging data. In particular, the Canada-France Imaging Survey (CFIS) and the Legacy Survey of Space and Time (LSST) data in $u$-band will be beneficial for further cleaning the catalogues of GC candidates derived from \Euclid observations from non-GCs.

\begin{figure*}[htbp!]
\begin{center}
\includegraphics[trim={0 0 0 0},angle=0,width=\linewidth]{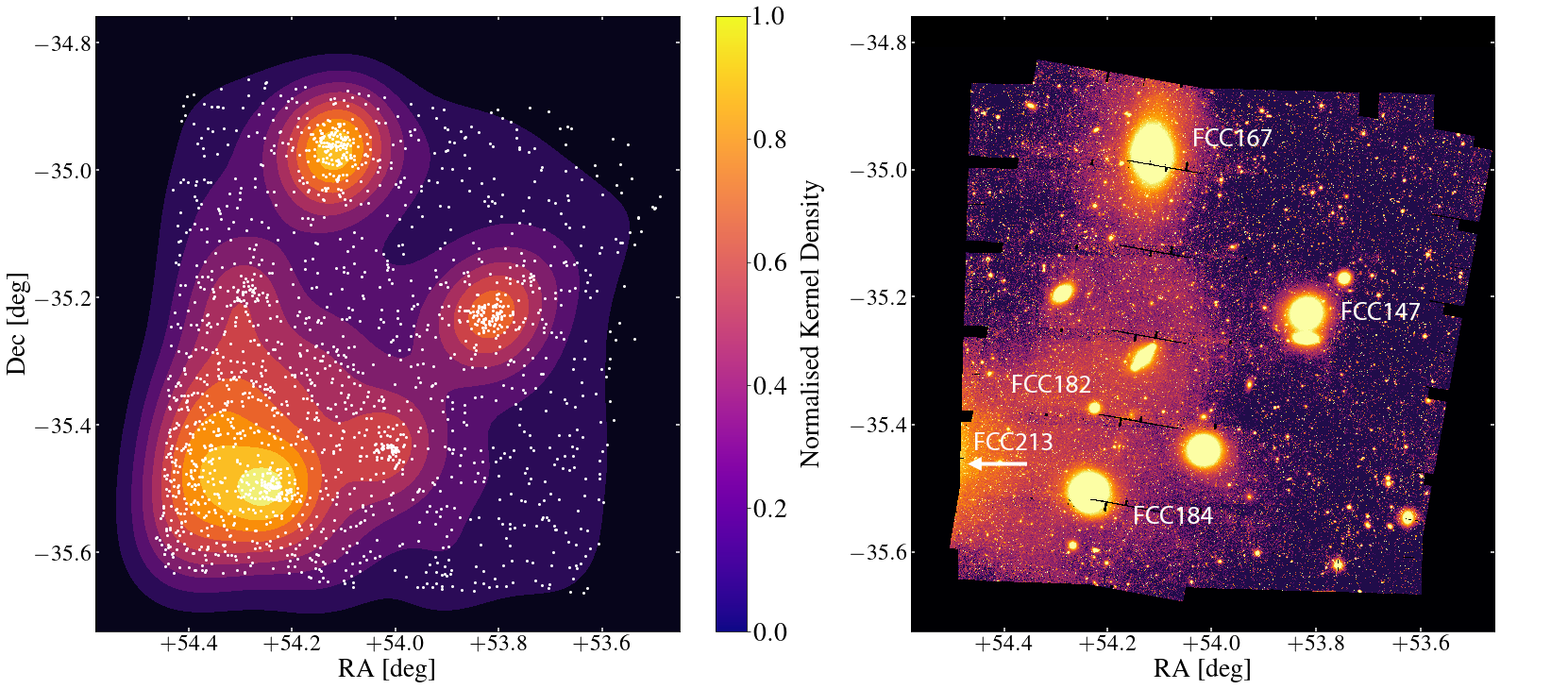}
\caption{Projected distribution and density map of GC candidates brighter than $\IE = 23.5$ (left panel) versus the diffuse in $\IE$ (background image in the right panel) in the ERO-F field of view in the Fornax cluster.}
\label{icgc}
\end{center}
\end{figure*}

\subsection{\label{sc:icgcs}The spatial distribution of GCs}

Due to the magnitude limits of the spectroscopic searches for GCs in the Fornax cluster, such surveys have only covered the brightest end of the GCLF and are highly incomplete at the faint end. The ACSFCS survey is deep but limited to small fields around the galaxies in their sample. Therefore, the distribution of the intracluster GCs (ICGCs) is less known outside the core of the Fornax cluster. On larger spatial scales, the most recent ground-based photometric searches of GCs within the virial radius of the cluster \citep{Cantiello2020,Saifollahi2021b} are expected to be contaminated by non-GCs, as discussed in the previous subsection. In the \Euclid data, we automatically identify GC candidates independent of their location and thus also in the intra-cluster region. Since our sample is expected to be about 70\% complete, we can examine the spatial distribution of intra-cluster GCs.

Figure\,\ref{icgc} shows the projected distribution and the density contours of the selected GCs brighter than $\IE=23.5$ across the FoV in the left-hand panel. In the right-hand panel, the four-image $\IE$ stacked frame is shown. In this frame, the main galaxies and extended diffuse intracluster light (ICL) within the Fornax cluster can be easily seen. At first glance when considering Fig.\,\ref{icgc}, the density distribution of the GCs appears to share some patterns with the ICL. In particular, it seems that both the GC distribution and the ICL show a characteristic excess in the south-east corner of the FoV, which is expected because this is the direction of FCC\,213 (NGC\,1399), the central dominant galaxy of the Fornax cluster that is located just outside the field of view \citep{dabrusco2016,Cantiello2020,diego2023}. Here, we discuss and investigate this possible connection between the ICL and the GC density distribution in more depth.

\subsubsection{\label{sc:icl}Intracluster light within the Fornax cluster}

The apparent distribution of the diffuse light in the ERO-F data must be interpreted with caution because of low-level straylight in some areas of the ERO-F image. \citet{Iodice-2016} used FDS deep images to highlight the extended diffuse halo of FCC\,213, the western part of which is also clearly seen in the south-east corner of our FoV. \citet{Iodice-2017} subtracted galaxy light and they were able to describe a much more extended diffuse-light distribution, which reaches (in projection) from FCC\,213 to FCC\,184 and to the smaller elliptical galaxy FCC\,182, and even to the edge-on spiral FCC\,170. The 4-image $\IE$ stack from ERO-F confirms the presence of diffuse light on these scales. The diffuse light is clearly significant from FCC\,213 to FCC\,184 and FCC\,167, but only marginally significant north of FCC\,170. The future EWS data, in which a correction for straylight in $\IE$ will be implemented that is not yet available, will be needed to check whether or not diffuse light really extends from the central regions of the galaxy cluster to FCC\,167 in the north of our FoV, or to FCC\,147 towards the west. Such extensions are not seen in the ground-based data of \citet{Iodice-2017}, but a connection to FCC\,167 would not be too surprising considering the current understanding of the three-dimensional structure of this area of the Fornax cluster. Indeed, according to the surface brightness fluctuation (SBF) estimates of \citet{Blakeslee2009}, recalled in Table\,\ref{gal-catalogue-m},  FCC\,213 and FCC\,167 are at similar distances, respectively, $(20.9\pm0.9)\,\mathrm{Mpc}$ and $(21.2\pm0.7)\,\mathrm{Mpc}$. The projected separation between these two galaxies is only about $250\,\mathrm{kpc}$. On the contrary, \citet{Blakeslee2009} place the spiral FCC\,170 about 1\,Mpc further ($21.9$\,Mpc), and FCC\,147 about 1\,Mpc closer to us ($19.6$\,Mpc).

\begin{figure*}[htbp!]
\begin{center}
\includegraphics[width=\linewidth]{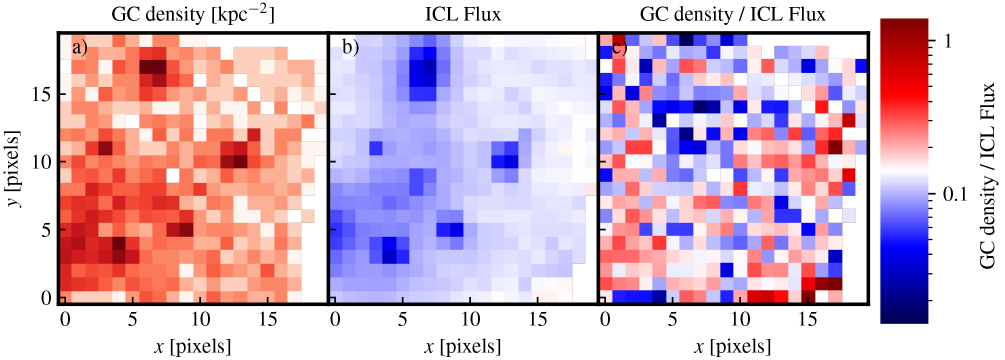}
\caption{(a): Normalised GC density (arbitrary unit) in the Fornax FoV binned into a 20 $\times$ 20 grid. (b): Normalised flux (arbitrary unit) of diffuse light within the same cells as panel (a), with all bright galaxies and other bright sources masked. (c): Ratio between the normalised GC density and the normalised flux of the diffuse light. The colour-bar on the left corresponds to the ratios shown in this panel.}
\label{fig:gc_density}
\end{center}
\end{figure*}

\begin{figure}
\begin{center}
\includegraphics[width=\linewidth]{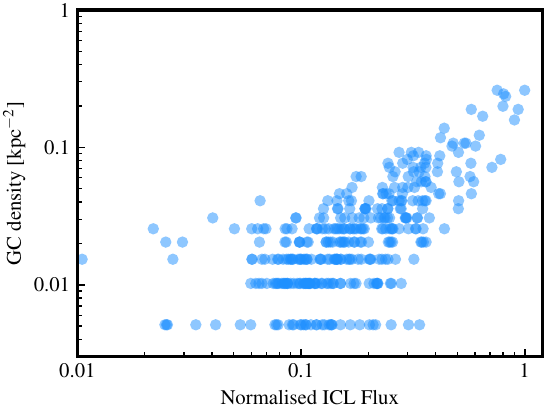}
\caption{The normalised diffuse light flux (Fig.\,\ref{fig:gc_density}b) plotted against the normalised GC density (Fig.\,\ref{fig:gc_density}a) for each bin of the 2D-histogram separately.}
\label{fig:gc_flux_den}
\end{center}
\end{figure}

\subsubsection{\label{sc:gc-vs-icl} Distribution of GCs vs. ICL}

To further investigate this possible connection between the ICL and the GC density, we divide the FoV of the ERO-F data into a 20 $\times$ 20 grid and calculate the GC density, the average ICL flux (in ADU), and the ratio between these two in each cell. This is shown in Fig.\,\ref{fig:gc_density}. To calculate the ICL flux, we mask the pixels in the image above a threshold that corresponds to roughly 4 times the half-light radius for the bright galaxies. This is necessary so that the calculation of the average flux is not dominated by bright galaxies and stars in the cells. As seen in Fig.\,\ref{fig:gc_density}, the area around the brightest galaxies in the FoV produces several local peaks in both GC density and ICL flux maps. Outside the proximity of the bright galaxies, we see substantial ICL flux, as well as GC density. To calculate the ratio between GC density and ICL flux, we use the same mask used earlier. As shown in Fig.\,\ref{fig:gc_density}c, the ratio between GC density and ICL flux scatters around the mean and there are no areas with obvious lack of GCs compared to what would be expected from the ICL, indicating that the ICL mostly follows the GC distribution.

This correlation between the ICL and GC density is visualised in Fig.\,\ref{fig:gc_flux_den}. Here we plot the normalised ICL flux against the normalised GC density for each cell of the two-dimensional histogram, to investigate whether GC density correlates with the ICL flux. We find a correlation between the average ICL flux and GC density indicating that indeed most of the GCs follow the ICL distribution. The empty area in the top-left of the figure implies that for a given density of GCs, there is always a minimum flux in ICL light that is significantly above the background. This implies that for a certain threshold in GC density, there is always a corresponding minimum diffuse light level, supporting the idea that this correlation is real and not an artefact (e.g., stray light).

Thus, while there is a scatter in the relation, the ERO-F data suggests that GCs trace the ICL light in galaxy clusters. A more global correlation between diffuse light and the density distribution of GCs was also found in ultra-deep JWST observations of a massive cluster at $z=0.4$ \citep{diego2023,martis2024}, and in the \Euclid observations of the Perseus galaxy cluster (\citealp{EROPerseusICL}). A combined view of GCs and ICL within the Fornax cluster emphasizes \Euclid's excellent view of the compact sources and low-surface brightness features at 20\,Mpc, despite the non-optimal data used in this work.

\begin{figure*}[htbp!]
\includegraphics[trim={0 0 0 0},angle=0,width=0.24\linewidth]{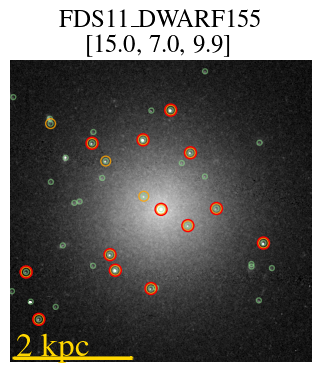}
\includegraphics[trim={0 0 0 0},angle=0,width=0.24\linewidth]{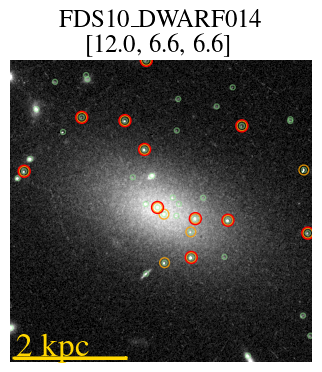}
\includegraphics[trim={0 0 0 0},angle=0,width=0.24\linewidth]{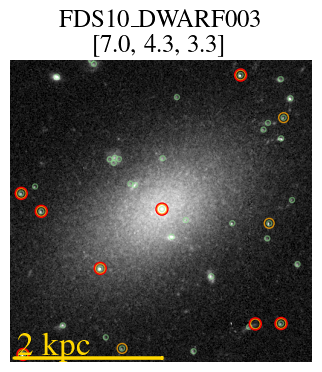}
\includegraphics[trim={0 0 0 0},angle=0,width=0.24\linewidth]{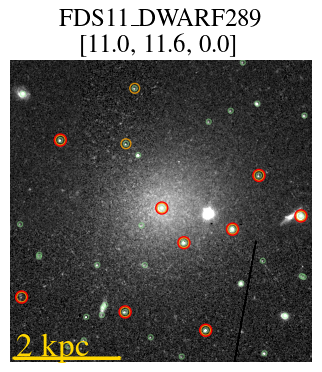}
\includegraphics[trim={0 0 0 0},angle=0,width=0.24\linewidth]{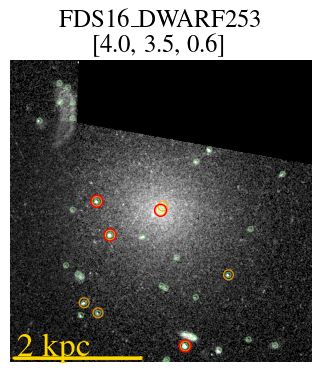}
\includegraphics[trim={0 0 0 0},angle=0,width=0.24\linewidth]{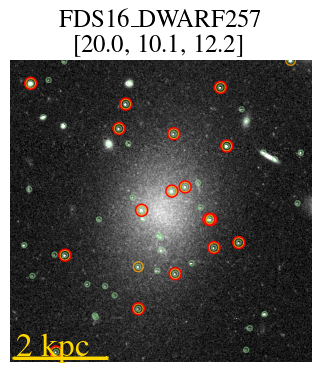}
\includegraphics[trim={0 0 0 0},angle=0,width=0.24\linewidth]{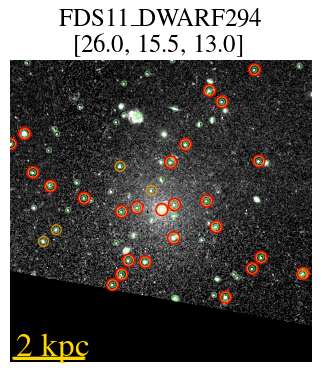}
\includegraphics[trim={0 0 0 0},angle=0,width=0.24\linewidth]{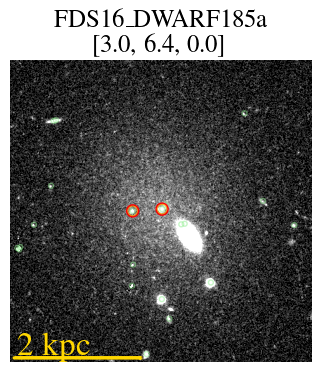}
\includegraphics[trim={0 0 0 0},angle=0,width=0.24\linewidth]{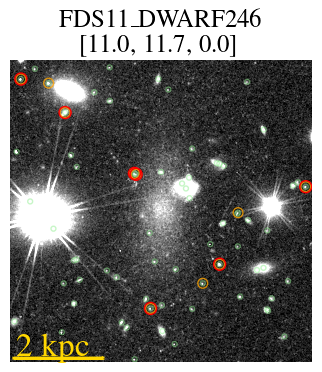}
\includegraphics[trim={0 0 0 0},angle=0,width=0.24\linewidth]{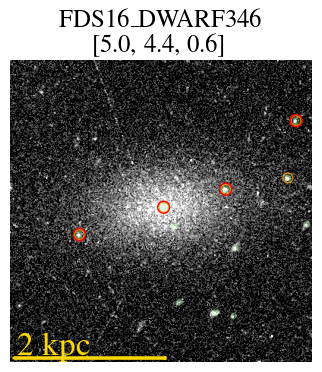}
\includegraphics[trim={0 0 0 0},angle=0,width=0.24\linewidth]{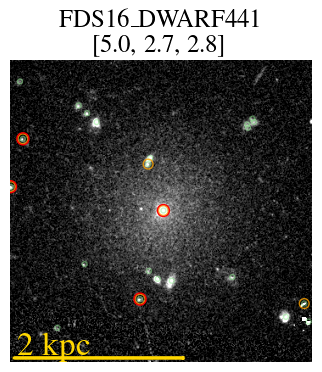}
\includegraphics[trim={0 0 0 0},angle=0,width=0.24\linewidth]{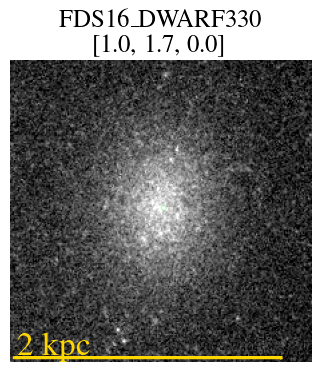}
\includegraphics[trim={0 0 0 0},angle=0,width=0.24\linewidth]{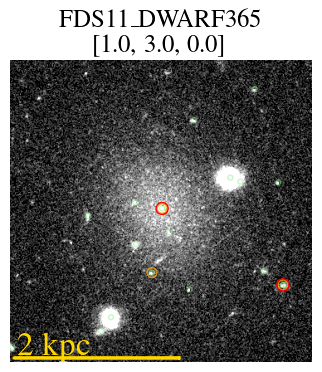}
\includegraphics[trim={0 0 0 0},angle=0,width=0.24\linewidth]{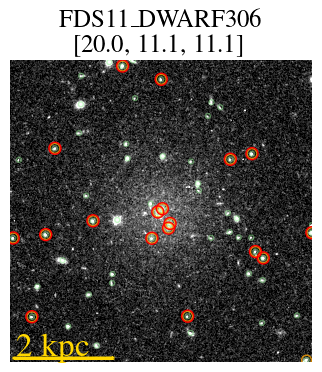}
\caption{\Euclid VIS image cut-outs centred on 14 of the most massive dwarf galaxies in the Fornax cluster, with stellar masses of $M_{*} > 10^7 M_{\odot}$, from the highest stellar mass (top-left) to the lowest stellar mass (bottom-right). The cutouts correspond to 4 times the half-light radius (of dwarf galaxies) on each side. North is up, east to the left. The red circles mark the final GC candidates that were selected based on their compactness and colours. Those candidates that pass the compactness criteria, but do not satisfy colour selection, are shown with orange circles. The green circles show all the sources identified around these dwarf galaxies. The three numbers below the name of each dwarf galaxy correspond to the number of GC candidates within 3$R_{\rm e}$, the number of GC candidates in the background normalised to the area within 3$R_{\rm e}$, and the estimated number of GCs corrected for incompleteness, respectively.}
\label{gc-dwarf}
\end{figure*}

\begin{figure*}[htbp!]
\begin{center}

\includegraphics[trim={0 0 0 0},angle=0,width=0.24\linewidth]{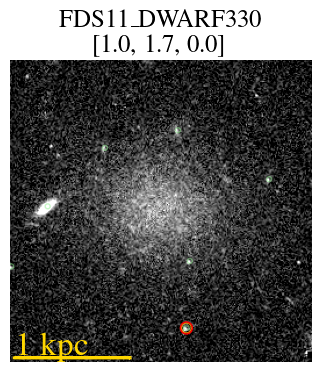}
\includegraphics[trim={0 0 0 0},angle=0,width=0.24\linewidth]{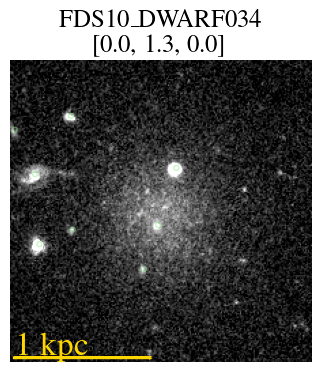}
\includegraphics[trim={0 0 0 0},angle=0,width=0.24\linewidth]{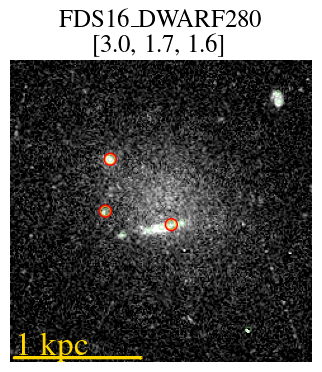}
\includegraphics[trim={0 0 0 0},angle=0,width=0.24\linewidth]{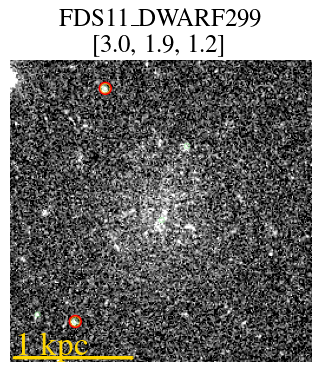}
\includegraphics[trim={0 0 0 0},angle=0,width=0.24\linewidth]{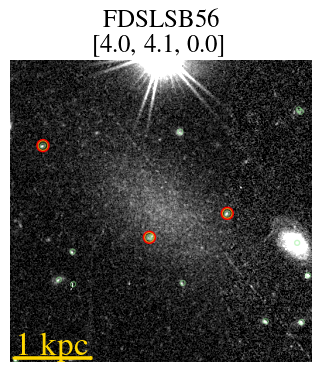}
\includegraphics[trim={0 0 0 0},angle=0,width=0.24\linewidth]{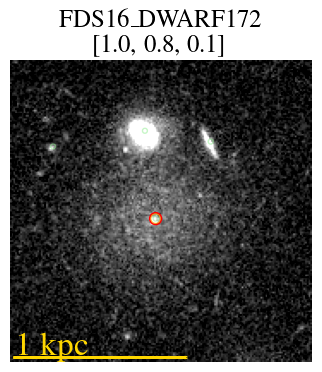}
\includegraphics[trim={0 0 0 0},angle=0,width=0.24\linewidth]{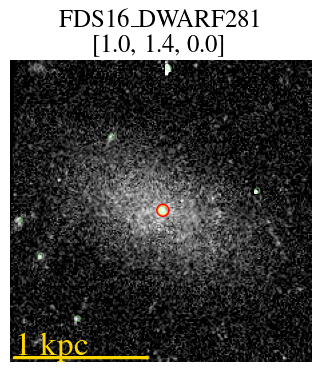}
\includegraphics[trim={0 0 0 0},angle=0,width=0.24\linewidth]{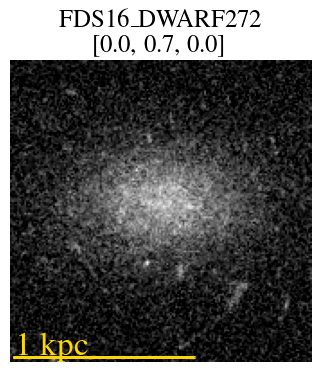}
\includegraphics[trim={0 0 0 0},angle=0,width=0.24\linewidth]{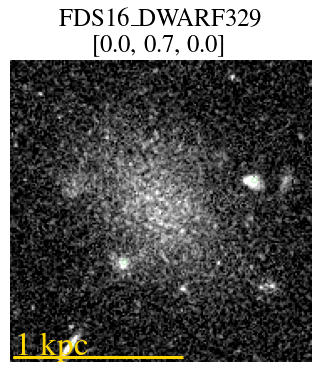}
\includegraphics[trim={0 0 0 0},angle=0,width=0.24\linewidth]{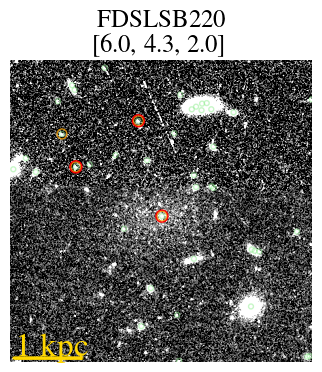}
\includegraphics[trim={0 0 0 0},angle=0,width=0.24\linewidth]{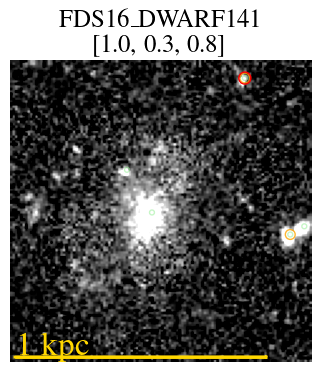}
\includegraphics[trim={0 0 0 0},angle=0,width=0.24\linewidth]{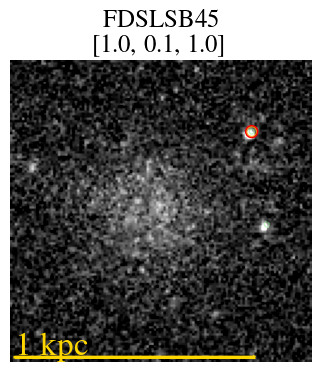}
\includegraphics[trim={0 0 0 0},angle=0,width=0.24\linewidth]{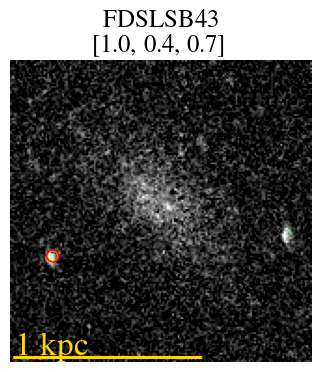}
\includegraphics[trim={0 0 0 0},angle=0,width=0.24\linewidth]{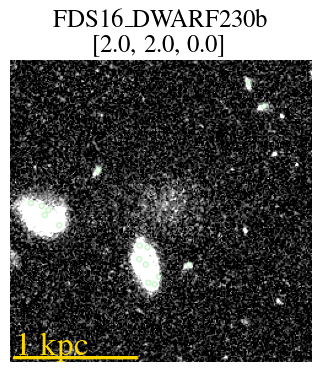}
\includegraphics[trim={0 0 0 0},angle=0,width=0.24\linewidth]{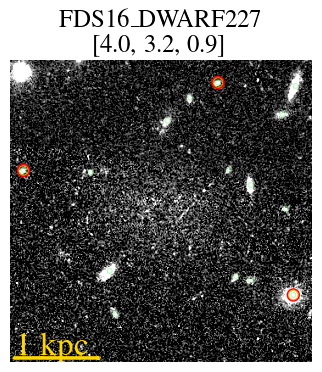}
\includegraphics[trim={0 0 0 0},angle=0,width=0.24\linewidth]{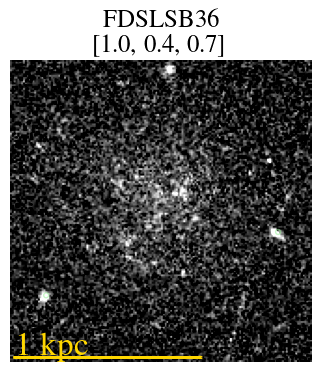}
\caption{Similar to Fig.\,\ref{gc-dwarf} but for dwarf galaxies with stellar mass $M_{*} < 10^7 M_{\odot}$, from the highest stellar mass (top-left) to the lowest stellar mass (bottom-right). On average, the galaxies in this stellar mass range do not host GCs; however, there are a few exceptions with GC number counts greater than zero. Considering the detections and possible contaminants, statistically one expects the dwarf galaxies FDSLSB45 (with $M_*=5.89 \times 10^5 M_{\odot}$), FDSLSB43 (with $M_* = 5.83 \times 10^5 M_{\odot}$), FDS16$\_$DWARF227 (with $M_* = 5.78 \times 10^5 M_{\odot}$), and FDSLSB36 (with $M_* = 5.76 \times 10^5 M_{\odot}$) with 1.0, 0.7, 0.9, and 0.7 GCs, respectively.}
\label{gc-dwarf+}
\end{center}
\end{figure*}

\begin{figure*}[htbp!]
\begin{center}
\includegraphics[trim={0 0 0 0},angle=0,width=1\linewidth]{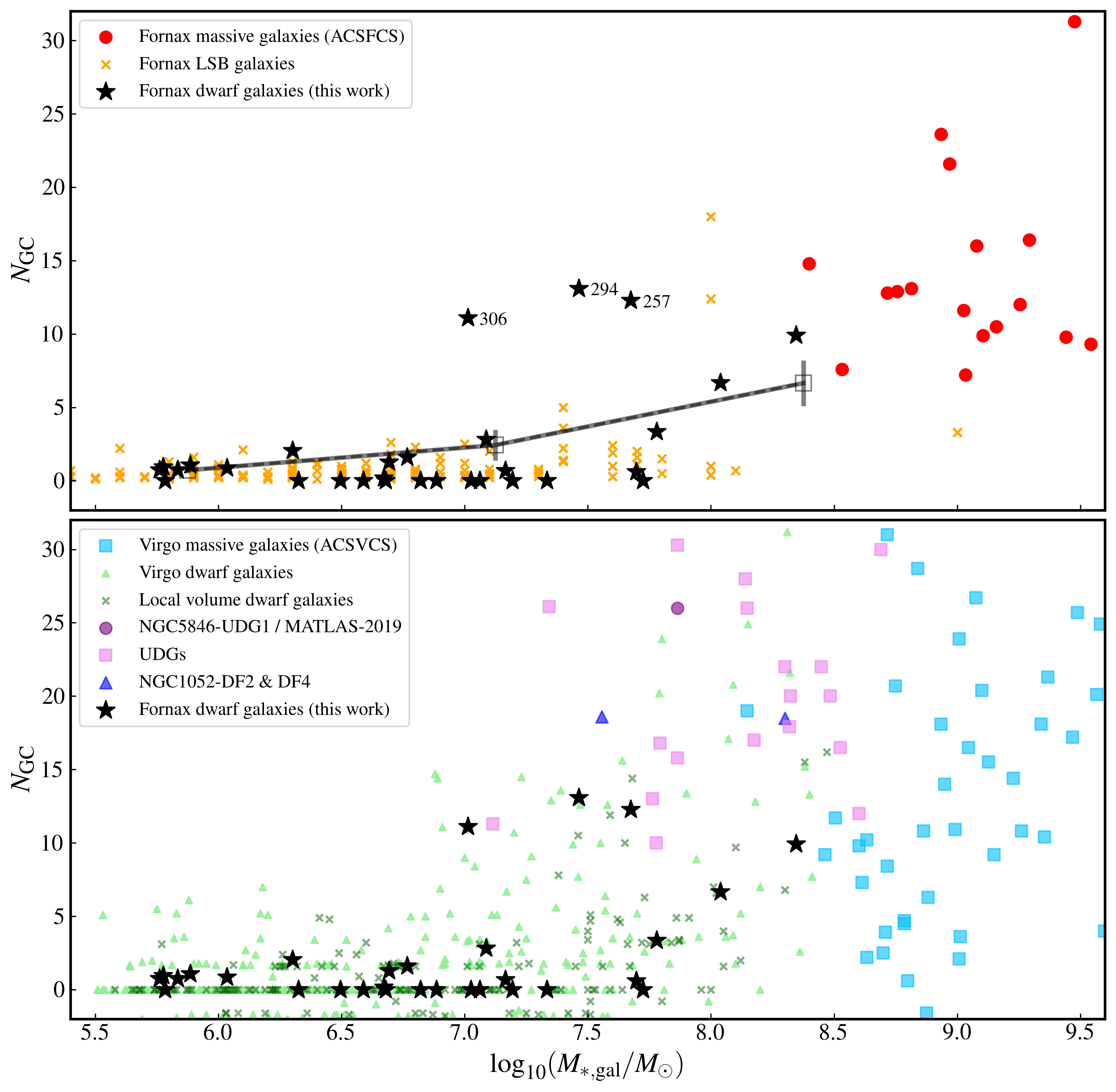}
\caption{Total number of GCs as a function of the stellar mass of the host for the dwarf galaxies in this work (black stars).  \textit{Top:} comparison with other studies in the Fornax cluster. The total GC numbers of massive galaxies from ACSFCS from \citet{Liu_etal19_ACSFCS} are shown as red disks and those of the Fornax cluster central LSB galaxies from \citet{prole2018} as yellow crosses. The black curve shows the average GC number for a given mass range of dwarf galaxies in this work (30 galaxies). Error bars correspond to the uncertainties on the mean. The three relatively GC-rich dwarf galaxies above this curve are indicated by the last three digits of their FDS name. \textit{Bottom:} GC numbers for galaxies in other environments. Massive galaxies from the ACS Virgo Cluster Survey \citep{peng2008} are shown as blue squares, Virgo cluster dwarf galaxies from \citet{carleton2021} as green triangles, dwarf galaxies in the local Volume from  \citet{carlsten2022} as green crosses, NGC5846-UDG1/MATLAS2019 from \citet{muller21} with a purple disk (the value of 54 GCs reported by \citealp{matlas-danieli} for this galaxy is beyond the range displayed), UDGs from \citet{saifollahi2022} and \citet{anna23} (see references therein) as pink squares, and NGC1052 DF2 and DF4 from \citet{zili-gclf} as dark blue triangles.}
\label{gc-dwarf-result3}
\end{center}
\end{figure*}

\begin{figure}[htbp!]
\begin{center}
\includegraphics[angle=0,width=\linewidth]{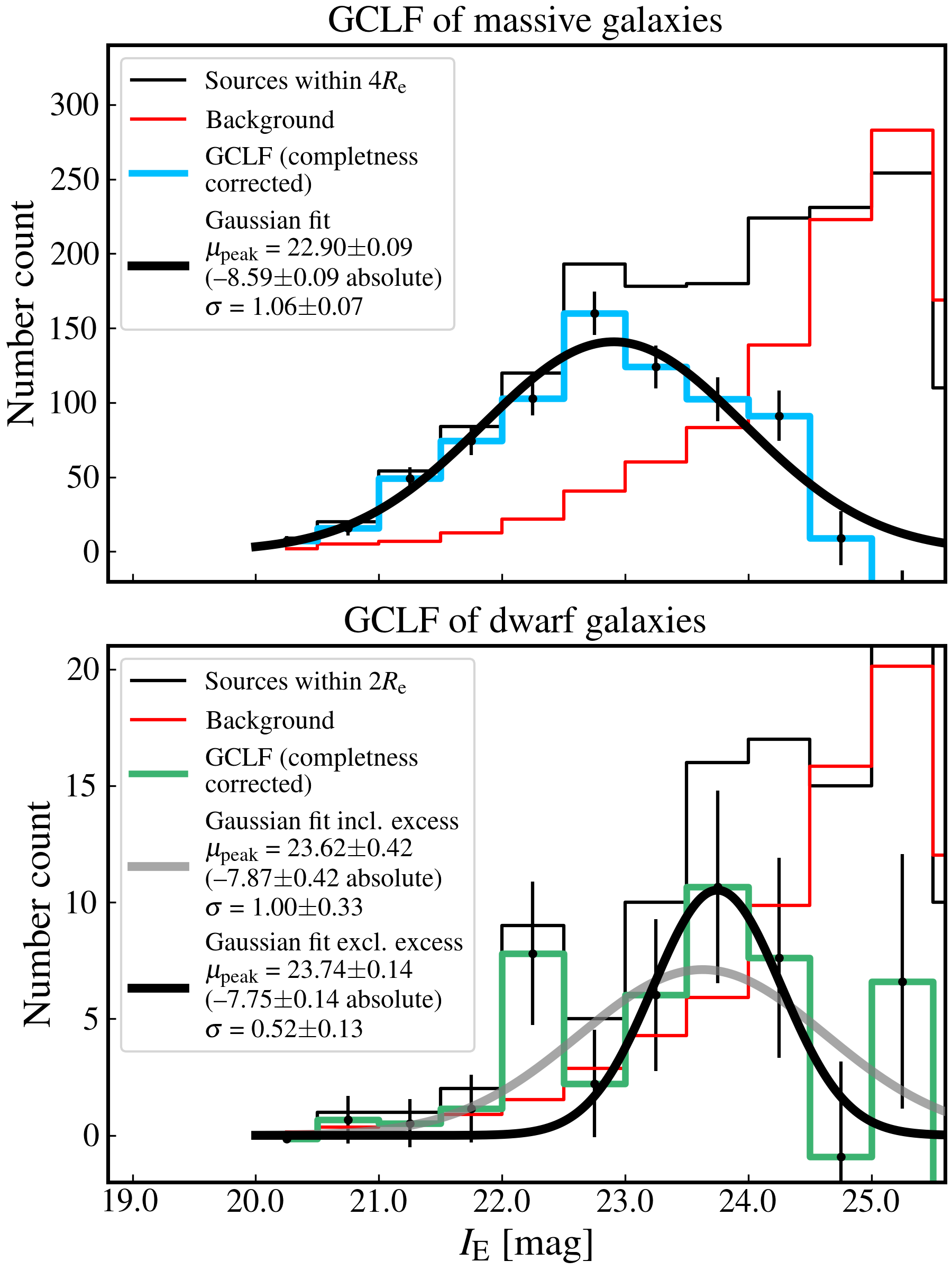}
\caption{The stacked GC luminosity function (GCLF) of massive galaxies (top) and dwarf galaxies (bottom). The GCLF is derived for the final GC candidates in this work (selected based on their compactness, colour and ellipticity). The black histogram shows all GC candidates within 4\,$R_{\rm e}$ and 2\,$R_{\rm e}$ of a given host galaxy for massive and dwarf galaxies, respectively. The red histogram shows the magnitude distribution of background sources, and the blue histogram is the background-corrected histogram. The best fit Gaussian to the background-corrected histogram is shown as the black line. The histograms are completeness-corrected. The errorbars represent the Poisson uncertainties for the number of GC within the magnitude bin and take into account the Poisson uncertainties from the (subtracted) background. The displayed best Gaussian fit for the GCLF of dwarf galaxies is done including and excluding the GC excess between $\IE = 22.0$ and $22.5$., shown with grey and black curves, respectively.}
\label{gc-dwarf-result1}
\end{center}
\end{figure}

\subsection{\label{sc:Results-gc-prop-of-dwarfs}GC properties of dwarf galaxies}

The dwarf galaxies we study here (30 galaxies) were taken from the FDS dwarf catalogue (\citealp{venhola2018,venhola2022}), where the authors apply size and colour criteria for discerning member dwarf galaxies from background galaxies. We present the list of these dwarf galaxies and their properties in Table\,\ref{gal-catalogue-d}. The results of exhaustive searches for new dwarf galaxies in the ERO-F data will be discussed elsewhere. Preliminary tests on these dwarf galaxies have shown that we can use the SBF method to estimate distances for the galaxies in the field, including several dwarfs. This will provide a great tool to establish group membership of the dwarf galaxies in the future.

In Sect.\,\ref{sc:Results-vs-archive} we discussed that the GC sample of dwarf galaxies in the ERO-F FoV is expected to be 80\% complete. Our sample, unlike the majority of previous works, extends to GCs well below the GCLF turn-over magnitude. However, at those faint limits, the contribution from intracluster GCs, as well as contamination (foreground stars and background galaxies) become increasingly important which one must carefully take into account. Figure\,\ref{gc-dwarf} shows the brightest dwarf galaxies in the sample and their GCs identified in this work. By visual inspection of the $\IE$ images of the dwarf sample, it seems that the dwarf galaxies with a stellar mass less than $M_*= 10^{7} M_{\odot}$, on average host no GCs, although with a few exceptions (Fig.\,\ref{gc-dwarf+}). 

Our search for GCs around dwarf galaxies in the ERO-F data also identifies the nuclear star clusters (NSCs, \citealp{turner2012}) of several dwarf galaxies in the sample as GC candidates. This is reasonable, given the similarity of the majority of NSCs of dwarf galaxies to bright GCs, in terms of compactness and colours. The suggested formation scenario of NSCs in dwarf galaxies is that NSCs are formed after inspiraling into the centres of galaxies (\citealp{yasna2018,Sanchez-Janssen2019,johnston,fahrion2022,roman2023}). We find that 47\% of all the dwarf galaxies in this study host an NSC, which is consistent with predictions for the nucleation fraction at that mass range \citep{Sanchez-Janssen2019}. Furthermore, for four dwarf galaxies in the sample, we identify a previously undetected NSC, namely galaxies FDS11$\_$365, FDS10$\_$034, FDS16$\_$172, and FDSLSB220.  This proves the capabilities of \Euclid for detecting the faintest NSCs, which is necessary for studying the nucleation fraction of dwarf galaxies. Considering the newly discovered NSCs, we find a nucleation fraction of 85\% for galaxies with a stellar mass between $M_*= 10^{7} M_{\odot}$ and $M_*= 10^{8} M_{\odot}$ (13 galaxies), which is higher than the expected range (50--60\%, \citealp{Sanchez-Janssen2019}). The NSCs of dwarf galaxies will be studied in detail in a future paper.

In this section, we focus on the properties of GCs around these dwarf galaxies. The intracluster GC candidates are a major source of contamination when studying GCs of dwarf galaxies since there is no way for us to determine whether the candidates of a given dwarf are bound to it. Because our observations are close to the core of the cluster, the density of intra-cluster GCs relative to the GCs of dwarf galaxies is high. Due to this and the fact that dwarfs host a small number of GCs, we use the stacked average properties of GCs around these dwarf galaxies in order to be less affected by individual contamination. 

\subsubsection{Total GC number}

The GC numbers are estimated by counting all the GC candidates as faint as $\IE = 25.0$ within a 480\,arcsec box around each galaxy. We count the GC candidates within 3$R_{\rm e}$ (dwarf-GC count), and between 5$R_{\rm e}$ and 15$R_{\rm e}$ (background count), and normalize the counts to the area within 3$R_{\rm e}$. Then, we subtract the latter from the former and correct the result for incompleteness (a factor of 1.25, considering the overall completeness of 80\% in GC identification), which then gives us an estimate of the total number of GCs ($N_{\rm GC}$). Note that this is different from the approach taken in the majority of recent works on GCs of dwarf galaxies that count GCs up to the turn-over magnitude of GCLF and correct the GC total number for the faint end of the GCLF (a factor of 2). Instead, we count all the GCs, considering that our observations are expected to reach the faint end of GCLF. Additionally, we justify the choice of 3$R_{\rm e}$ later in this section, where we study the radial distribution of GCs around their host dwarf galaxies. This choice is consistent with the GC radial distribution previously observed for dwarf galaxies (\citealp{carlsten2022}). Ideally, one needs to study the GCLF and the GC distribution of individual objects to estimate the GC number. However, this is not possible for low GC count objects such as the dwarf galaxies in our sample. 

The total number of GCs in a given galaxy correlates with its total dark matter halo mass (\citealp{harris2013,burkert2020}) as a consequence of the hierarchical formation of massive galaxies (\citealp{elbadry}). However, for dwarf galaxies, while the observational evidence supports such a correlation, more studies are needed to understand the physics behind this scaling relation. We present the total number of GCs for a given galaxy in Fig.\,\ref{gc-dwarf-result3}. For dwarf galaxies less massive than $M_* =  10^7 M_{\odot}$, the average number of GCs per galaxy is consistent with zero, except in a few cases. In particular, for a few dwarf galaxies with a stellar mass of less than $M_* = 10^6 M_{\odot}$, namely FDSLSB45 (with $M_*=5.89 \times 10^5 M_{\odot}$), FDSLSB43 (with $M_* = 5.83 \times 10^5 M_{\odot}$), FDS16$\_$DWARF227 (with $M_* = 5.78 \times 10^5 M_{\odot}$), and FDSLSB36 (with $M_* = 5.76 \times 10^5 M_{\odot}$) with 1.0, 0.7, 0.9, and 0.7 GCs, respectively. The dwarf galaxies with a stellar mass larger than $M_* = 10^7 M_{\odot}$ host between 0 and 13 GCs, while 17\% of them have a GC number of 5 and more.

Figure\,\ref{gc-dwarf-result3} compares the total GC numbers of dwarf galaxies in this work with the ones in the literature in the Fornax cluster (upper panel) and in other environments (lower panel). The upper panel of Fig.\,\ref{gc-dwarf-result3} presents the total GC numbers of \citet{Liu_etal19_ACSFCS} for the massive galaxies in Fornax using the ACSFCS observations, and those of \citet{prole2018} for LSB galaxies in the central regions of Fornax using the data of FDS. We also present the average GC number of dwarf galaxies in this work for four stellar mass bins (black curve). Overall, the GC numbers of galaxies seem to follow a continuous relation from the massive galaxies (red circles) to dwarf galaxies (back stars). Some of the dwarf galaxies show a higher than average GC number, namely FDS16$\_$DWARF257, FDS11$\_$DWARF294, and FDS11$\_$DWARF306 with estimated GC numbers of 12.2, 13.0 and 11.1 (background-subtracted and completeness-corrected). These GC-rich dwarf galaxies are indicated in the upper panel of Fig.\,\ref{gc-dwarf-result3} and are located above the black curve. 

In the lower panel of Fig.\,\ref{gc-dwarf-result3}, along with the total GC numbers in this work, we present the GC numbers of the massive (\citealp{peng2008}) and dwarf galaxies (\citealp{carlsten2022}) of the Virgo cluster, and of local Volume dwarf galaxies (\citealp{carlsten2022}). In addition to these GC samples, we show the results for some GC-rich ultra-diffuse galaxies (UDGs, \citealp{vd15}) in the literature studied using HST \citealp{saifollahi2022,anna23}.\footnote{\citet{anna23} provides spectroscopic properties as well as GC numbers of UDGs and it is compilation of several works. Here we use the GC numbers from the catalogue, except the GC numbers of some of the Coma cluster UDGs DF44, DFX1, DF07, and DF17. For these Coma cluster UDGs, we use the more recent GC number estimates in \citet{saifollahi2022}.} We also included NGC5846-UDG1/MATLAS2019 in the NGC5846 group (\citealp{muller21,matlas-danieli}), and NGC1052-DF2 and NGC1052-DF4, two UDGs lacking dark matter in the NGC1052 group (\citealp{vd-df2,vd-df4}). In this context, The GC numbers of the three GC-rich dwarf galaxies (FDS16$\_$DWARF257, FDS11$\_$DWARF294, FDS11$\_$DWARF306) seem to be an extension of the GC-rich UDGs with 2 to 3 times more GCs than average at a given stellar mass. These dwarf galaxies are a UDG by definition (FDS11$\_$DWARF306 with $R_{\rm e}=1.45$\,kpc is marginally a UDG). The galaxy FDS11$\_$DWARF246 is another UDG in the sample with zero GCs. This dwarf galaxy is located close to the FCC\,213 and therefore, the estimated GC number for this object is highly affected by the ICGS. In the future, with such information on the GC contents of dwarf galaxies as well as their stellar populations, we will be able to investigate the various formation scenarios of UDGs and dwarf galaxies (\citealp{anna23,buzzo24}). The latter would be possible using \Euclid's data combined with the ground-based surveys.

\subsubsection{GC luminosity function (GCLF)} 
\label{sc:GCLF}

The GCLF of massive galaxies typically follows a near-Gaussian profile in logarithmic units of luminosity with a peak around $M_V=-7.5$ (\citealp{rejkuba2012}) and a standard deviation of $\sigma=1.2\,\mathrm{mag}$. Trends with host galaxy properties have been identified, such as a decrease in the width of the GCLF with decreasing luminosity \citep{villegas2010} (hereafter V+10). The study of the GCLF of low-mass dwarf galaxies requires larger galaxy numbers than for massive galaxies, and the GCLF properties in this regime remain controversial. While earlier works showed that the GCLF is Gaussian for dwarf galaxies (\citealp{Georgiev}), more recent work on UDG and LSB galaxies has suggested that there might be a wide range of GCLFs for dwarf galaxies (\citealp{df2-gc}). This could be a consequence of small galaxy samples, as well as small numbers of GCs within dwarf galaxies, or bias towards certain environments and conditions.

Before producing the GCLF, we repeat the photometry of GC candidates in $\IE$ using an aperture with a radius of 3 times the FWHM in $\IE$, about 5.7\,pixels. This aperture is twice the size of the aperture used in Sect.\,\ref{sc:Methods-photom} and provides a better estimate of the total luminosity of the GCs. This choice of aperture size is particularly important for studying the GCLF at the bright end, where GCs are typically larger. For these GCs, our earlier photometry with smaller apertures would underestimate their total luminosity by about 10\% for the largest GCs, while with the larger aperture, the fraction of missing light is below 1\%. Using a smaller aperture for photometry of GCs would lead to more accurate magnitude and colour estimates for fainter and smaller GCs, which is critical for GC identification and that is why it was used earlier in Sect.\,\ref{sc:Methods-photom}.

Here, we produce the GCLF using all GC candidates within $4\,R_{\rm e}$ of their host for massive galaxies, and within $2\,R_{\rm e}$ for dwarf galaxies. We exclude the central $0.1\, R_{\rm e}$ of dwarf galaxies to remove the NSCs, and the central $0.2\, R_{\rm e}$ of massive galaxies where GC selection becomes too difficult. The different selection radii are implemented because the radial extent of GC systems relative to the effective radius of their host is known to scale with the host galaxy mass \citep{debortoli2022}, and because a selection radius larger than $2\,R_{\rm e}$ for dwarfs is not warranted by our data (see Sect.\,\ref{subsc:gc-radial}). For massive galaxies, our selection radius is about 15\,\% larger than the typical effective radius of the GC systems found by \citet{hudson2018}.

The stacked GCLF of dwarf galaxies includes dwarf galaxies with a stellar mass larger than $M_*= 10^{7} M_{\odot}$ (14 dwarf galaxies, shown in Fig.\,\ref{gc-dwarf}), which have GC numbers larger than 1. We sum up all the GC candidates for these 14 dwarf galaxies and derive the stacked luminosity function of GC candidates of dwarf galaxies. Next, we estimate the average contamination from the background and subtract it from the luminosity function of GC candidates. The background includes intracluster GCs, as well as any foreground stars or background galaxies that passed the GC-selection criteria. We estimate its contribution by selecting GC candidates between 5$R_{\rm e}$ and 15$R_{\rm e}$ from the 14 dwarf galaxies used for this analysis, normalised to the total area of the background and multiplied by the total area within $2.5\,R_{\rm e}$ of the 14 dwarf galaxies. Finally, we correct the background-subtracted luminosity function for the incompleteness of detection, based on the results of Sect.\,\ref{sc:Methods-gcsim}. We use the same background and completeness correction procedure for deriving the GCLF of massive galaxies. The outcome is the stacked GCLF of galaxies shown in Fig.\,\ref{gc-dwarf-result1}. For the massive galaxies (top panel), we do indeed recover a Gaussian GCLF for our GC candidates. It has a turnover at $\IE=22.90 \pm 0.09$, which corresponds to an $\IE$ absolute magnitude of $-8.59 \pm 0.09$. A conversion between $\IE$ and $V$ for old and metal-poor GCs leads to $M_V = -8.06$\, mag. The width of the GCLF is $\sigma = 1.06 \pm 0.07$. The derived GCLF parameters are consistent with the GCLF parameters reported in V+10 using the ACSFCS data.

The GCLF of dwarf galaxies is shown in the bottom panel of Fig.\,\ref{gc-dwarf-result1}. Its appearance is more irregular than that of the massive galaxies; despite stacking we remain in a regime of small number statistics with this single-field observation. The initial dwarf GC-candidate sample contains 112 objects. After statistical correction for the background, this number is reduced to about 38 objects, which we distribute in 12 bins. The statistical errors are estimated with Poisson statistics before background subtraction and then propagated. We have obtained best-fit Gaussian estimates of the intrinsic GCLF of dwarf galaxies in two ways. Firstly, we computed heteroscedastic maximum-likelihood parameters for the {\em binned}\ background-corrected distribution (Fig.\,\ref{gc-dwarf-result1}). Secondly, we computed parameters using {\em individual}\ measurements (before any background subtraction), for a model that combines a representation of the background magnitude distribution, a Gaussian GCLF for the dwarfs, the completeness correction, and a prior on the proportion of contaminants that is compatible with the statistical background subtraction of the first method. 
In the second approach, the parameters are the mean and standard deviation of the Gaussian and the relative scaling of the Gaussian relative to the background. 
This latter method takes into account the individual photometric uncertainties. Details for it are provided in Appendix\,\ref{ap:gclf}. Both methods can be repeated, with variations for instance in the exact bin boundaries for the first one, or in the exact range of magnitudes taken into account for the second. 

Using the first method, we find that the best-fit Gaussians have a mean magnitude of $\mu_{\text{peak}} = 23.6 \pm 0.42$, and a GCLF width of $\sigma = 1.00 \pm 0.33$. The second method based on individual measurements gives $\mu_{\text{peak}} = 23.2 \pm 0.2$, and a GCLF width of $\sigma = 0.9 \pm 0.2$. The uncertainties quoted here account for changes in the best-fit parameters depending on the details of the fitting method. However, the maxima of the posterior probability distributions are quite flat, and a broader range of parameters has probabilities within one $e$-fold of the best fit (contours in Fig.\,\ref{figApp:GCLF_4panels}c). The errors for $\mu$ and $\sigma$ are correlated (same figure). Extrapolating the results of ACSFCS to lower luminosities similar to the dwarf galaxies in this work, the estimated GCLF peak is, within uncertainties, consistent with that of V+10, however we find a wider GCLF. Extrapolating the results from V+10 suggests a narrow GCLF with $\sigma=0.5$ for dwarf galaxies of the same luminosity as in our sample, while our probability analysis shows that such a small $\sigma$ is unlikely to describe the observed GCLF. We estimate that the ratio between the probabilities of having a broad GCLF with $\sigma=0.9$ and a narrow GCLF width with $\sigma=0.5$ is more than 40. This broadening of the GCLF of dwarfs can not arise from a distance difference: based on the SBF distance for the Fornax cluster galaxies (\citealp{Blakeslee2009}), the bright galaxies in ERO-F are at $(20 \pm 1)$\,Mpc. This $\pm$\,5\% scatter in distance translates into roughly $\pm$\,0.1\,mag on the distance modulus, which could contribute only about $\pm$\,0.1\,mag to the width of the GCLF.

The larger GCLF width found here seems to arise from a population of bright GCs with $\IE$ between 22 and 22.5, corresponding to $M_{V}$ between $-9$ and $-8$. In particular, the binned GCLF of Fig.\,\ref{gc-dwarf-result1} displays a peak for $\IE$ between 22 and 22.5. The aspect of the peak depends to some extent on binning choice, but a clear jump at $\IE=22.4$ is seen even in the distribution of individual GC-candidate magnitudes. Although this peak represents only a 1.5 to 2\,$\sigma$ deviation from the bin-value expected from a smooth model, its presence has the effect of favouring larger values of $\sigma$. Given the location of this second GCLF peak at $M_{V}=-9$ this GCLF is similar to the stacked GCLF of two UDGs in the NGC1052 galaxy group lacking dark matter, NGC1052-DF2 and NGC1052-DF4 (\citealp{zili-gclf}). Such massive GCs around low-mass dwarf galaxies are valuable objects for studying star cluster formation. Repeating the fitting after excluding the excess of bright GCs using the first and second methods results in a lower $\sigma$, about $\sigma = 0.52 \pm 0.13$ and $\sigma = 0.5 \pm 0.2$ (see Fig.\,\ref{figApp:GCLF_4panels+}), respectively. These values are consistent with the above-mentioned expectations for the GCLF of dwarf galaxies. These results overall show that the GCLF of dwarf galaxies is not a single Gaussian distribution.

The relevance of the bright peak in the dwarf GCLF will have to be re-investigated once \Euclid has covered the whole Fornax cluster. The effect of the bright GCs on the fit parameters justifies a closer examination of those objects. We find that the majority of the GC candidates composing the peak are around only three of the 14 dwarf galaxies (FDS11$\_$DWARF155, FDS10$\_$DWARF014, FDS11$\_$DWARF294) and each contribute two GCs in the bright GC excess. The first two out of these three dwarf galaxies are the most massive ones in the sample, with stellar masses equal to or larger than $M_*= 10^{8} M_{\odot}$. Additionally, all three dwarf galaxies have a bright and dominant NSC, and a relatively high number of GCs among the dwarf sample, with 9.9, 6.6, and 13.0 GCs. The presence of the bright GC components might be connected to the massive halos of these galaxies, since these three dwarf galaxies are among the ones with the largest GC number (\citealp{forbes2024}) and brightest NSCs. Two of them are 100\,kpc from the centre of the Fornax cluster while the third one is at 300\,kpc. We will re-investigate this finding using the future \Euclid data for the Fornax cluster and other nearby systems.

\begin{figure}[htbp!]
\begin{center}
\includegraphics[trim={0 0 0 0},angle=0,width=1\linewidth]{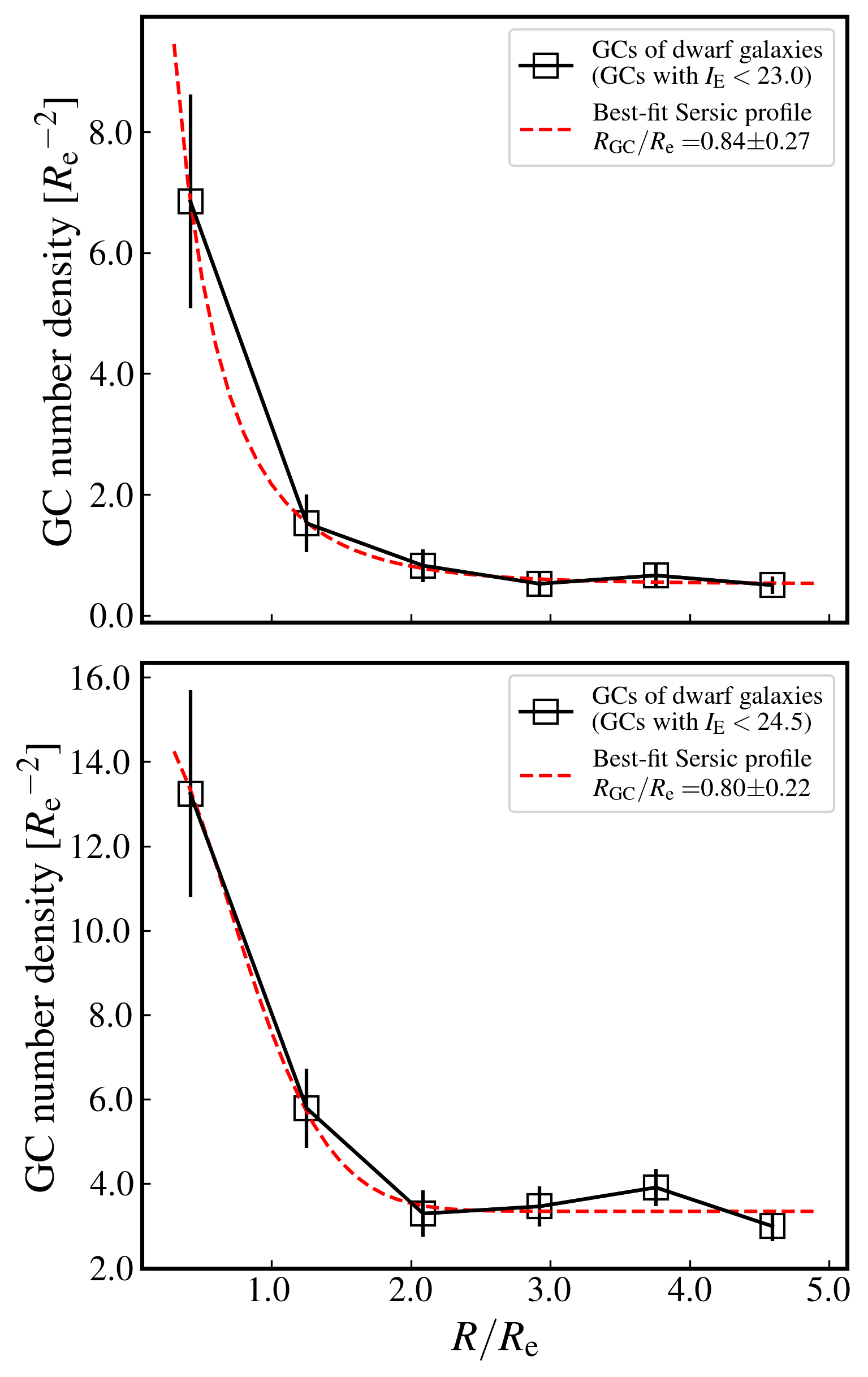}
\caption{Stacked radial distribution of the GC number density of dwarf galaxies (14 galaxies with stellar masses of $M_{*} > 10^7 M_{\odot}$), for GCs brighter than $\IE=23$ (upper panel) and GCs brighter than $\IE=24.5$ (lower panel). The GC radial distance from the centres of their host galaxies (x-axis) is normalised to the host galaxy's effective radius ($R_{\rm e}$). The GC number density (y-axis) represents sum of GCs around all the 14 dwarf galaxies has the unit of $R_{\rm e}^{-2}$. For a typical dwarf galaxy with $R_{\rm e} = 1$\,kpc this unit corresponds to kpc$^{-2}$. The red-dashed curve demonstrates the best-fit S\'ersic profile for the binned data and the displayed error bars. The errors correspond to Poisson uncertainties of the number of GCs normalised to the unit area.}
\label{gc-dwarf-result4}
\end{center}
\end{figure}

\subsubsection{GC radial distribution}
\label{subsc:gc-radial}

With the \Euclid data and the GC sample presented here, one can inspect the stacked radial profile of GCs. We study this profile for the dwarf galaxies. For that purpose, we normalize the galactocentric distances of GCs to their host galaxy to the effective radius of the host ($R_{\rm e}$). Additionally, we only consider dwarf galaxies with a stellar mass of $M_*= 10^{7} M_{\odot}$ and larger (14 galaxies) because they seem to have a few GC candidates. Stacking the GC radial distributions of these dwarf galaxies, we aim to enhance the low GC number statistics and reduce the statistical effect of contamination. 

Figure\,\ref{gc-dwarf-result4} presents the GC radial profile of these dwarf galaxies. As is seen in this figure, GCs are distributed out to 3$R_{\rm e}$ from their host dwarf galaxies. Beyond 3$R_{\rm e}$, the GC distribution reaches a plateau (i.e., has reached the ICGCs), consistent with being due to a uniform background. This is valid for the GCs brighter than $\IE = 23.0$ (upper panel) and $\IE = 24.5$ (lower panel). We fit a S\'ersic profile to the stacked GC radial distribution to characterize this distribution. We quantify the radial profile of GCs by using the ratio between GC half-number radius and galaxies' effective radius ($R_{\rm GC}/R_{\rm e}$). We estimate $R_{\rm GC}/R_{\rm e} = 0.84 \pm 0.27$ and $R_{\rm GC}/R_{\rm e} = 0.80 \pm 0.22$ for the GCs brighter than $\IE = 23.0$ and $\IE = 24.5$, respectively. Note that the GC radial profile of dwarf galaxies presented here is produced excluding the NSCs. The GC radial profiles in Fig.\,\ref{gc-dwarf-result4} are based on 67 and 305 GC candidates within 5$R_{\rm e}$, for GCs brighter than $\IE = 23.0$ and $\IE = 24.5$, respectively. However, about half of these GCs are around three GC-rich dwarf galaxies in the sample (FDS16$\_$DWARF257, FDS11$\_$DWARF294, and FDS11$\_$DWARF306). 
Therefore, the GC radial profile might be highly biased to these systems. However, we do not see any difference in GC profiles when we divide the three GC-rich systems and the rest of the dwarf galaxies. These values imply that the GCs of dwarf galaxies, regardless of being GC-rich or GC-poor, and UDG or non-UDG, have a more compact radial distribution compared to the GC-rich dwarf galaxies. This could imply that the dwarf galaxies have gone through a similar evolutionary path as the UDGs, but slightly different because they do not host a higher number of GCs than average while UDG have higher (2--3 times more) GC numbers.

These results do not support the common view of GC profile of dwarf galaxies with $R_{\rm GC}/R_{\rm e} = 1.5$ (\citealp{vd17,lim2018}). This ratio has been studied in a few more recent works using ground-based data of dwarf galaxies (\citealp{carlsten2022}) and space-based data for GC-rich UDGs (\citealp{muller21,montes21,saifollahi2022,janssens}). \citet{carlsten2022} studied dwarf galaxies in the local Volume, in galaxy groups and also in the Virgo cluster and found $R_{\rm GC}/R_{\rm e}=1.06$ and $R_{\rm GC}/R_{\rm e}=1.25$, respectively. Within uncertainties, the estimated $R_{\rm GC}/R_{\rm e}$ in this work is consistent with \citet{carlsten2022}, but also indicates that GC distribution is even more compact. \citet{carlsten2022} was based on optical ground-based imaging and therefore the slightly higher values could arise from higher contamination in ground-based GC samples. However, our estimated $R_{\rm GC}/R_{\rm e}$ is consistent with $R_{\rm GC}/R_{\rm e}<1$ found for GC-rich UDGs using the HST. The fact that these objects are GC-rich is the reason that the authors could estimate $R_{\rm GC}/R_{\rm e}$ for these objects.

Furthermore, based on the best-fit S\'ersic function, the stacked radial distribution of GCs (Fig.\,\ref{gc-dwarf-result4}) reaches the background level $(0.5 \pm 0.1)$\,$R_{\rm e}^{-2}$ and $(3.3 \pm 0.3)$\,$R_{\rm e}^{-2}$ for the GCs brighter than $\IE = 23.0$ and $\IE = 24.5$, respectively. On average, this corresponds to 0.04\,$R_{\rm e}^{-2}$ and 0.24\,$R_{\rm e}^{-2}$ GCs for each galaxy. Assuming an average $R_{\rm e} = 1$\,kpc for the dwarf galaxies in the sample, these values convert to 0.04\,${\rm kpc}^{-2}$ and 0.24\,${\rm kpc}^{-2}$. In total, we estimate, 0.4 and 2.1 GCs within 3$R_{\rm e}$ of each dwarf galaxy. The GC background originates from ICGCs and non-GCs. Therefore, the above-mentioned estimations provide an upper limit on the purity of the GC candidates and an upper limit on the contamination from non-GCs for their corresponding magnitude range.

\section{\label{sc:Summary}Summary and conclusions}

In this paper, we have presented the \Euclid ERO data of a field in the Fornax cluster to investigate the potential of \Euclid data for studying GCs, in particular around dwarf galaxies.

We have followed a step-by-step careful analysis to maximize the efficiency of source detection and the accuracy of the source photometry. We have identified a catalogue of candidate GCs in the Fornax FoV studied using the \Euclid data by modelling the PSF, performing source detection, and finally aperture photometry of all the detected sources. We then applied cuts in colour and compactness to select the most likely sources for our final GC candidate catalogue. We estimated the completeness of GC detection by injecting artificial GCs into the data. 

Additionally, we assessed the performance of the GC detection using the spectroscopically known GCs, as well as GC candidates of the ACSFCS. Overall, we recover approximately 80\% of GCs in $\IE$ and $\YE$ down to magnitude $\IE=25.2$. This is 1.7\,mag fainter than the typical turn-over magnitude of the GCLF at $M_V = -7.5$. The completeness limit is driven by the NISP images, which in these early \Euclid observations lacked ancillary data needed for the more precise calibrations that became possible a few weeks later; EWS data will be more complete. 

The purity of the derived GC candidate catalogue is much harder to estimate because there is no easy way to know how many foreground and background contaminants remain after our colour and completeness cuts. Photometrically selected GC samples tend to have contamination and only spectroscopic follow-up will show exactly what the purity of the sample is. Here we assessed the purity of the bright GCs and showed that about 80\% of the GC candidates for GCs brighter than $\IE = 21.5$ are true GCs, based on the $uiK_{\rm s}$ and $gri$ colour-colour diagrams of GC candidates.

We used the final GC candidate catalogues to study in particular the properties of the GCs surrounding dwarf galaxies in the Fornax cluster as well as the spatial distribution of the intracluster GCs. Our main findings are as follows.

\begin{itemize}

\item We identified more than 5000 new GC candidates in the Fornax cluster and within the ERO-F FoV, brighter than $\IE = 25$ ($M_{V} = -6$). This magnitude corresponds to 1.5\,mag fainter than the typical GCLF turn-over magnitude ($M_{V} = -7.5$) at the distance of the Fornax cluster. We showed that the ICGC sample is 70\% complete to this magnitude. Furthermore, we discussed that the identified ICGCs follow a similar distribution to the ICL within the FoV. \\

\item Overall, dwarf galaxies in the Fornax cluster have $N_{\rm GC}$ values consistent with the expectations for dwarf galaxies of their stellar mass. three dwarf galaxies (UDG by definition) seem to have more GCs than the average, being GC-rich. Dwarf galaxies less massive than $M_* = 10^7 M_{\odot}$ ($M_{r}>-13.5$) on average have no GCs, except in a few cases. In particular, we identify a few dwarf galaxies with a stellar mass less than $M_* = 10^6 M_{\odot}$ that have about 1 GC each. \\
    
\item The GCLF of dwarf galaxies do not seem to follow a Gaussian distribution. However, assuming that it is a Gaussian distribution, the GCLF has a turn-over magnitude at about $\IE=23.62 \pm 0.42$ ($M_{V}=-7.38\pm 0.42$) consistent with values in the literature. However, the width of GCLF of dwarf galaxies, $\sigma = 1.00 \pm 0.33$ is larger than the expected value. This broadening of the GCLF arises from bright GCs with $\IE$ between 22 and 22.5 ($M_{V}=-9$ and $-8.5$) which the majority are associated with three of the dwarf galaxies in our sample. Excluding these bright GCs, we find a turn-over magnitude $\IE=23.74 \pm 0.14$ ($M_{V}=-7.75\pm 0.14$) and a narrower Gaussian distribution with $\sigma = 0.52 \pm 0.13$ which is more consistent with the expectations for dwarf galaxies of the same stellar mass regime.\\

\item We find the ratio between GC half-number radius and dwarf galaxies' half-light radius $R_{\rm GC}/R_{\rm e} = 0.84 \pm 0.27$ and $R_{\rm GC}/R_{\rm e} = 0.80 \pm 0.22$ for GCs brighter than $\IE=23.0$ and $\IE = 24.5$, respectively. These numbers imply that in dwarf galaxies, GCs follow the same (or more compact) radial distribution as stars. These results do not support the current view of the GC profile of dwarf galaxies (with $R_{\rm GC}/R_{\rm e}=1.5$) and indicate that the GC radial profile of dwarf galaxies is more compact. Furthermore, we discussed that this is similar to what has been observed for some of the UDGs. However, compared to those UDGs, the majority of dwarf galaxies studied here are poor in GCs (for their stellar mass), while those UDGs host 2--3 times more GCs.\\

\end{itemize}

We showed in this paper that \Euclid imaging is well suited to study dwarf galaxies in order to explore low GC-count systems such as dwarf galaxies and study the \textit{Galaxy-GC-Halo} connection. With the upcoming Euclid Wide Survey that will span 14\,000\,deg$^{2}$ of the sky \citep{EuclidSkyOverview} such studies will be possible for all the nearby galaxies and galaxy clusters besides the Fornax cluster.

\begin{acknowledgements}
\AckERO \AckEC  
TS, KV, AL, and PAD acknowledge support from Agence Nationale de la Recherche, France, under project ANR-19-CE31-0022. 
MC support from the project INAF Exploration of Diffuse Galaxies with Euclid" (INAF-EDGE, 2022, P.I. Leslie K. Hunt).
MAR acknowledges funding from the European Union’s Horizon 2020 research and innovation programme under the Marie Skłodowska-Curie grant agreement No 101066353 (project ELATE).
AV and MP funding from the Academy of Finland grant n:o 347089.
DM thanks the Fundacti\'on Jes\'us Serra visiting programme and acknowledges financial support from PRIN-MIUR-22, CHRONOS (PI S. Cassisi).
CT acknowledges the INAF grant 2022 LEMON (Italy).
JHK acknowledges grant
PID2022-136505NB-I00 funded by MCIN/AEI/10.13039/501100011033 and EU, ERDF.
This research has made use of the SIMBAD database (\citealp{simbad}), operated at CDS, Strasbourg, France, the VizieR catalogue access tool (\citealp{vizier}), CDS, Strasbourg, France (DOI : 10.26093/cds/vizier), and the Aladin sky atlas (\citealp{aladin1,aladin2}) developed at CDS, Strasbourg Observatory, France and SAOImageDS9 (\citealp{ds9}). This work has been done using the following software, packages and \textsc{python} libraries: 
\textsc{Topcat} (\citealp{Topcat2005}); 
\textsc{Numpy} (\citealp{numpy}), \textsc{Scipy} (\citealp{scipy}), \textsc{Astropy} (\citealp{astropy}), \textsc{Scikit-learn} (\citealp{scikit-learn}). 
\end{acknowledgements}


%
%

\bibliography{mybib_EROF, Euclid, EROplus}

%

\begin{appendix}

\onecolumn 

\section{Fitting the GCLF using individual measurements.}
\label{ap:gclf}

\begin{figure*}[htb!]
\begin{center}
\includegraphics[clip=,width=0.7\textwidth]{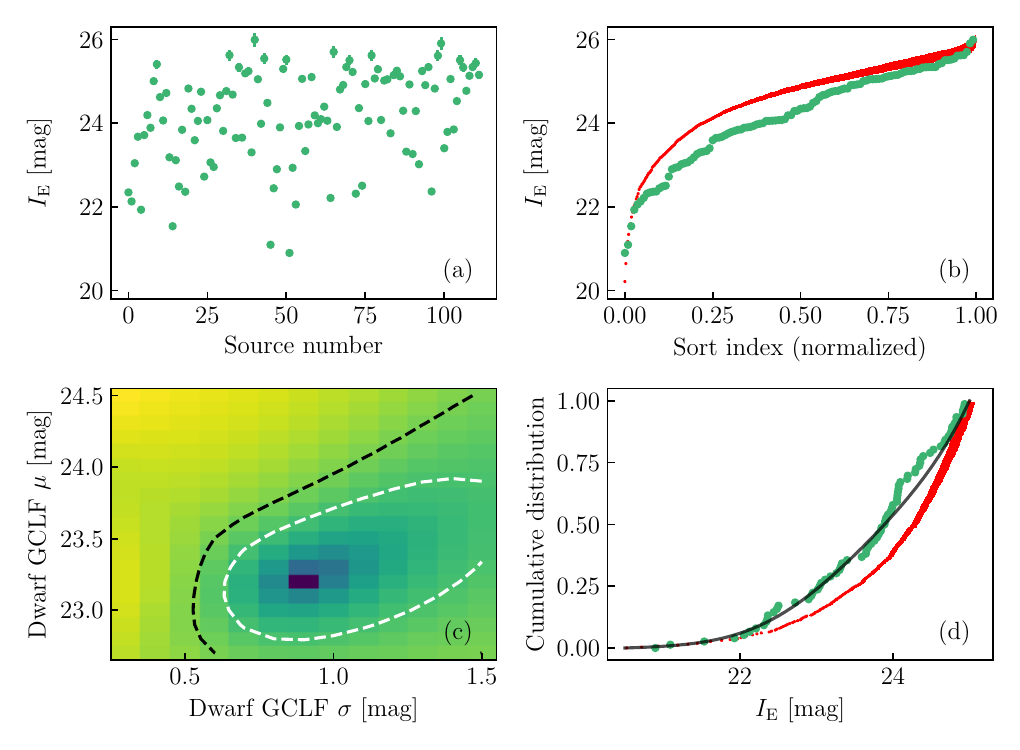}
\end{center}
\caption[]{Data set of individual $\IE$ magnitudes of GC candidates and fits to the GCLF. (a): photometry of objects within $2\,R_{\rm e}$ of the centres of 14 dwarf galaxies. (b): same data after sorting (green), compared with the sorted magnitudes of all the sources in background rings around dwarf galaxies (red). (c): map of posterior probabilities obtained when, for each $(\mu,\sigma)$-pair, the value of $\beta$ that maximizes the posterior probability is retained. The white and black contours correspond, respectively, to an e-fold and a 10-fold decrease in probability compared to the best fit. (d): cumulative distributions of observed magnitudes together with the cumulative distribution of the best-fit model (black line).
}
\label{figApp:GCLF_4panels}
\end{figure*}

\begin{figure*}[htb!]
\begin{center}
\includegraphics[clip=,width=0.7\textwidth]{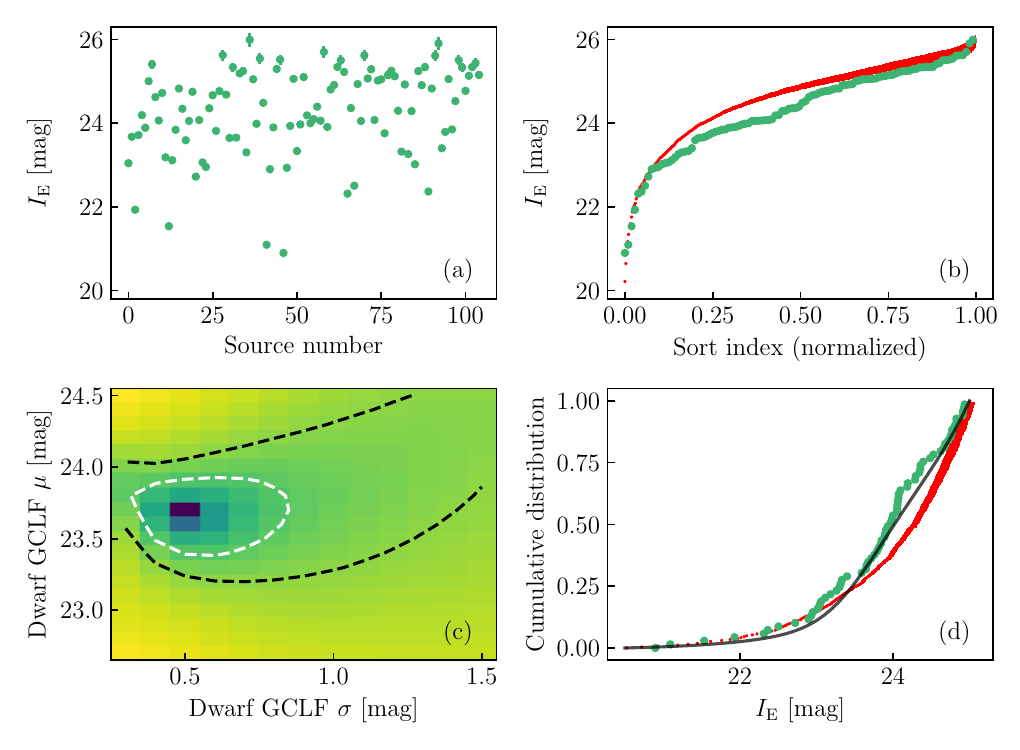}
\end{center}
\caption[]{Similar to Fig.\,\ref{figApp:GCLF_4panels} but after removing about two-thirds of the GC candidates (7) between $\IE = 22$ and $22.5$.
}
\label{figApp:GCLF_4panels+}
\end{figure*}

We fit the stacked distribution of $\IE$ magnitudes of GC-candidates located within $2\,R_{\rm e}$ of dwarf galaxy centres, with two components: a Gaussian with mean magnitude $\mu$ and standard deviation $\sigma$, and an exponential that matches the distribution of $\IE$ of GC-candidates in background areas around the dwarfs down to $\IE=25$. The background counts are already the result of nature combined with the completeness function $c(m)$ of Fig.\,\ref{completeness-all} (top panel). The Gaussian model of the dwarf component of the GCLF must still be multiplied by $c(m)$. The resulting probability for any random GC to be detected within $\text{d}m$ of magnitude $m$ is written 

\begin{equation}
p(m)\,\text{d}m \ = \ \ \underbrace{ \left[ \, 
   c(m)\,\exp \frac{-(m-\mu)^2}{2\sigma^2}
   \ + \ \beta \,\exp \frac{(m-25)}{1.0} \right]}_{{\textstyle f(m;\, \mu,\sigma, \beta)}}\ \text{d}m 
   \ / \ N(\mu,\sigma,\beta),
   \label{eq1}
\end{equation}
where $ N(\mu,\sigma,\beta) $ is the integral 
of $f(m;\, \mu,\sigma, \beta)$ over the full
range of magnitudes $m$ that are being considered (in practice $20.5 \leqslant \IE \leqslant 25$),
and 
\begin{equation}
c(m) = 0.5 \, \frac{0.86-1.0\,(m-26.31)}{%
\sqrt{1+1.0^2\,(m-26.31)^2} }.
   \label{eq2}
\end{equation}
The denominator of the argument of the second exponential in $p(m)$ results from a fit, and varying it by $\pm$5\,\% has little effect on results. Similarly, the factors $1.0$ in $c(m)$ result from a fit (see Eq.\,\ref{eq:Fleming}). The photometric measurements have errors, the standard deviations $s(m)$ of which we represent with 
\begin{equation}
s(m) = 0.01 + 0.07 \exp\, \left[ (m-25)/1.2 \right].
\label{eq3}
\end{equation}
That function implements a minimum uncertainty of $0.01$ at magnitudes brighter than about $21.5$ then a progressive increase so that the $1\,\sigma$ uncertainties amount to $0.05$ at $\IE=24.2$ and $0.1$ at $\IE=25.3$. The stacked $\IE$ distribution in the magnitude range of interest contains $N$ objects with individual magnitudes $m_i$ and individual uncertainties $s_i$. For a given $(\mu, \sigma, \beta)$, the probability of observing one object within $s_i$ of $m_i$ is evaluated as 
\begin{equation}
p_i = \int_{m_i-s_i}^{m_i+s_i} p(m)\,\text{d}m,
\label{eq4}
\end{equation}
and, assuming independence between the data points, the likelihood of the data set is 
$\mathcal{L}(\mu, \sigma, \beta) = \prod_{i} p_i$.,A maximum likelihood fit would assume flat priors for $\mu$, $\sigma$ and $\beta$. However, the GC-candidate counts on dwarf galaxies and around them provide prior information on $\beta$. In the background areas, we have 3718 GC candidates; in the on-dwarf areas, which cover 50 times less sky, 
we counted 112. After renormalisation, we find that the on-dwarf areas typically contain 
$3718 / 50 = 74$ contaminants and $38$ true dwarf members. The ratio is
$ R_{\text{ref}} = 74 / 38 = 1.95$. We can allow for stochasticity roughly by assuming Poisson statistics.
$$ \frac{\delta R}{R}  \simeq \frac{\sqrt{78}}{78} + \frac{\sqrt{38}}{38} \simeq 0.27
 \quad \text{hence} \quad \delta R \, \simeq \, 0.27 \, R_{\text{ref}} = 0.53$$
If we let $G$ denote the integral of the first term of $f(m)$, and $\beta B$ the integral of the second 
term, the ratio $\beta\, B/G$ should not be too far from $R_{\text{ref}}$. This defines a first guess for $\beta$ for any choice of $\mu$ and $\sigma$\,:
$\overline{\beta} = 1.95 \ G/B$. 
The prior on $\beta$ is modelled with a Gaussian probability distribution,
$$ p(\beta \, | \, \mu,\sigma) \propto 
  \frac{1}{\sqrt{2\pi \sigma_{\beta}^2}} \ \exp \left[ - \frac{(\beta-\overline{\beta} )^2}{2\sigma_{\beta}^2} \right], $$
truncated to keep only $\beta>0$, and normalized. We adopt
$\sigma_{\beta} = 0.53 \ G/B$. For our fits, we maximize a posterior probability that accounts for this prior.

The data and an example of the fit results are shown in Fig.\,\ref{figApp:GCLF_4panels}. The same figure for results after removing two-thirds of the GC candidates within $\IE=22$ and $22.5$ is shown in Fig.\,\ref{figApp:GCLF_4panels+}. In these figures, panel (a) presents the photometry of objects within $2\,R_{\rm e}$ of the centres of 14 dwarf galaxies, with the errors just described. Panel (b) shows the same data after sorting and compares that distribution with the sorted magnitudes of all the sources in background rings around dwarf galaxies. Panel (c) shows the map of posterior probabilities obtained when, for each $(\mu,\sigma)$-pair, the value of $\beta$ that maximizes the probability is retained. We checked that the range explored for $\beta$ was sufficient, by verifying that it does not reach the extremes defined in the code for any $(\mu,\sigma)$-pair in the region of interest. The white contour corresponds to an $e$-fold decrease in likelihood compared to the best fit, the black contour a 10-fold decrease. Finally, panel (d) shows the cumulative distributions of observed magnitudes together with the cumulative distribution of the best-fit model. The best fits are not very good fits: a KS-test returns a $p$-values around 0.26 and 0.30, respectively (note that the KS-test does not take into account photometric errors). These $p$-values mean that deviations from the modelled cumulative distribution as large as observed would result from random realization more than one time out of four; in other words, the Gaussian model for the dwarf GCLF cannot be safely excluded.

\begin{figure*}[htb!]
\begin{center}
\includegraphics[clip=,width=0.8\textwidth]{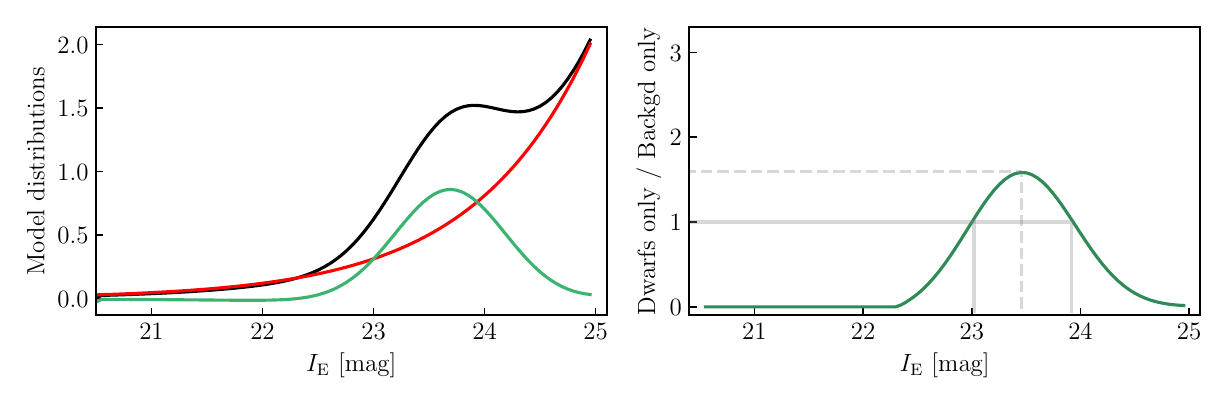}
\end{center}
\caption[]{Left panel: inferred magnitude distributions, for the fit carried out within the range $20.5 < \IE < 25$ excluding two-thirds of the bright GCs (as in Fig.\,\ref{figApp:GCLF_4panels+}). 
The distributions for all GC candidates, for the background GCs (exponential) and for the dwarf-galaxy GCs (Gaussian) are shown in black, red, and green, respectively.
Right panel: ratio of the dwarf GCLF to the background GCLF in the model, with ratios of 1 and 1.6 highlighted.
}
\label{figApp:contam}
\end{figure*}

The inferred magnitude distributions for the background objects and the dwarf GCs are shown separately in Fig.\,\ref{figApp:contam}, for the same fitting parameters as in Fig.\,\ref{figApp:GCLF_4panels+} (that is after excluding two-thirds of the bright GCs within $\IE=22.0$ and $22.5$). Although the peak of the dwarf GCLF model occurs at 23.7, the contrast with respect to background sources is highest at about 23.5. We recall that the sample of GC candidates examined here results from selection criteria that are more permissive than, for instance, the selection used to investigate the spatial distribution (see last paragraph of Sect.\,\ref{sc:GCselection}); the modelling done here handles the contamination by explicitly including that component.

\newpage
\section{Colours and colour transformations for old stellar populations}
\label{ap:colors}

\begin{figure*}[htb!]
\begin{center}
\includegraphics[clip=,width=0.495\textwidth]{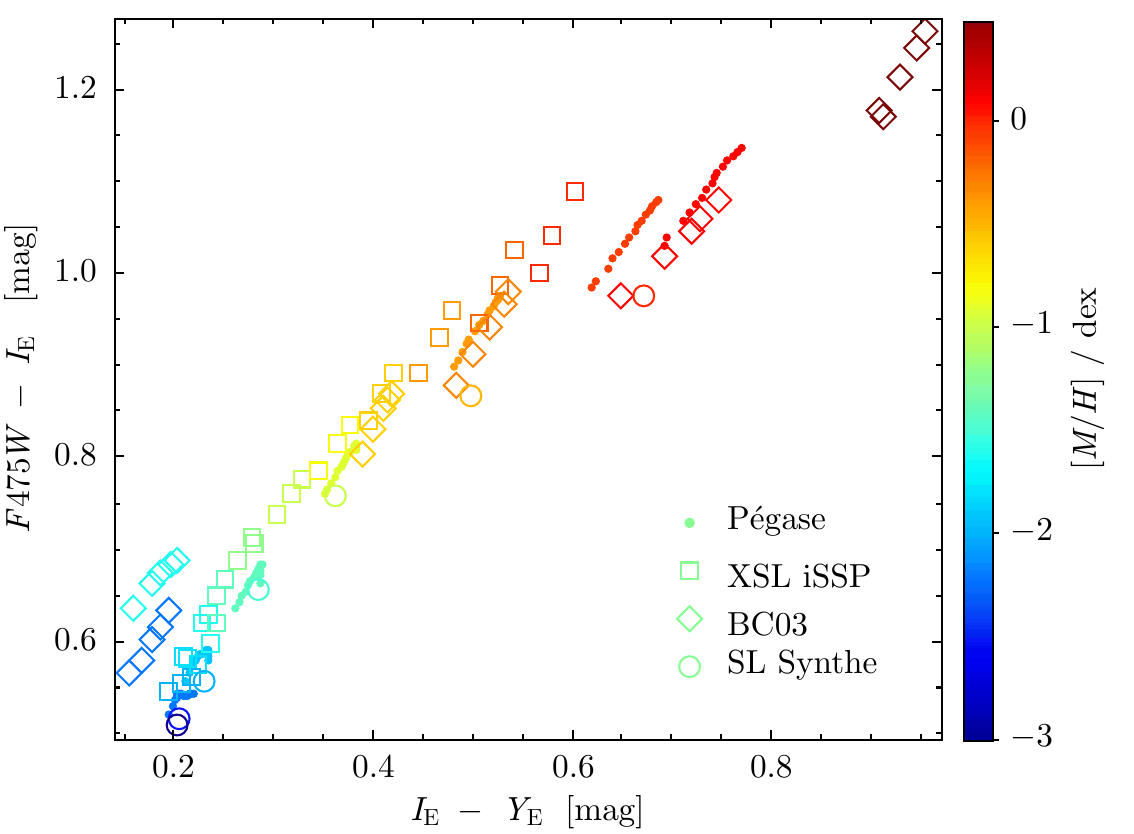}
\includegraphics[clip=,width=0.495\textwidth]{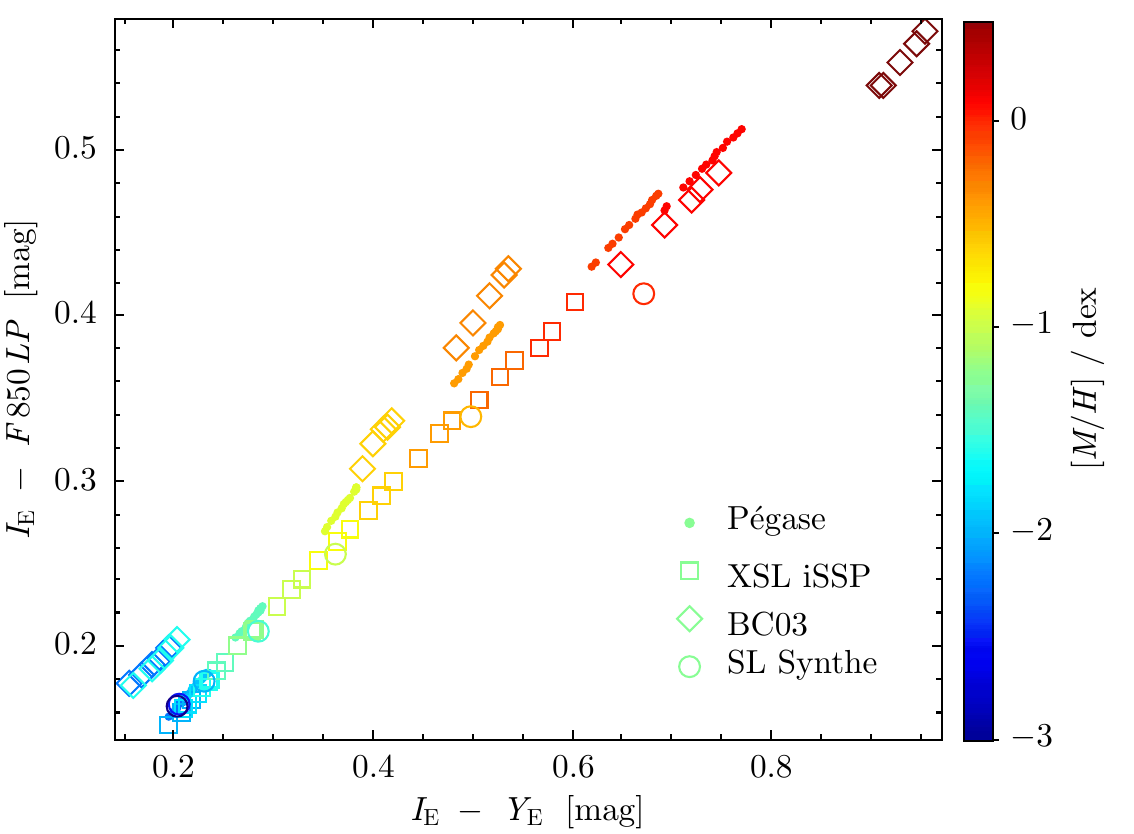}\\
\includegraphics[clip=,width=0.495\textwidth]{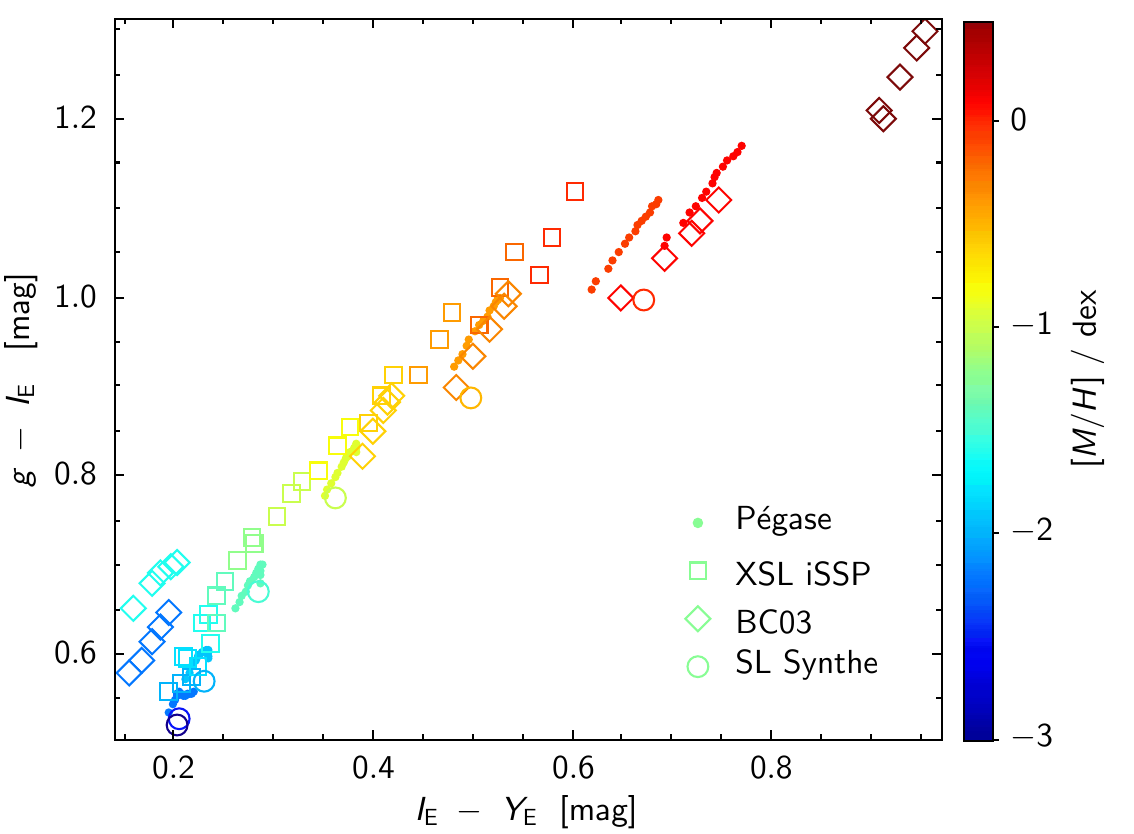}
\includegraphics[clip=,width=0.495\textwidth]{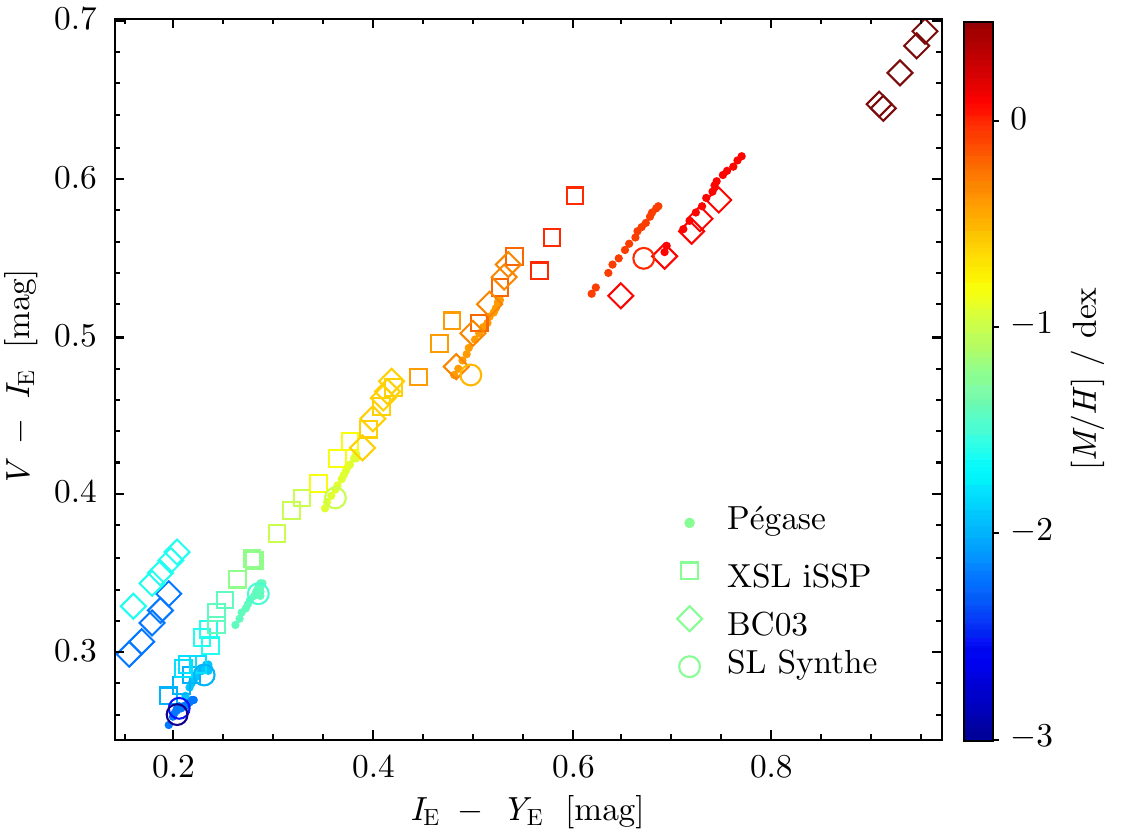}
\end{center}
\caption[]{Colour-transformations based on synthetic photometry carried out on model spectra produced for old single stellar populations by various population synthesis codes (see text for details). The F$475$W and F$850$LP passbands in the top panels are for filters of the Advanced Camera for Surveys on HST, as used in the ACSFCS survey of the Fornax cluster. In the bottom left panel, the $g$ passband is from the OmegaCAM instrument on VST (European Southern Observatory), as used in the FDS survey. The passbands used include the filter, the instrument, and the detector.
}
\label{figApp:coltransf}
\end{figure*}

\begin{figure*}[htb!]
\begin{center}
\includegraphics[clip=,width=0.495\textwidth]{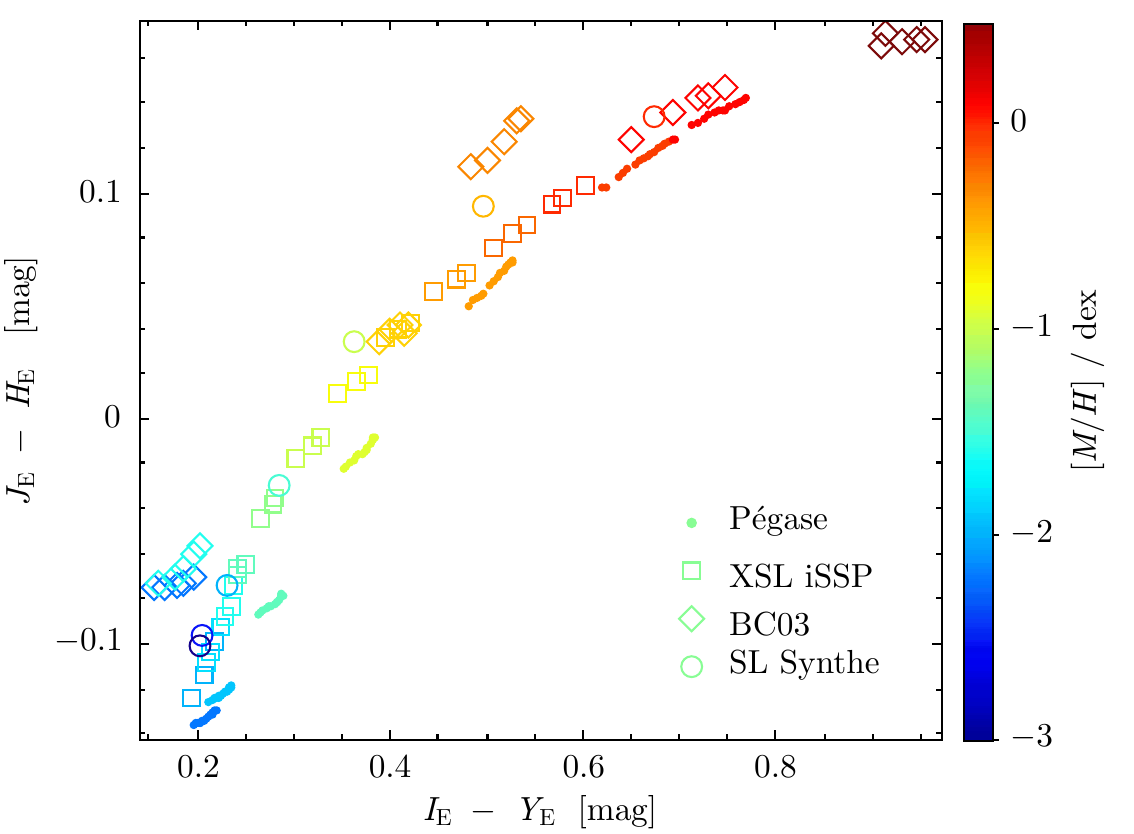}
\end{center}
\caption[]{\Euclid colour-colour plane for the same model stellar populations as in Fig.\,\ref{figApp:coltransf}.
}
\label{figApp:colcol}
\end{figure*}

For convenience, Fig.\,\ref{figApp:coltransf} displays the colour transformations obtained using a variety of population synthesis models, all restricted to single stellar populations with ages older than about 7\,Gyr (the exact sampling of ages depends on the model family). Figure\,\ref{figApp:colcol} displays $\JE - \HE$ vs. $\IE - \YE$, for comparison with the figure used in the main text to implement a colour-selection for GC candidates. The models used and their main ingredients are the following.\\

-- P\'egase (\citealt{FiocRV1997}, hereafter FRV97; \citealt{leborgne2004}): stellar evolution tracks mainly from the Padova group (published 1993--1996), but extended through the thermally pulsing asymptotic giant branch (AGB) and to the post-AGB as described in FRV97; isochrones computed on-the-fly; and BaSeL spectral library \citep{lejeune1997, lejeune1998}.\\

-- XSL iSSP \citep{VerroSSP2022} : PARSEC isochrones version 2S with the TP-AGB extension of the COLIBRI model \citep{marigoCOLIBRI2013,Pastorelli2020}; and empirical  spectra from the X-shooter Spectral Library \citep{VerroXSL2022}. \\

-- BC03 (\citealt{BC03}, 2016 version\footnote{\small \tt http://www.bruzual.org/bc03/Updated\_version\_2016/}): Padova isochrones (1994+); and BaSeL3.1 spectral library \citep{Westera2002}. \\

-- SL Synthe: Models computed using BaSTI isochrones \citep{Hidalgo2018,Pietrinferni2021} and synthetic spectra produced with ATLAS or MARCS stellar atmospheres \citep{Kurucz1970,Gustafsson2008} and the SYNTHE and Turbospectrum spectral synthesis codes \citep{Kurucz1981,Kurucz2005,Alvarez1998}. 
See \citet{soren2022} for details.\\

In future data releases, a refined photometric calibration of \Euclid data over the sky will help discriminate between various families of population synthesis models.

\end{appendix}


\end{document}